\documentclass[aps,prd,12pt,preprint,superscriptaddress,nofootinbib,floatfix]{revtex4-1}
\usepackage{latexsym}
\usepackage{amsfonts}
\usepackage[latin1]{inputenc}
\usepackage{subfig}
\usepackage{graphicx}
\usepackage{mathrsfs}
\usepackage{amssymb}
\usepackage{amsmath}
\usepackage{verbatim}
\def\D0{\slash\!\!\!\!\!\!\!\!\!\:D0}

\begin{document}

\begin{flushleft}
{SHEP-10-40}\\
\today
\end{flushleft}

\title{The Higgs sector of the minimal $B-L$ model \\
at future Linear Colliders}
\author{Lorenzo Basso}
\affiliation{
School of Physics \& Astronomy, University of Southampton,\\
Highfield, Southampton SO17 1BJ, UK}
\affiliation{
Particle Physics Department, Rutherford Appleton Laboratory, \\Chilton,
Didcot, Oxon OX11 0QX, UK}
\author{Stefano Moretti}
\affiliation{
School of Physics \& Astronomy, University of Southampton,\\
Highfield, Southampton SO17 1BJ, UK}
\affiliation{
Particle Physics Department, Rutherford Appleton Laboratory, \\Chilton,
Didcot, Oxon OX11 0QX, UK}
\author{Giovanni Marco Pruna}
\affiliation{
School of Physics \& Astronomy, University of Southampton,\\
Highfield, Southampton SO17 1BJ, UK}
\affiliation{
Particle Physics Department, Rutherford Appleton Laboratory, \\Chilton,
Didcot, Oxon OX11 0QX, UK}
\begin{abstract}
{\small \noindent
We investigate the phenomenology of the Higgs sector of the minimal $B-L$ extension of the Standard Model at a future $e^+e^-$ Linear Collider. We consider the discovery potential of both a sub-TeV and a multi-TeV machine. We show that, within such a theoretical scenario, several novel production and decay channels involving the two physical Higgs states, precluded at the LHC, could experimentally be accessed at such machines. Amongst these, several Higgs signatures have very distinctive features with respect to those of other models with enlarged Higgs sector, as they involve interactions of Higgs bosons between themselves, with $Z'$ bosons as well as with heavy neutrinos. In particular, we present the scope of the $Z'$ strahlung process for single and double Higgs production, the only suitable mechanism enabling one to access an almost decoupled heavy scalar state (therefore outside the LHC range).}
\end{abstract}
\maketitle

\newpage


\section{Introduction}
\label{Sec:Intro}

The Standard Model (SM) of strong and Electro-Weak (EW) interactions has passed most of the experimental and theoretical tests performed so far. Among the very few aspects it fails to explain there is the observed pattern of neutrino masses and mixing angles \cite{Fogli:2005cq} and the way the EW symmetry is broken. The SM means to generate masses for the EW gauge bosons (but the photon) is via EW Symmetry Breaking (EWSB) through the Higgs mechanism. In the minimal realisation
of the SM, the Higgs mechanism employs a complex doublet field and it predicts the presence of one scalar particle: the Higgs boson. Up to now, this is an undetected particle of the SM and its properties are still unknown. Therefore, it is not unreasonable to think at modifications of the scalar sector that are 
still compatible with the precision tests undergone experimentally. The most economical way to modify the scalar sector of the SM is to include one (or more) scalar singlets, either real \cite{O'Connell:2006wi,BahatTreidel:2006kx,Barger:2006sk} or complex \cite{Barger:2008jx}, whose phenomenology at hadronic and leptonic colliders has been studied in great details, as well as their impact on the precision observables, as the latter are able to constrain the viable parameter space of the extended Higgs sectors (see, e.g., \cite{Barger:2007im,Dawson:2009yx} and references therein).


Augmenting the scalar sector solely still does not provide an explanation for the observed pattern of the neutrino masses and mixings. A well motivated framework that remedies such flaws of the SM is its minimal $B-L$ gauged extension \cite{Jenkins:1987ue,Buchmuller:1991ce,Khalil:2006yi,BL_master_thesis}, based on the $SU(3)_C\times SU(2)_L\times U(1)_Y\times U(1)_{B-L}$ gauge group. The latter, with respect to the SM, consists of further three right-handed neutrinos (needed for the cancellation of the anomalies related to the new $U(1)$ group) and an additional complex Higgs singlet, responsible for giving mass to an additional $Z'$ gauge boson (see \cite{Basso:2008iv,Basso:2010jm} for details). By extending the fermion sector, as well as the  scalar sector, this model allows for a dynamical generation of the neutrino masses. The scalar sector is now made of two real {\it CP}-even scalars ($h_1$ and $h_2$, being $h_{1(2)}$ the lighter(heavier) scalar, remnant of the Higgs doublet and singlet fields after EWSB), that will mix together, as already well studied in the minimal extensions of the SM with scalar singlets discussed above. Notice that the presence of new coupled matters (neutrinos and a $Z'$) will alter the properties of the Higgs bosons. Moreover, the scalar mixing angle $\alpha$ is a free parameter of the model, and the light(heavy) Higgs boson couples to the new matter content proportionally to $\sin{\alpha}$($\cos{\alpha}$), i.e., with the complementary angle with respect to the interactions with the SM content -- as in the traditional literature of singlet scalar extended SM.
Finally, it is important to note that in this model ${B-L}$ breaking can take place at the TeV scale, i.e., far below that of any Grand Unification Theory (GUT). This $B-L$ scenario therefore has interesting phenomenological implications at present and future TeV scale colliders \cite{Emam:2007dy,Huitu:2008gf,Basso:2008iv,Emam:2008zz,Basso:2009hf,Basso:2010pe,Basso:2010yz}.

In this work we study the phenomenology of such a model at a future $e^+e^-$ Linear Collider (LC) 
\cite{Abe:2001gc,Abe:2001nnb,Abe:2001npb,Abe:2001nqa,Abe:2001nra,Accomando:1997wt,AguilarSaavedra:2001rg,Ackermann:2004ag},
with a view to outline the machine discovery potential of Higgs bosons. The results shown have been produced using CalcHEP \cite{Pukhov:2004ca}, where standard Initial State Radiation (ISR) functions are implemented, according to the formulae in Refs.~\cite{calchep_man,Jadach:1988gb,Basso:2009hf}. 
The model had been previously introduced via LanHEP \cite{Semenov:1996es} (see 
Refs.~\cite{Basso:2008iv,Basso:2010yz} for an exhaustive description).
We will present production cross sections for the $B-L$ Higgs bosons, 
highlighting the differences with respect to the SM case (and other extended models as well), and we will use these results to introduce new Higgs boson signatures at both an International Linear Collider (ILC) \cite{Djouadi:2007ik} and a Compact LInear Collider (CLIC) \cite{Assmann:2000hg}, with main focus on the gauge-Higgs boson interactions. In general, sub-TeV Centre-of-Mass (CM) energies ($\sqrt{s}=500$ GeV) will be suitable for an ILC, multi-TeV CM energies ($\sqrt{s}=3$ TeV) will be appropriate for CLIC while the case $1$ TeV may be appropriate to both. In all cases, results will be shown for some discrete choices of the $Z'$ mass and of the scalar mixing angle~$\alpha$. Their values have been chosen in each plot to highlight some relevant phenomenological aspects. 


This paper can be seen as the continuation of the work started in Refs.~\cite{Basso:2008iv,Basso:2009hf,Basso:2010pe}, where the authors dealt with the other new sectors of this model (i.e., the $Z'$ gauge boson and the heavy neutrino ones) and relies on the results of Refs.~\cite{Basso:2010jt,Basso:2010jm,Basso:2011na}, where the Higgs parameter space of the minimal $B-L$ model was studied in detail by accounting for all available experimental and theoretical constraints.
In this work we decided not to quote the latter, as we are only interested in portraying the gross features of the scalar sector and to highlight the interplay with the other sectors of the model.

In particular, for the discovery potential study we start from the results of Ref.~\cite{Basso:2009hf}, where the $Z'$ properties at a LC were studied in detail.
With reference to it, and as its continuation, the main result in this paper is the discovery potential analysis for the simplest new production mechanism for scalar particles in the $B-L$ model, the Higgs boson production in association with the $Z'_{B-L}$ boson, with the latter then decaying into muon pairs. Notice that this channel is not suitable for the Large Hadron Collider (LHC) \cite{Basso:2010yz}. Contrary to that study, we will point out here that backgrounds can be effectively reduced, being the $Z'\to  \mu^+\mu^-$ an essential channel, and we will show that a good range in scalar mixing angles can be probed in the first years of LC runs. The discovery potential of the other SM-like channels are not presented, as it is easily deducible from the studies for the SM case by simply rescaling the signal by the appropriate coupling reduction, as we will discuss.

This paper is organised as follows. In the next section 
we present our numerical results. For the most interesting case, the Higgs strahlung off $Z'$ boson, the LC discovery potential will be studied, for the case of single production and for the resonant double Higgs production. Finally, we conclude in section~\ref{Sec:Conclusions}.






\section{Results}\label{Sec:Results}

\begin{figure}[!h]
\scalebox{0.9}{
  \subfloat[]{ 
  \label{LC_h1_500}
  \includegraphics[angle=0,width=0.48\textwidth ]{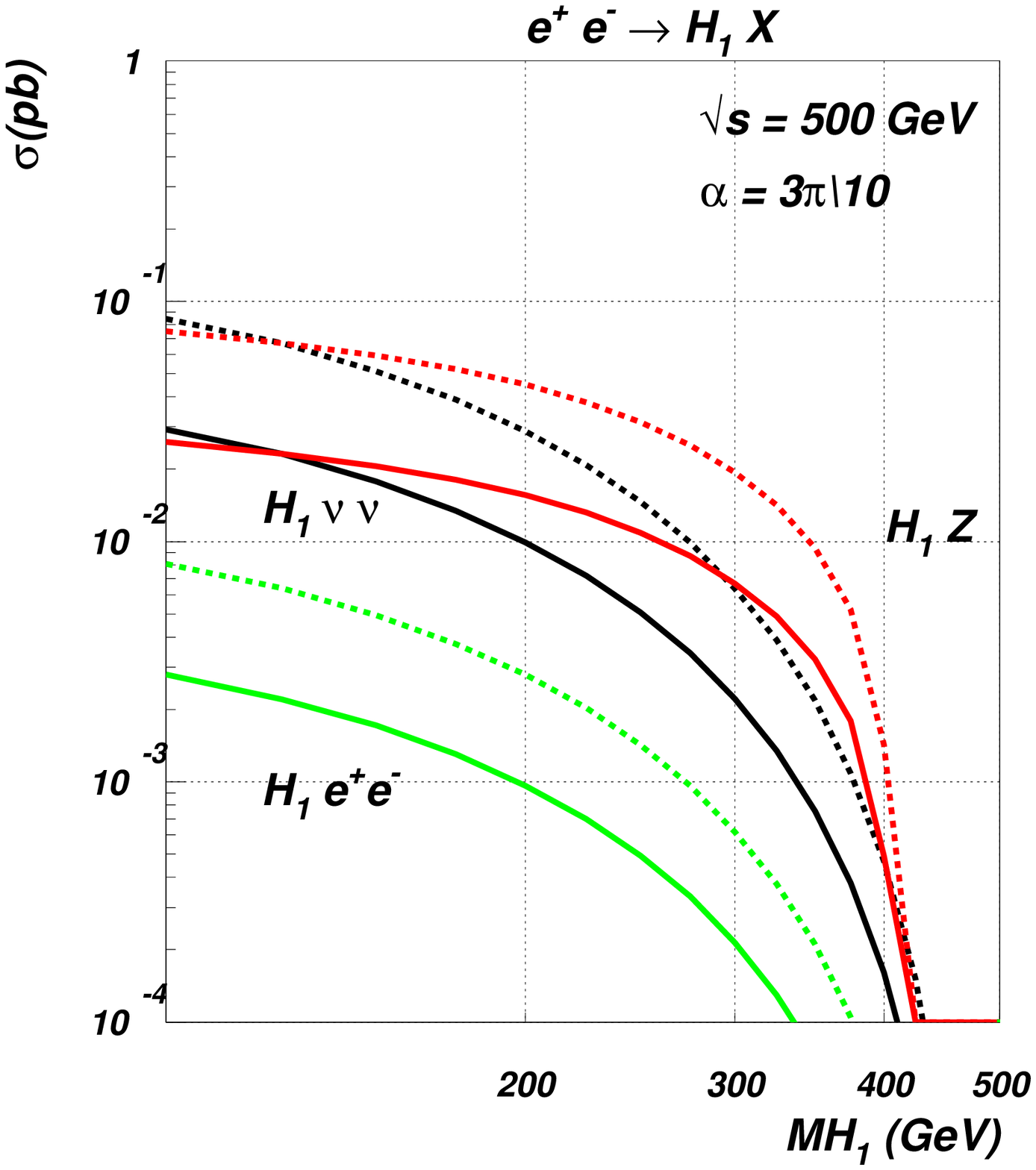}}
  \subfloat[]{
  \label{LC_h2_500}
  \includegraphics[angle=0,width=0.48\textwidth ]{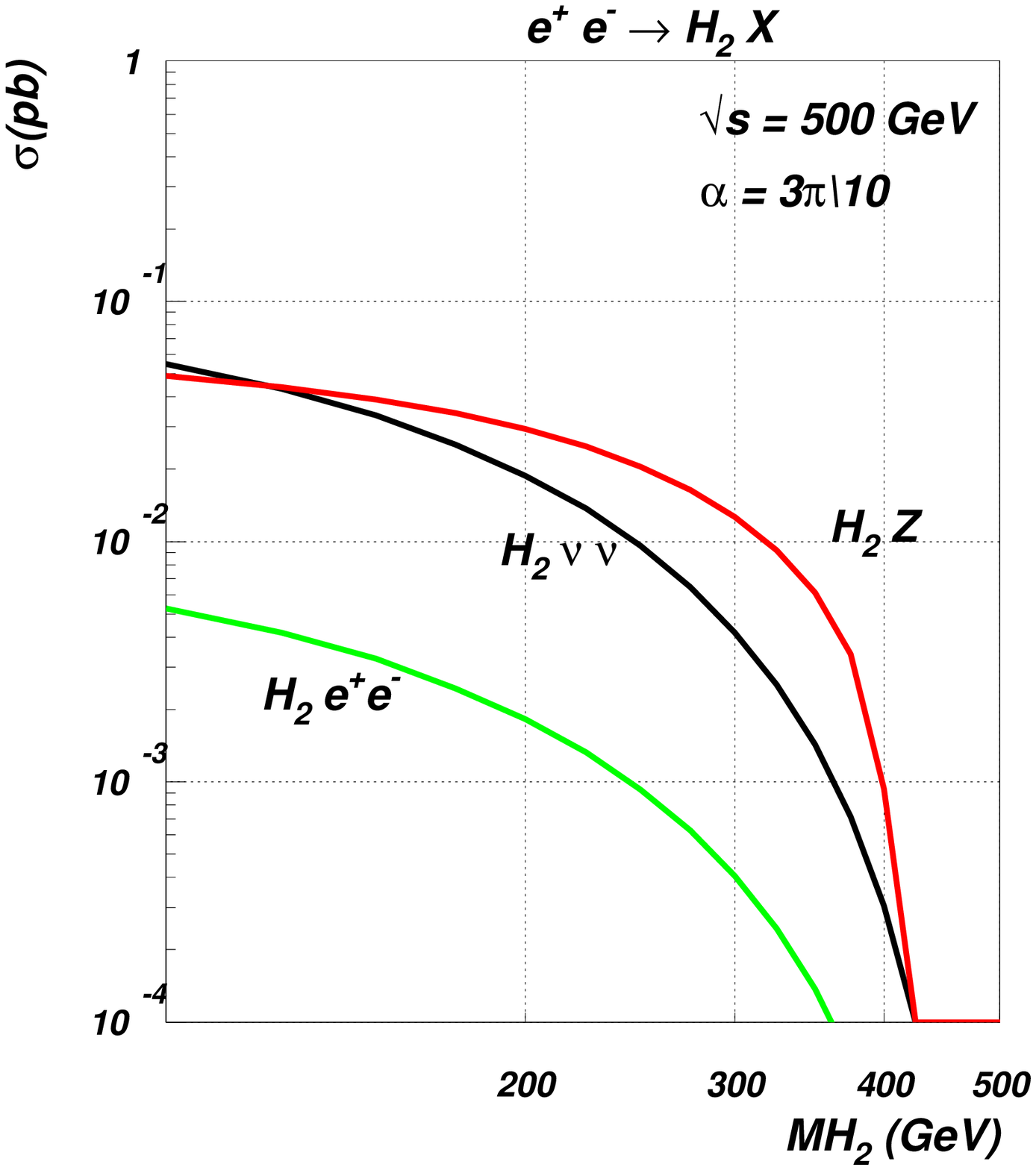}}  }  \\ \vspace{-0.3cm}
\scalebox{0.9}{  
  \subfloat[]{
  \label{LC_h1_1000}
  \includegraphics[angle=0,width=0.48\textwidth ]{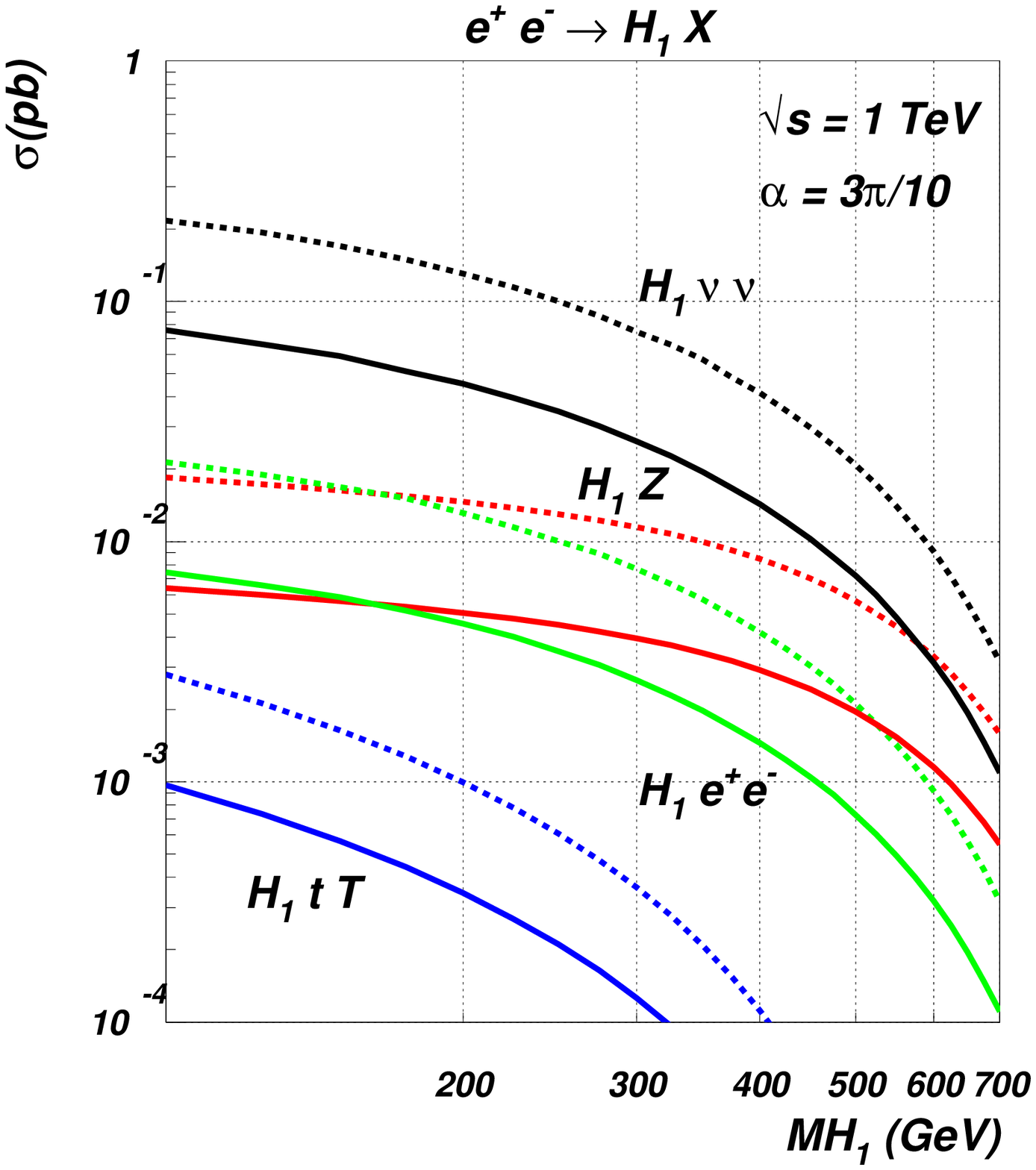}}
  \subfloat[]{
  \label{LC_h2_1000}
  \includegraphics[angle=0,width=0.48\textwidth ]{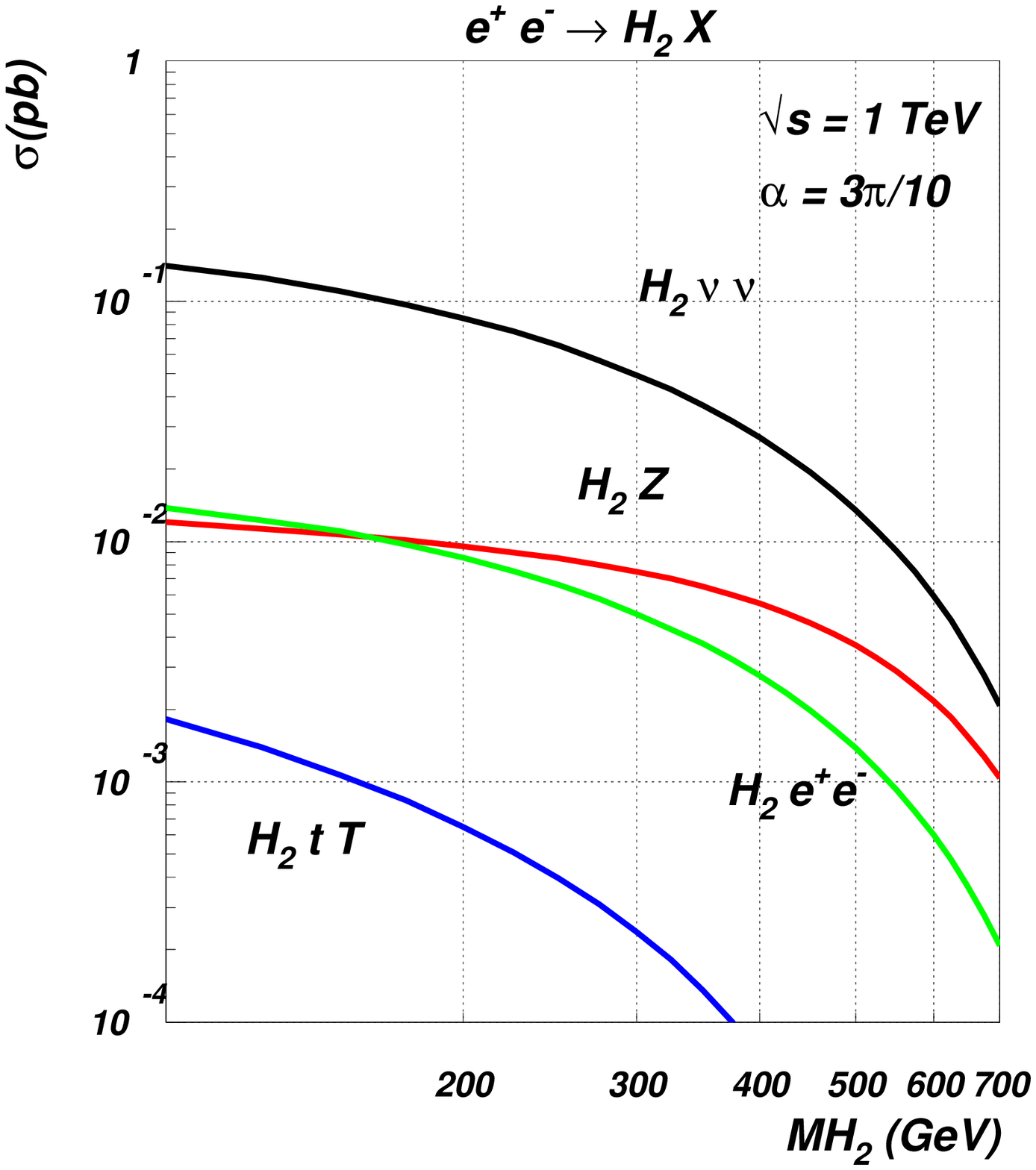}}
}  \\ \vspace{-0.3cm}
\scalebox{0.9}{
  \subfloat[]{
  \label{LC_h1_3000}
  \includegraphics[angle=0,width=0.48\textwidth ]{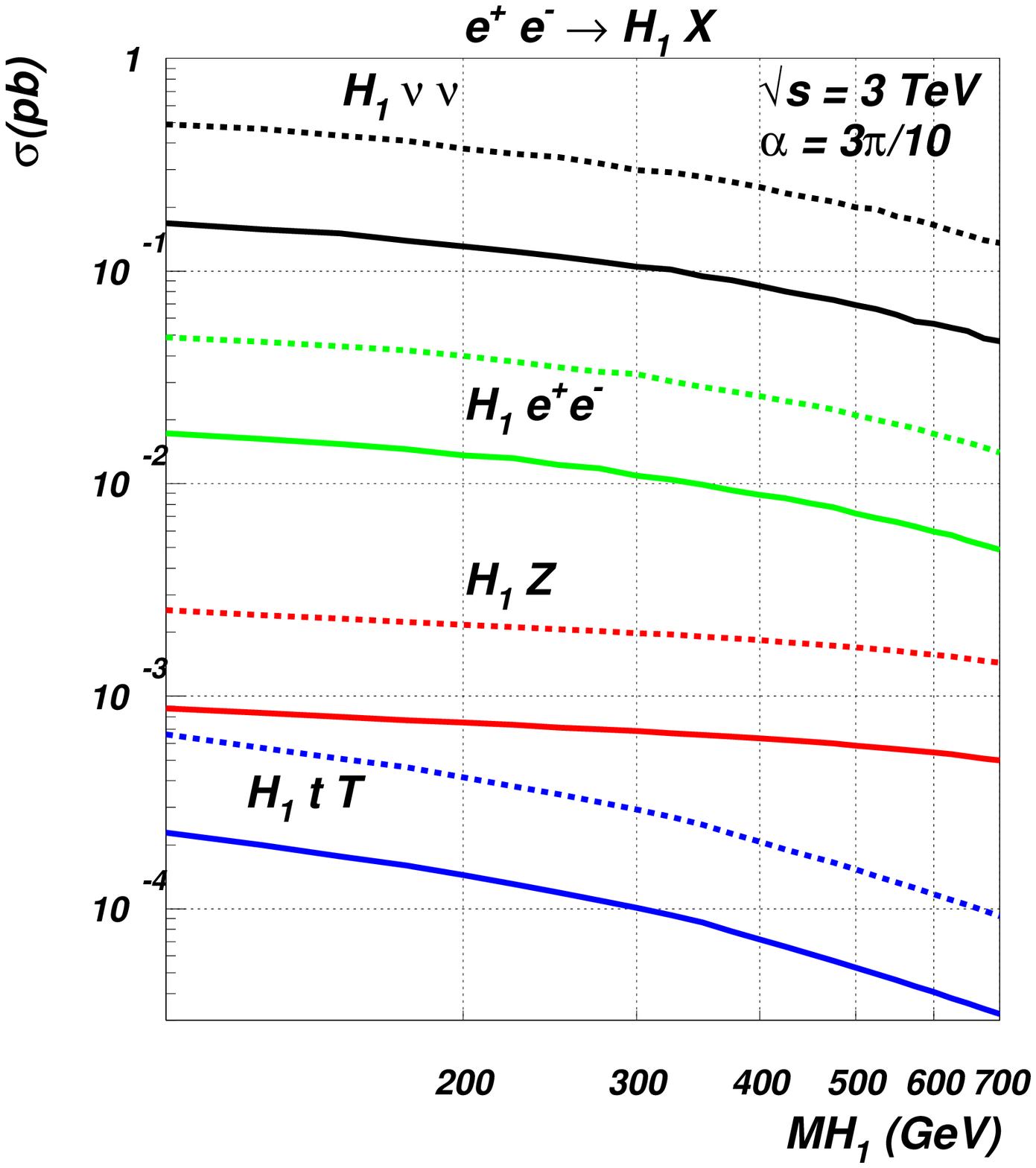}}
  \subfloat[]{
  \label{LC_h2_3000}
  \includegraphics[angle=0,width=0.48\textwidth ]{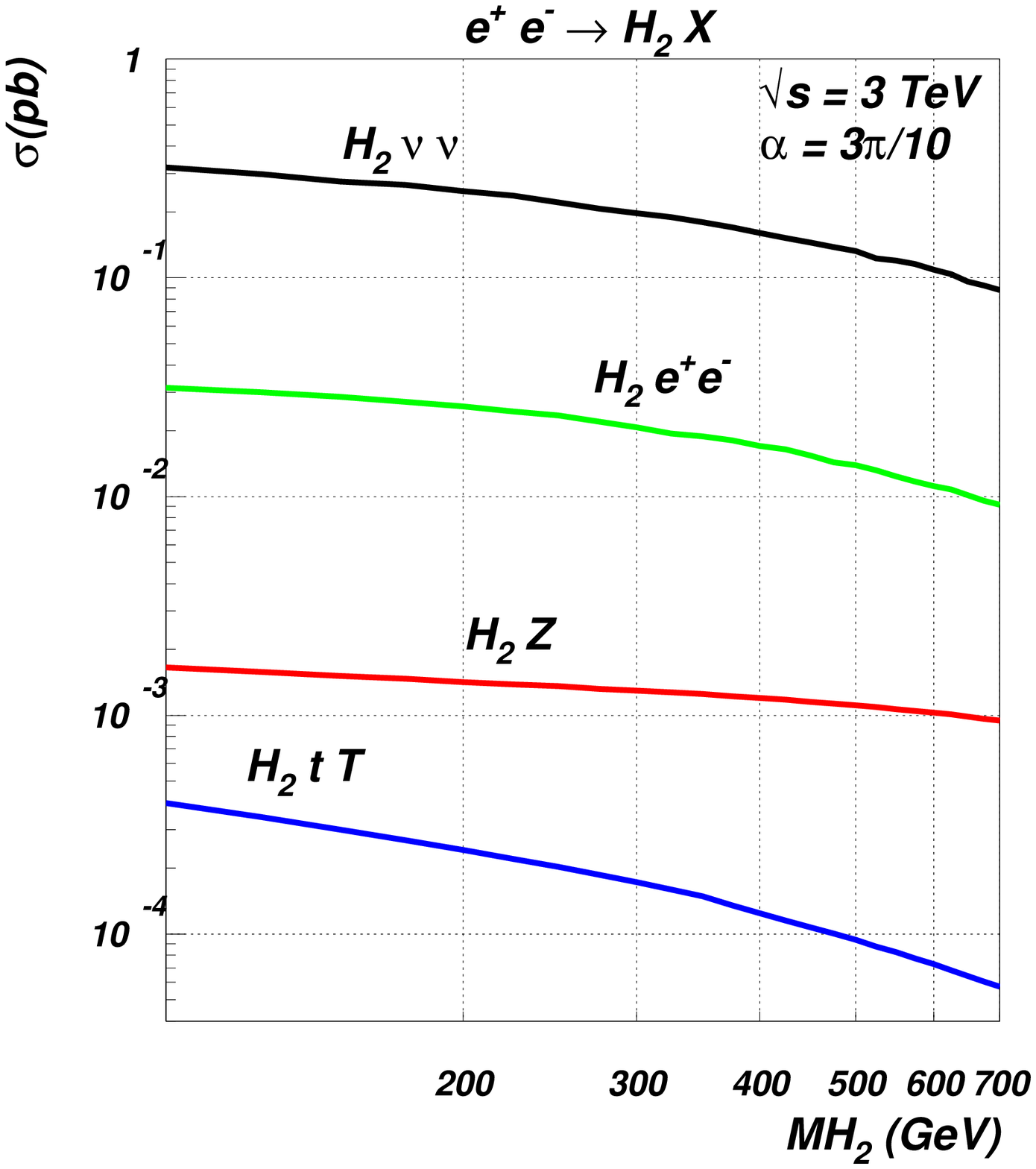}}
}
  \vspace*{-0.5cm}
  \caption{\it Cross sections for the process standard Higgs boson production mechanism as a function of the mass at the LC for $\alpha = 3\pi/10$ for (\ref{LC_h1_500}) $h_1$ and (\ref{LC_h2_500}) $h_2$  at $\sqrt{s}=500$ GeV, for (\ref{LC_h1_1000}) $h_1$ and (\ref{LC_h2_1000}) $h_2$  at $\sqrt{s}=1$ TeV and for (\ref{LC_h1_3000}) $h_1$ and (\ref{LC_h2_3000}) $h_2$  at $\sqrt{s}=3$ TeV. The dashed lines refer to $\alpha =0$. \label{ILC_stand_prod}}
\end{figure}

In this section we present our results for the scalar sector of the $B-L$ model at future LCs. 
Concerning the single Higgs production, we distinguish the standard production mechanisms (via SM gauge bosons, see~\cite{Djouadi:2005gi}) from the novel mechanisms present in the model under discussion (emphasising in particular the role of the $Z'$ gauge boson).
For completeness, figure~\ref{ILC_stand_prod} shows the cross section for the former, i.e., the 
vector boson fusion (via a gauge vector $V=W^\pm,Z)$, Higgs-strahlung from the SM-$Z$ boson and the associated production with a top quark pair, the latter being the least effective production mechanism, with cross sections of few fb at most. However, we will show in section~\ref{sect:NS_single_other} that this channel can be enhanced by the presence of the $Z'$ boson.

These channels are also suitable for producing a pair of Higgs bosons, although with much smaller cross sections. The observation of a Higgs boson pair is crucial to measure parameters of the scalar Lagrangian directly entering in the trilinear and quartic self-couplings \cite{Castanier:2001sf,Baur:2002rb,Baur:2009uw}, although this requires high statistics and large CM energy. Remarkable in this sense is the possible complementarity between the LHC and LCs, as shown in Refs.~\cite{Baur:2003gpa,Plehn:2005nk}. The analysis of the feasibility of these measurements goes beyond the scope of this paper (primarily focused on gauge-scalar interactions), thus this is left for future investigations. 

When dealing with Higgs boson pair production in the $B-L$ model, it should be noted that the $h_2 \rightarrow h_1 h_1$ decay is open for a large portion of the parameter space, contrary to the case of the Minimal Supersymmetric
Standard Model, for instance (where it is important only for very low values of $\tan{\beta}$, region that has been constrained at LEP~\cite{Schael:2006cr}). Hence, light Higgs boson pair production is enhanced by this channel, and not
only because it is resonant, but also because the $h_2-h_1-h_1$ coupling can be large.

\begin{figure}[!h]
  \subfloat[]{ 
  \label{LC_H1H1_1}
  \includegraphics[angle=0,width=0.48\textwidth ]{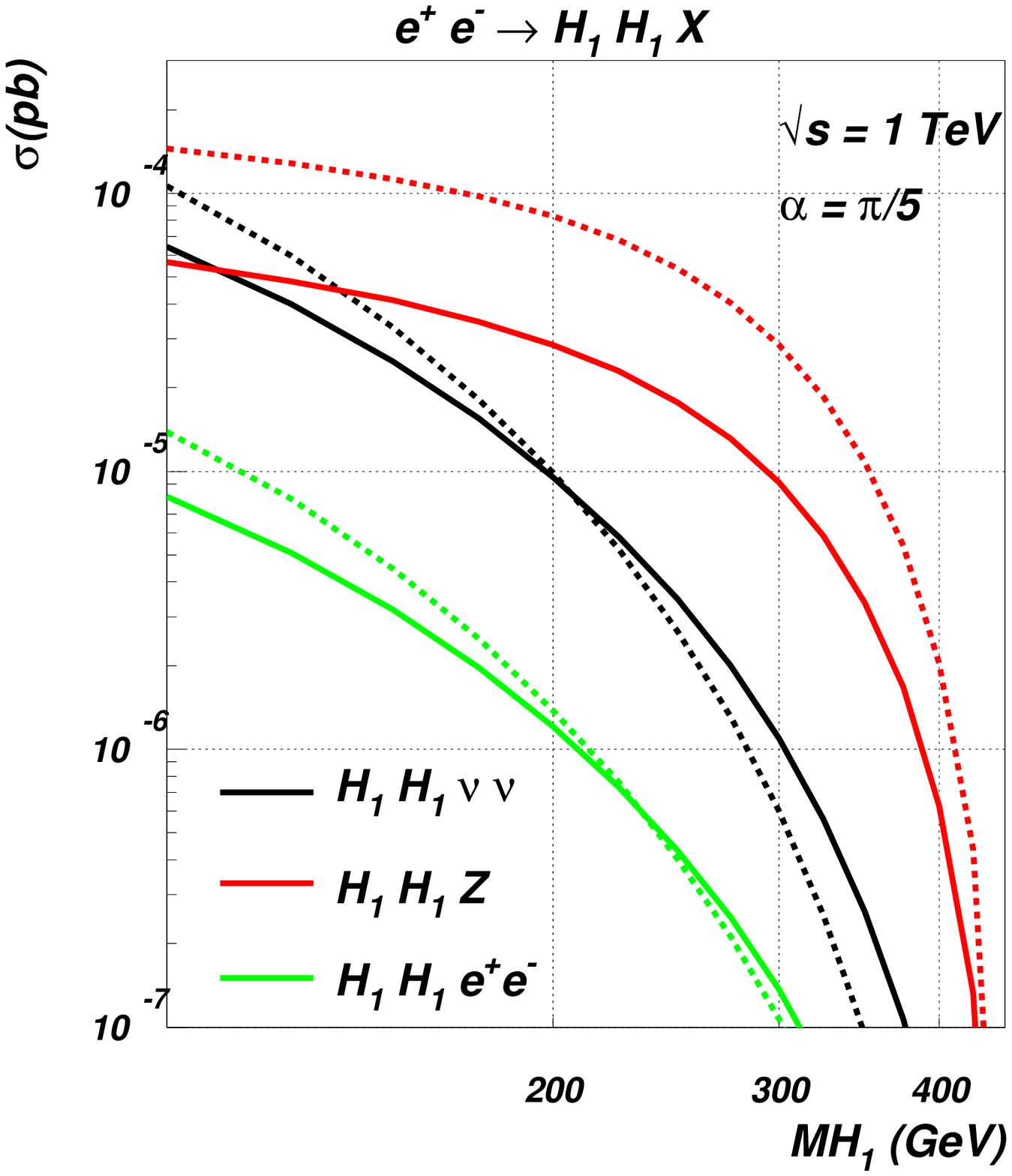}}
  \subfloat[]{
  \label{LC_H2-H1H1_1}
  \includegraphics[angle=0,width=0.48\textwidth ]{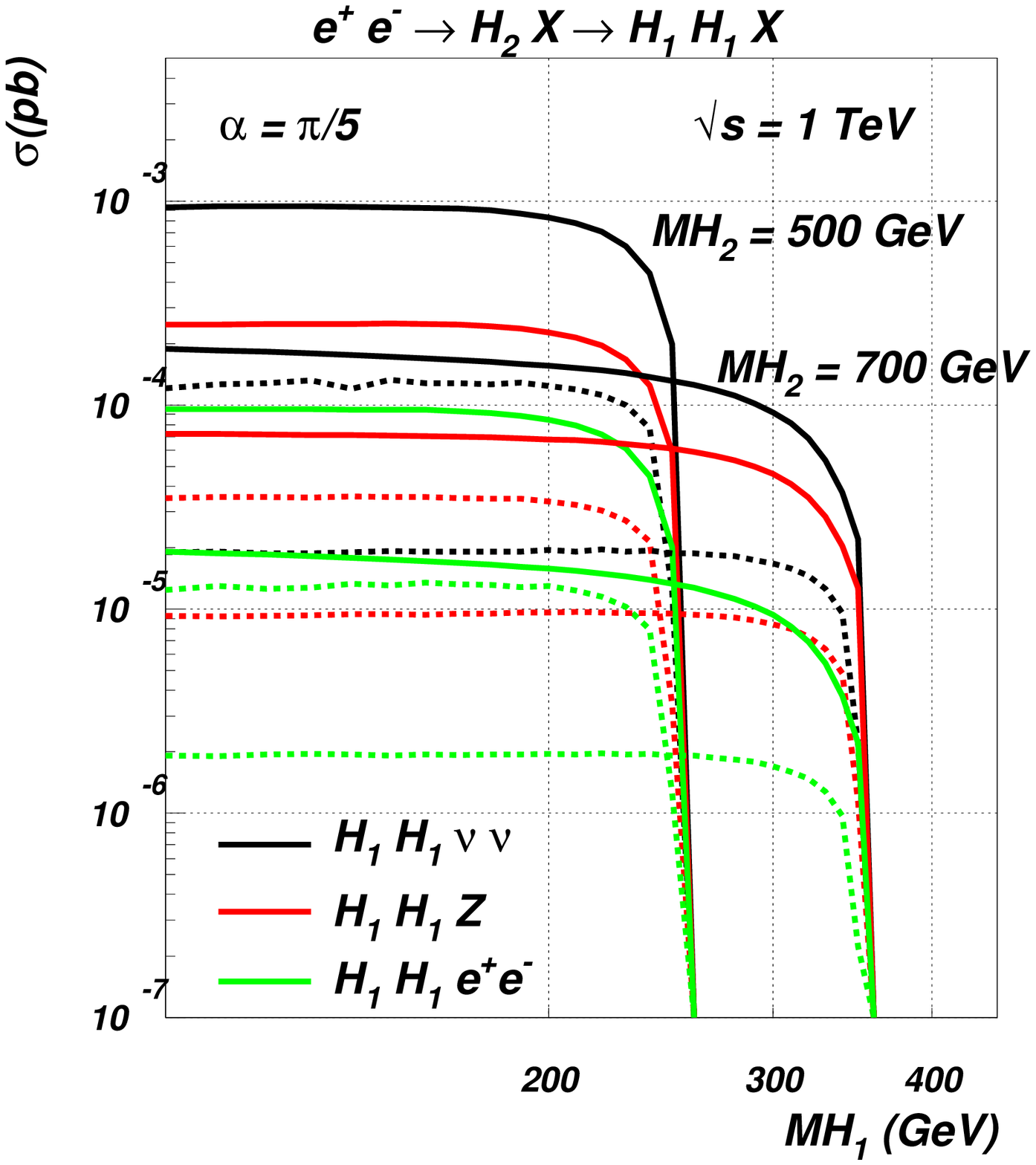}}\\
  \subfloat[]{ 
  \label{LC_H1H1_3}
  \includegraphics[angle=0,width=0.48\textwidth ]{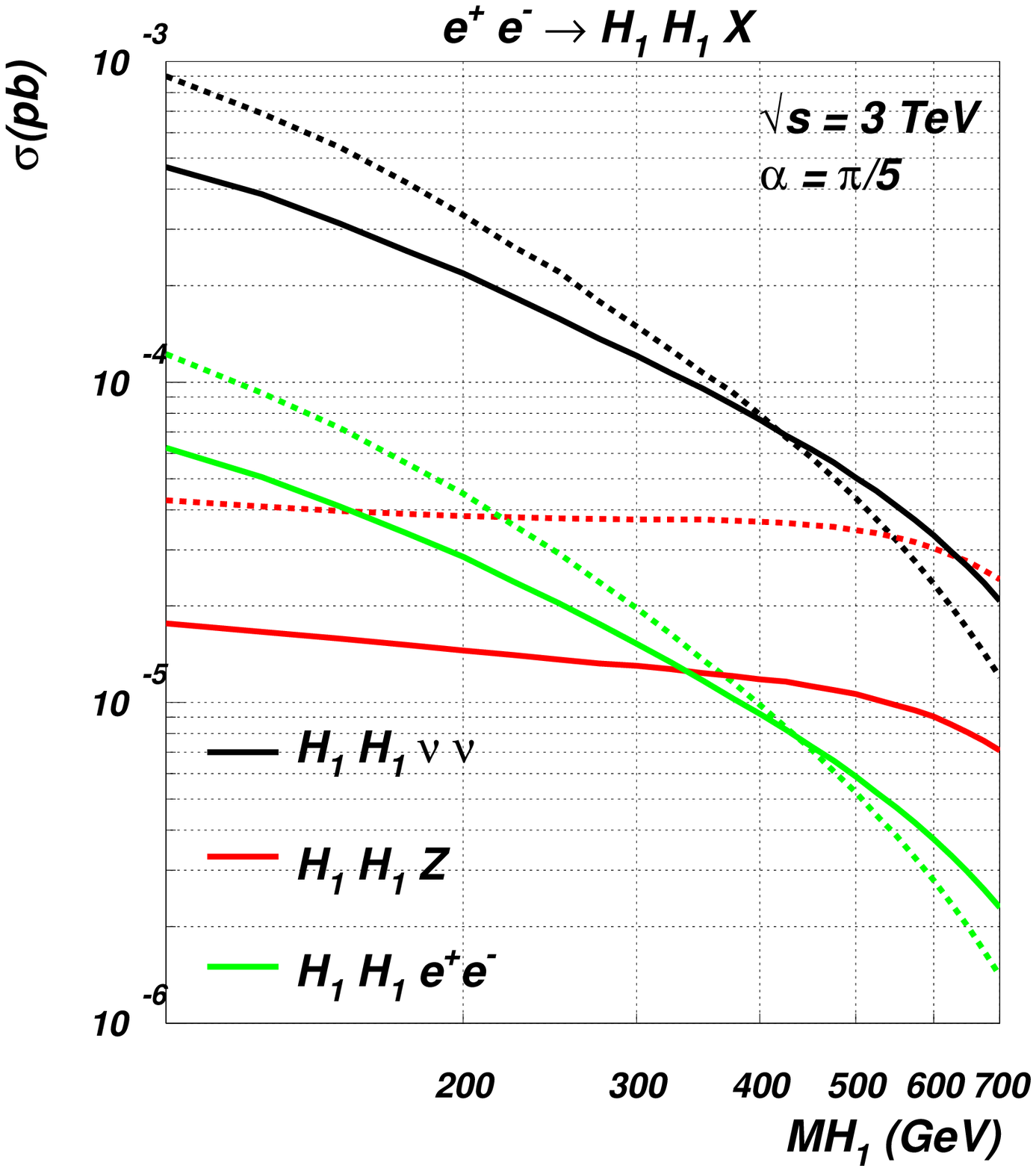}}
  \subfloat[]{
  \label{LC_H2-H1H1_3}
  \includegraphics[angle=0,width=0.48\textwidth ]{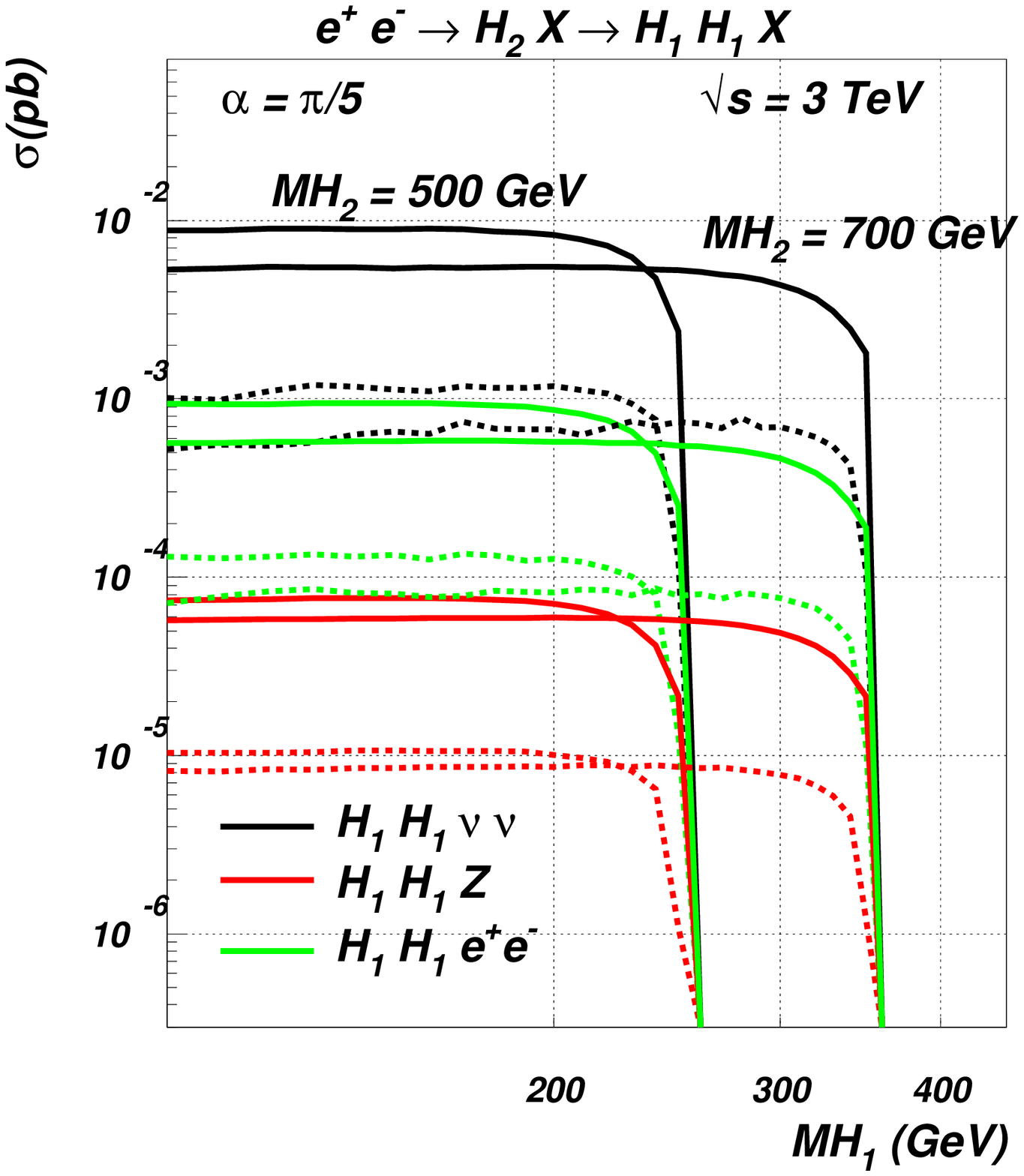}}
  \vspace*{-0.4cm}
  \caption{\it Cross sections for the double light Higgs boson production at the LC (\ref{LC_H1H1_1}) alone and (\ref{LC_H2-H1H1_1}) via $h_2$, for $\sqrt{s}=1$ TeV and (\ref{LC_H1H1_3}) alone and (\ref{LC_H2-H1H1_3}) via $h_2$, for $\sqrt{s}=3$ TeV. The dashed lines in figures~(\ref{LC_H1H1_1}) and (\ref{LC_H1H1_3}) refer to $\alpha =0$, while in figures~(\ref{LC_H2-H1H1_1}) and (\ref{LC_H2-H1H1_3}) they refer to $\alpha =\pi/20$.  \label{ILC_Standard_double}}
\end{figure}

Figure~\ref{ILC_Standard_double} shows the standard production mechanisms of a pair of light Higgs bosons. In the SM case (or when we neglect $h_2$), they are the same mechanisms for the single scalar production when a further Higgs boson is attached.
Noticeably, the light Higgs boson pair can also be originated from the decay of the heavy Higgs boson, with an enhancement of the cross sections of a factor $\mathcal{O}(10-100)$, depending on the Higgs boson masses.
Also, in the latter case the cross sections are constant with $m_{h_1}$ as long as the $h_2 \rightarrow h_1 h_1$ decay is allowed. This is a consequence of having chosen a specific value for $m_{h_2}$ and that BR($h_2 \rightarrow h_1 h_1) \approx 20\% $ is essentially constant for $m_{h_1} > M_W,\, M_Z$ \cite{Basso:2010yz}.


\subsection{Non-standard single-Higgs production mechanisms (and the role of the $\boldsymbol Z'$ boson)}\label{sect:NS_single}

In this section we discuss the novel mechanisms to produce a single Higgs boson (either the light one or the heavy one) in the $B-L$ model. All the new features arise from having a $B-L$-$Z'$ boson (henceforth also denoted as $Z'_{B-L}$) that interacts with both the scalar and fermion sectors, and, in particular, BR($Z' \rightarrow \ell ^+ \ell ^-) \simeq 15\%$, ($\ell = e,\,\mu$), which makes a lepton collider the most suitable environment for testing this model.

%

%
%
\subsubsection{The associated production of a $Z'$ boson and a Higgs boson}

\begin{figure}[!h]
  \subfloat[]{ 
  \label{H1Zp_1TeV}
  \includegraphics[angle=0,width=0.48\textwidth ]{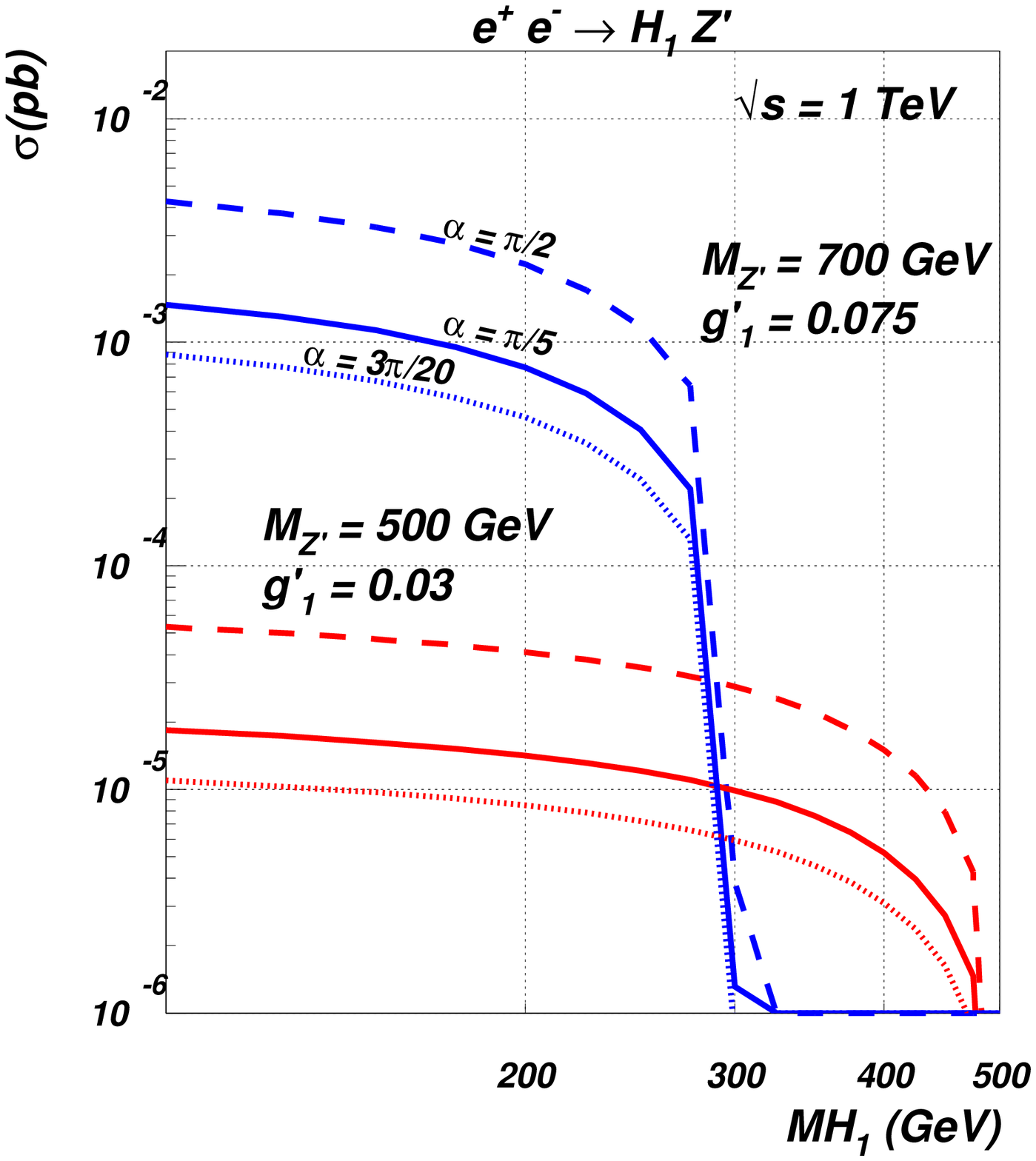}}
  \subfloat[]{
  \label{H2Zp_1TeV}
  \includegraphics[angle=0,width=0.48\textwidth ]{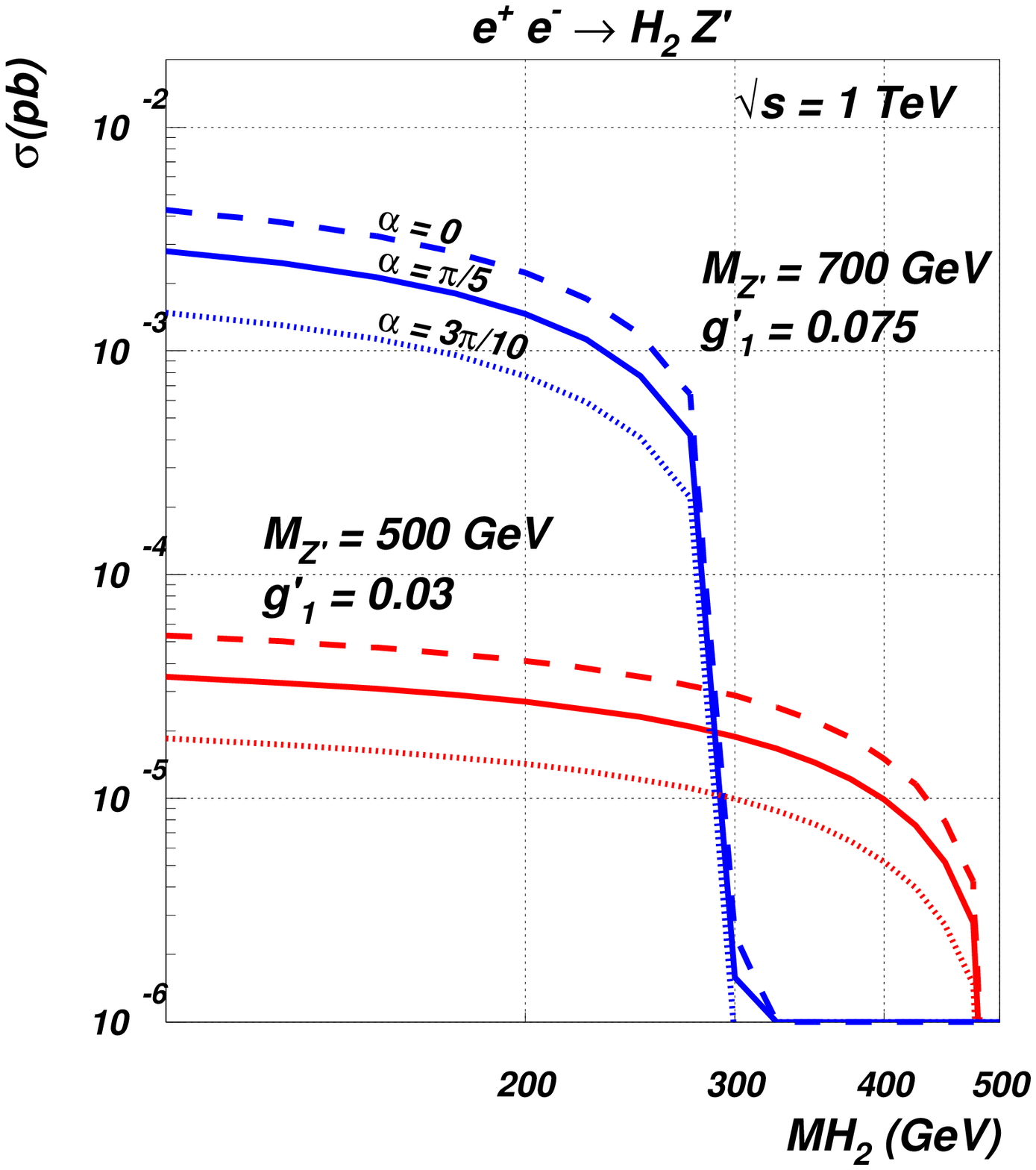}}\\
  \subfloat[]{ 
  \label{H1Zp_3TeV}
  \includegraphics[angle=0,width=0.48\textwidth ]{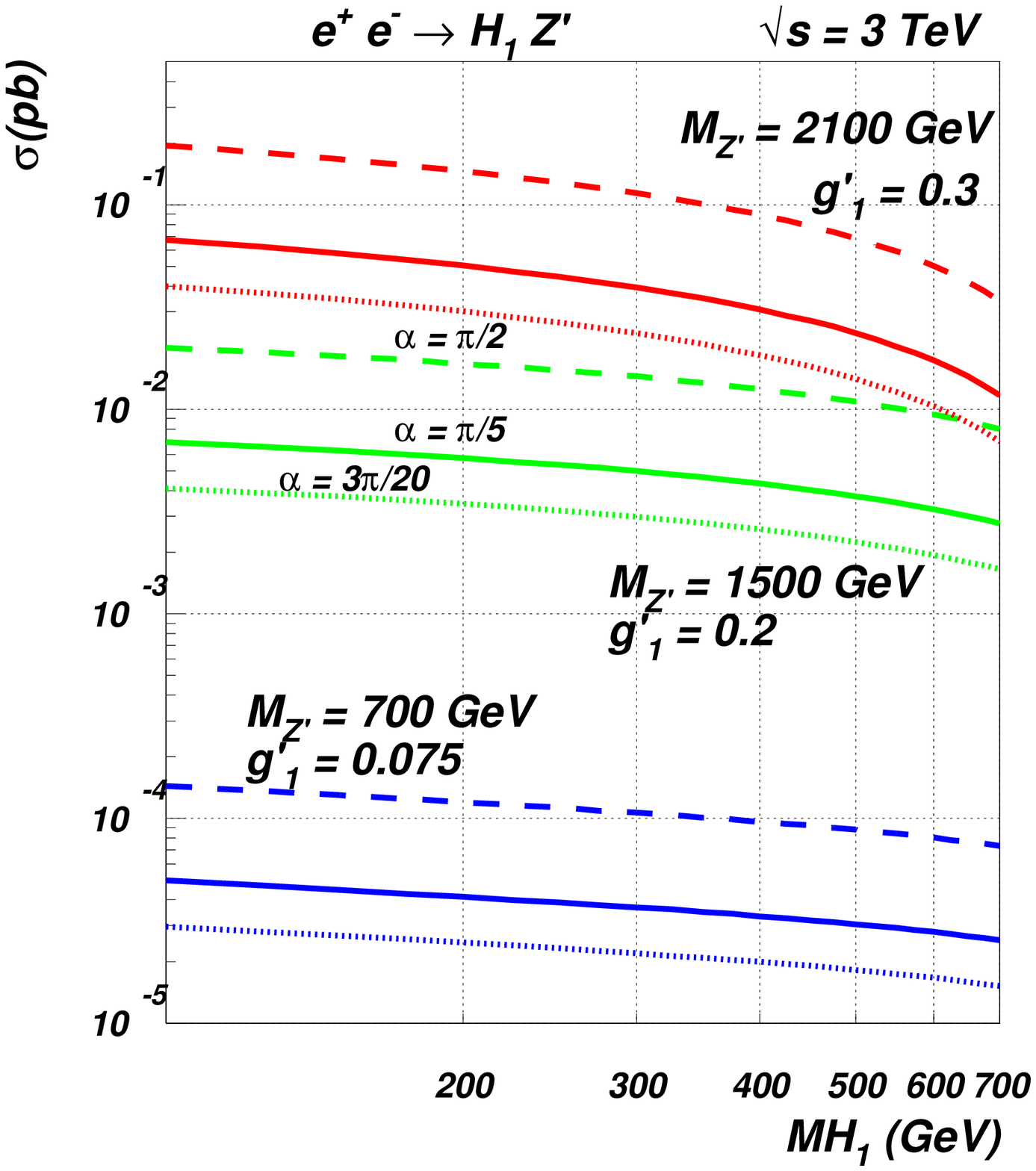}}
  \subfloat[]{
  \label{H2Zp_3TeV}
  \includegraphics[angle=0,width=0.48\textwidth ]{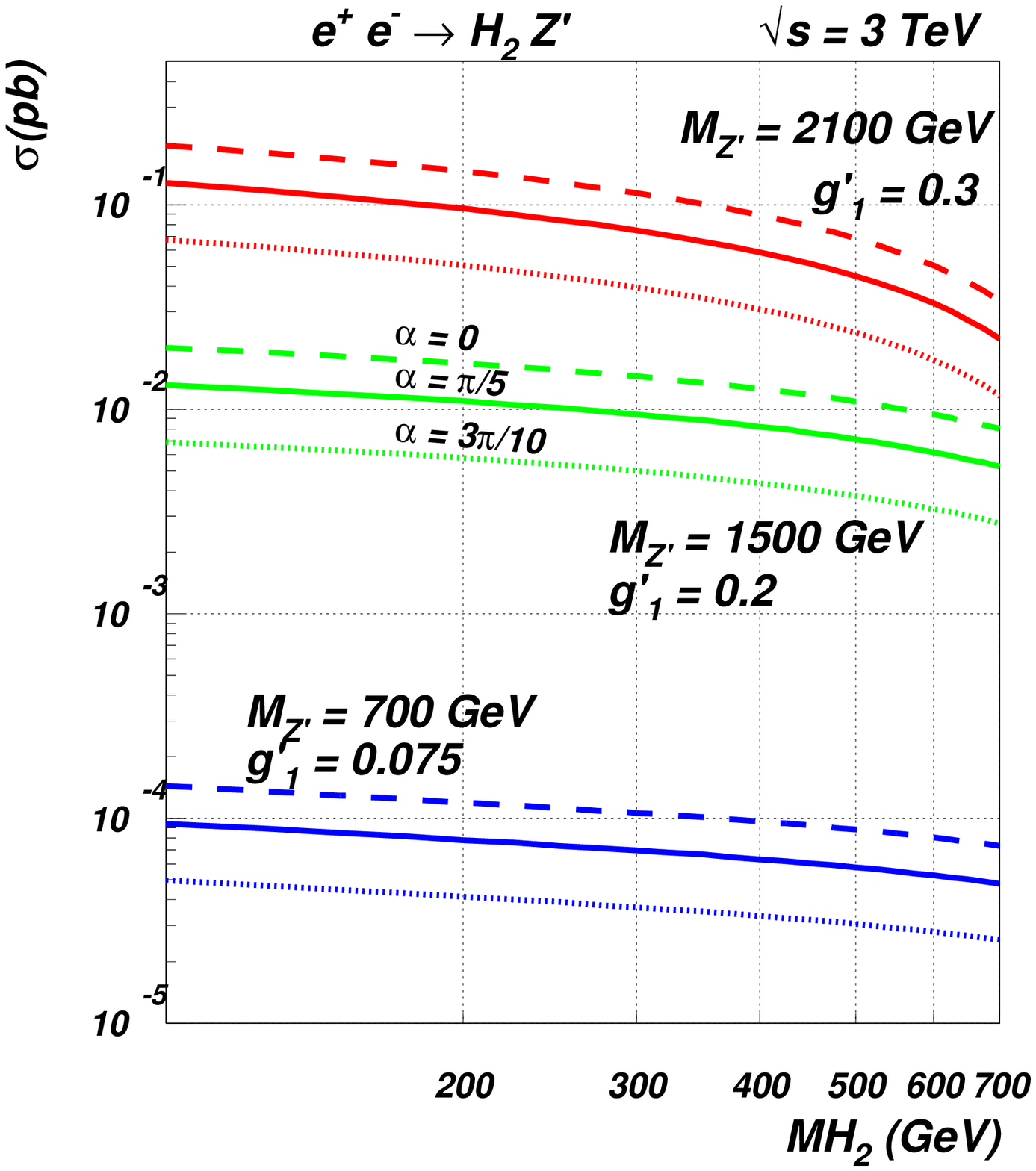}}  
  \vspace*{-0.4cm}
  \caption{\it Cross sections for the process $e^+e^-\rightarrow Z^{'\ast} \rightarrow H_{1(2)} Z'$ (\ref{H1Zp_1TeV}) for $h_1$ and (\ref{H2Zp_1TeV}) for $h_2$ at the LC at $\sqrt{s}=1$ TeV and (\ref{H1Zp_3TeV}) for $h_1$ and (\ref{H2Zp_3TeV}) for $h_2$ at the LC at $\sqrt{s}=3$ TeV. \label{ILC_Zp_strah}}
\end{figure}

We start by showing the cross section for the associated production of a Higgs boson and a $Z'$ boson,
\begin{equation}\label{Zp_strah}
e^+ e^- \to Z'\, h_{1,2}\, ,
\end{equation}
as in figure~\ref{ILC_Zp_strah}. Due to the stringent bounds on the $Z'$ boson mass and coupling to fermions, a sub-TeV CM energy collider is not capable of benefiting from this production mechanism, especially because of the naive kinematic limitation in the final state phase space.
 The CM energy is not sufficient to produce a $Z'_{B-L}$ and a Higgs boson, if both are on-shell. This is clear in figures~\ref{H1Zp_1TeV} and \ref{H2Zp_1TeV}, where a light $Z'$ boson (with mass of $500$ GeV) gives cross sections below $0.1$ fb. For a $Z'$ boson of $700$ GeV mass instead, the cross sections can be of the order of few fb, only for Higgs masses below $300$ GeV though, the kinematical limit.

The situation is considerably improved for a multi-TeV collider, not anymore limited in kinematics. As shown in figures~\ref{H1Zp_3TeV} and \ref{H2Zp_3TeV},  a Higgs boson can be produced in association with a $Z'$ boson 
of $1.5$ TeV mass with cross sections of $\sim 10$ fb in the whole range of the scalar masses considered, rising to $\mathcal{O}(100)$ fb if $M_{Z'}=2.1$ TeV is considered (and a suitable value for $g'_1$ is chosen). Although in the latter configuration the highest 
 Higgs boson mass that can be produced is smaller than for $M_{Z'}=1.5$ TeV, the cross sections for this process when the scalar mass is close to $700$ GeV (the maximum value considered here) are still above those when a $Z'$ boson of $1.5$ TeV mass is considered.
It is crucial to note that this is the only production mechanism that can 
potentially lead to the discovery of the heavy Higgs boson in the decoupling limit, i.e., for $\alpha \to 0$. As previously stated, the Higgs-strahlung off $Z'$ mechanism is not suitable for the LHC, making a multi-TeV linear collider possibly the ultimate chance for its discovery. This is dramatically different from scalar extensions of the SM in which the gauge content is not changed, as no SM matter couples strongly to the heavy Higgs boson in the decoupling limit, and we will comment on it at the end of this subsection.

Due to the importance of this channel, its discovery potential at future LCs is here presented\footnote{Regarding the SM-like production mechanisms, their discovery potential can be inferred from the analogous ones for the SM Higgs boson by rescaling the signal (and therefore the significance) by the sine(cosine) of the scalar mixing angle if considering the light(heavy) Higgs boson.}.

\begin{figure}[!h]
  \subfloat[]{
  \label{a15} 
  \includegraphics[angle=0,width=0.48\textwidth ]{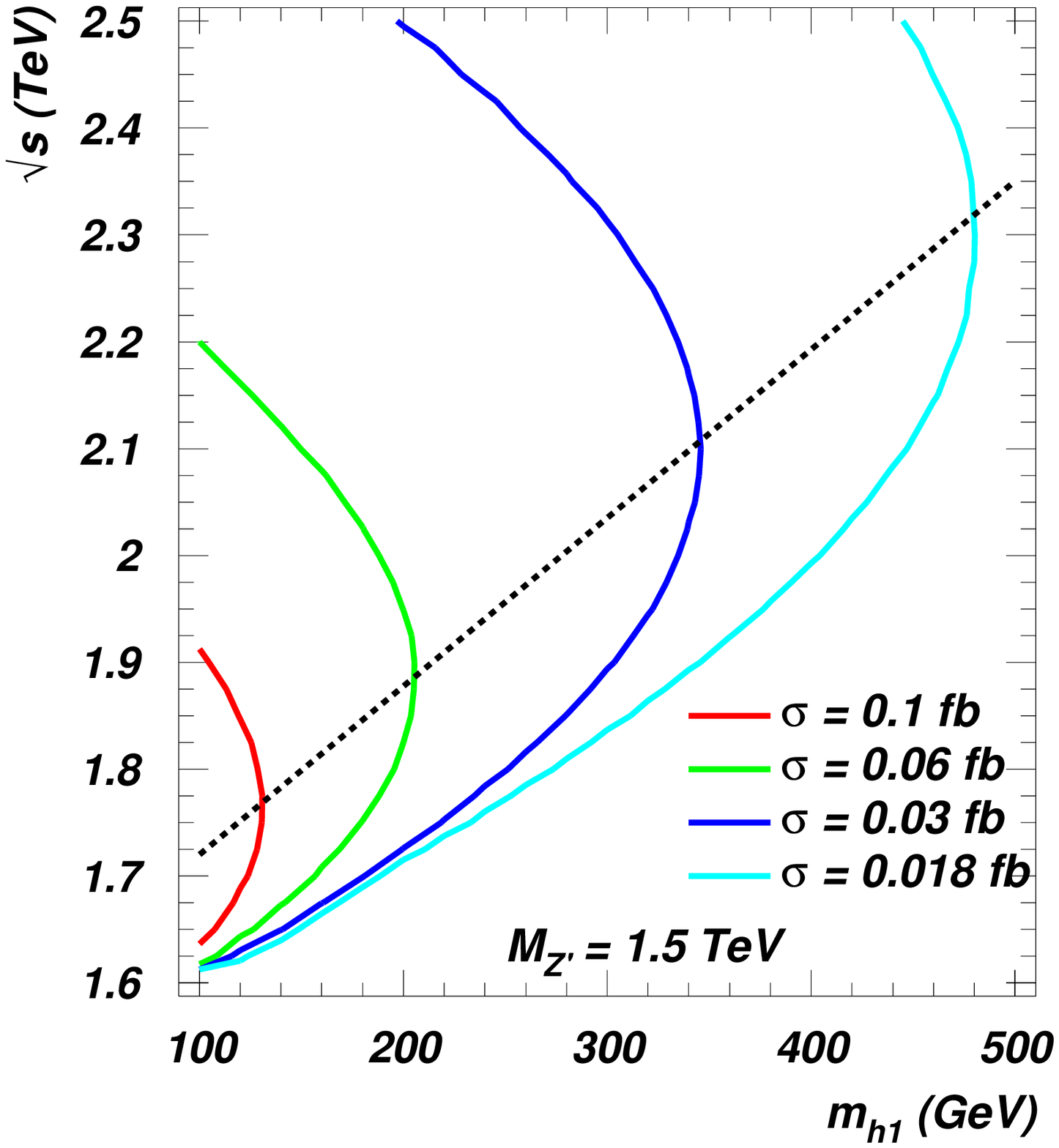}}
  \subfloat[]{
  \label{a21}
  \includegraphics[angle=0,width=0.48\textwidth ]{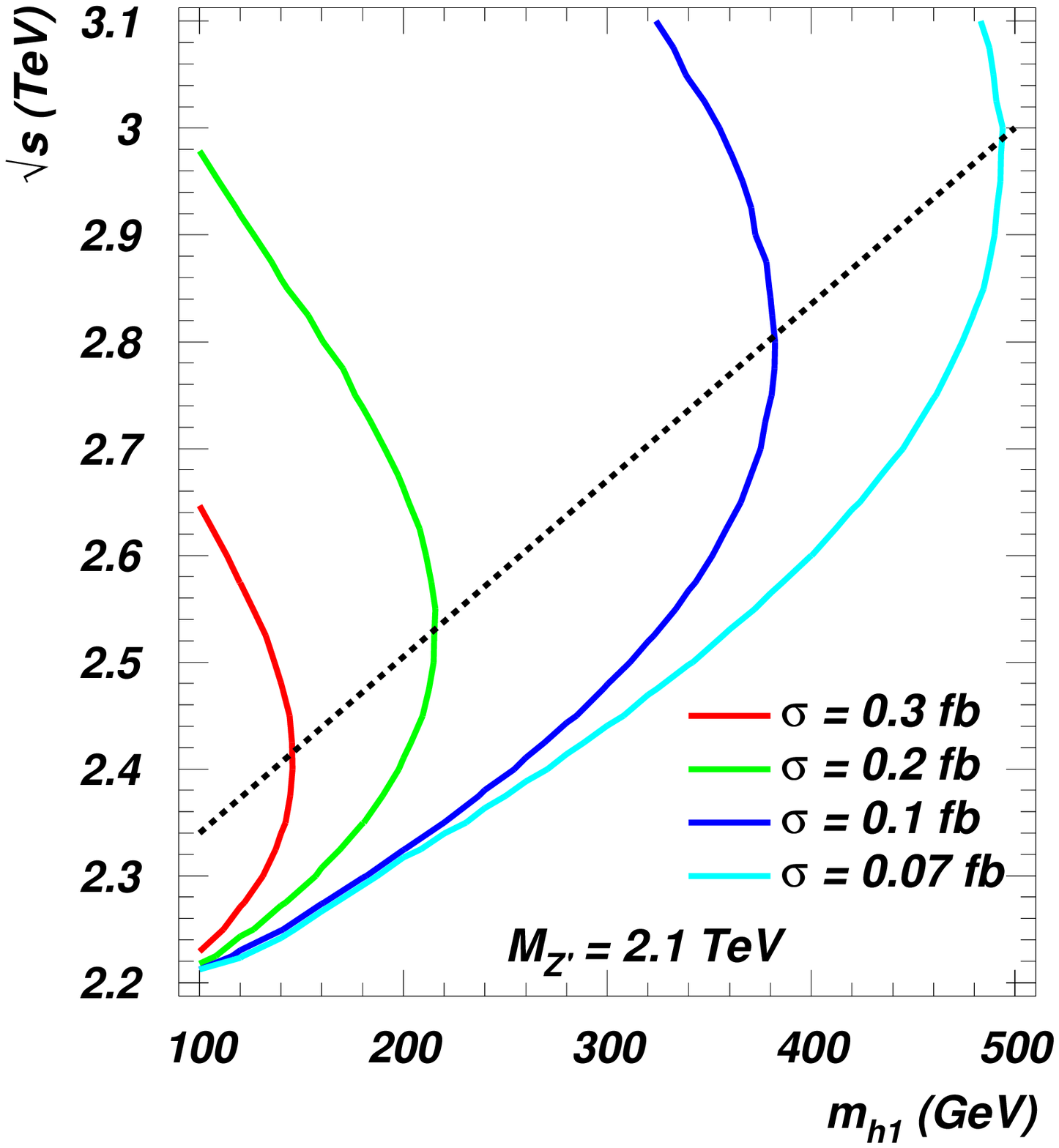}}
  \caption{\it ISR effect on the Higgs production from $Z'$ strahlung for (\ref{a15}) $M_{Z'}=1.5$ TeV and for (\ref{a21}) $M_{Z'}=2.1$ TeV . The dashed lines correspond to the $\sqrt{s}$ for which the cross section per fixed Higgs mass is maximised, according to eq.~(\ref{Zp_ISR_dep}).\label{Zp_ISR}}
\end{figure}

The $Z'$ boson mass is assumed to be known from Drell-Yan production (see, for example, Ref.~\cite{Basso:2009hf}).
Therefore, we first study the effect of ISR on the cross sections for the process of eq.~(\ref{Zp_strah}). 
Figure~\ref{Zp_ISR} clearly shows a linear dependence on $\sqrt{s_{MAX}}$, the CM energy that maximises 
the cross sections,
as a function of the Higgs mass only. Interpolating, we find
\begin{equation}\label{Zp_ISR_dep}
\frac{\sqrt{s_{MAX}}}{TeV}\approx \frac{M_{Z'}}{TeV}+0.1+1.5 \frac{m_H}{TeV} \qquad (H=h_1, h_2)\, .
\end{equation}
Per fixed Higgs and $Z'$ boson masses, the discovery potential can be maximised by fixing the CM energy to 
$\sqrt{s_{MAX}}$.
 We are left with the scalar mixing angle ($\alpha$) and the integrated luminosity as free parameters. 
Hence, this study shows what is the integrated luminosity required to start probing the values of $\alpha$ in 
the $B-L$ model.

We decided to analyse two different benchmark scenarios: a light Higgs boson of $120$ GeV of mass, 
decaying into $b$-quark pairs, and of $200$ GeV of mass, decaying into $W$-boson pairs 
(that we left undecayed, so that any particular decay mode can easily be implemented). In analogy with 
the study of Ref.~\cite{Basso:2009hf}, the decay of the $Z'$ boson into pairs of muons is considered as 
the most suitable. The total processes read
\begin{eqnarray}
e^+e^-\to Z'h_1 &\to& \mu^+\mu^- b\overline{b}\, , \\
e^+e^-\to Z'h_1 &\to& \mu^+\mu^- W^+W^-\, .
\end{eqnarray}
Only the light Higgs boson has been considered, being the case of a heavy Higgs boson with same mass just the 
symmetric one under $\alpha \to \pi/2 - \alpha$.
Two different $Z'$ boson masses have been chosen, $1.5$ TeV and $2.1$ TeV,
 while the gauge coupling $g'_1$ has been set to a discrete choice of values allowed by 
existing experimental constraints. Regarding the background, the relevant one is found to be 
$Z'Z/\gamma$ (where the $Z'$ boson decays to muons and the $b$-quark pair stems from a SM gauge boson) 
and $Z'W^+W^-$ (where, again, the muons come exclusively from the $Z'$ boson). The pure EW background 
(i.e., $ZZ$, $Z\gamma$, and $\gamma\gamma$ for the $b$-quark final state; 
$W^+W^-Z/\gamma$ for the $W$-boson final state) is two orders of magnitude below them, hence it is 
neglected here.

For both the signal and the background, we have assumed standard acceptance cuts (for muons and quarks) at a LC~\cite{Assmann:2000hg}
\begin{eqnarray}\label{mu_cut}
\mbox{muons:}\qquad E^\mu &>& 10~{\rm GeV},\qquad |\cos{\theta}^\mu|<0.95\, , \\ \label{b_cut}
\mbox{$b$ quarks:}\qquad E^b &>& 10~{\rm GeV},\qquad |\cos{\theta}^b|<0.9\, ,
\end{eqnarray} 
and a window in the invariant mass distribution of the Higgs decay products has been taken as large as $20$ GeV and centred at the Higgs mass, independently of considering the $b$-quark pair or the $W$-boson pair final state, to naively simulate the detector resolution. For the muons, we require them to reconstruct the $Z'$ boson mass within $3$ intrinsic widths, always wider than the di-muon resolution for the values of the gauge coupling here considered~\cite{Basso:2008iv,Assmann:2000hg}. Finally, regarding the $b$-quark tagging efficiency, it has been assumed to be $62\%$, according to Ref.~\cite{Abe:2001pea}. The $W$-boson reconstruction efficiency has been set to $1$ for simplicity.

\begin{figure}[!h] 
  \subfloat[]{
  \label{lumi120_1500}
  \includegraphics[angle=0,width=0.48\textwidth ]{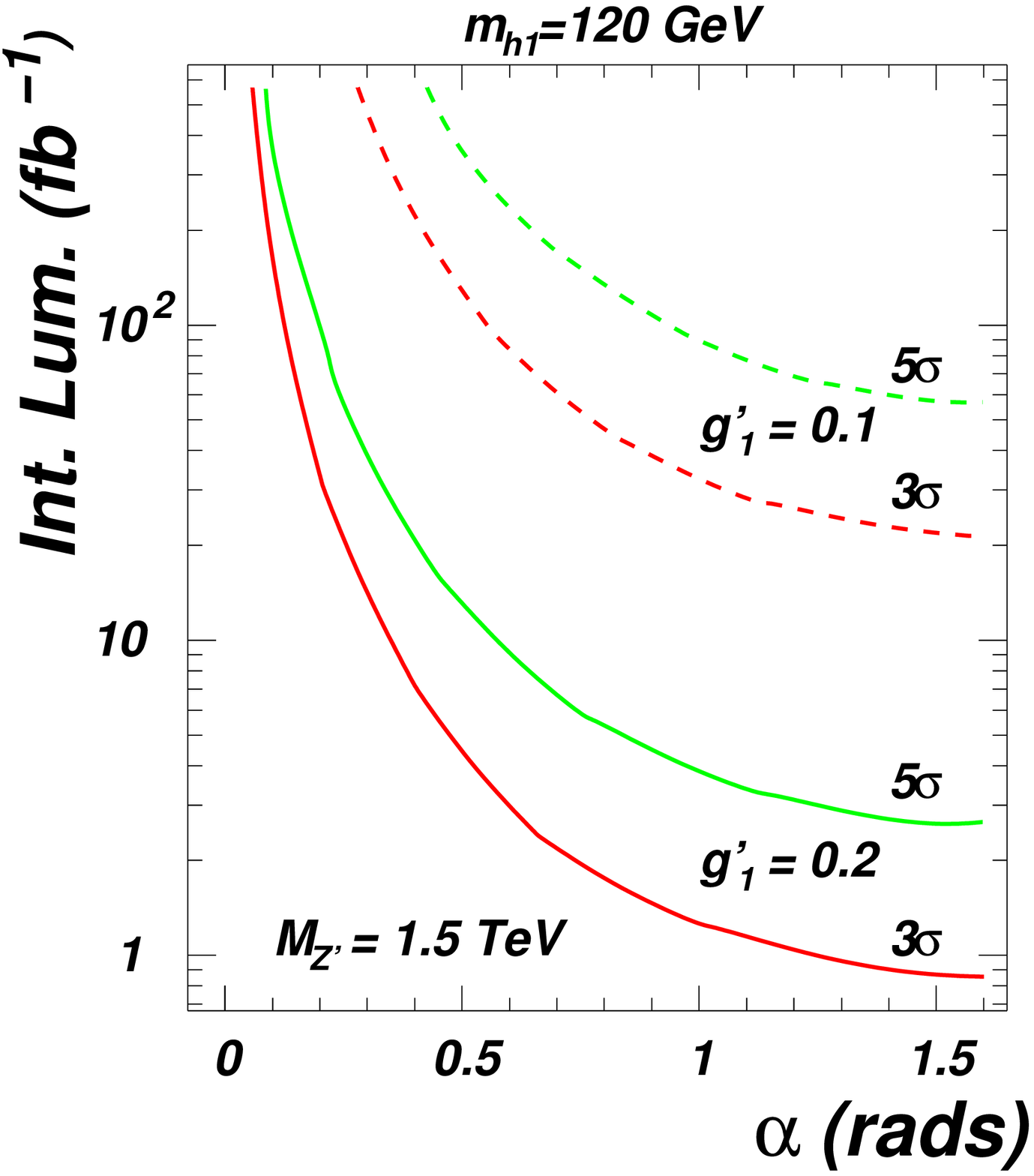}}
  \subfloat[]{
  \label{lumi200_1500}
  \includegraphics[angle=0,width=0.48\textwidth ]{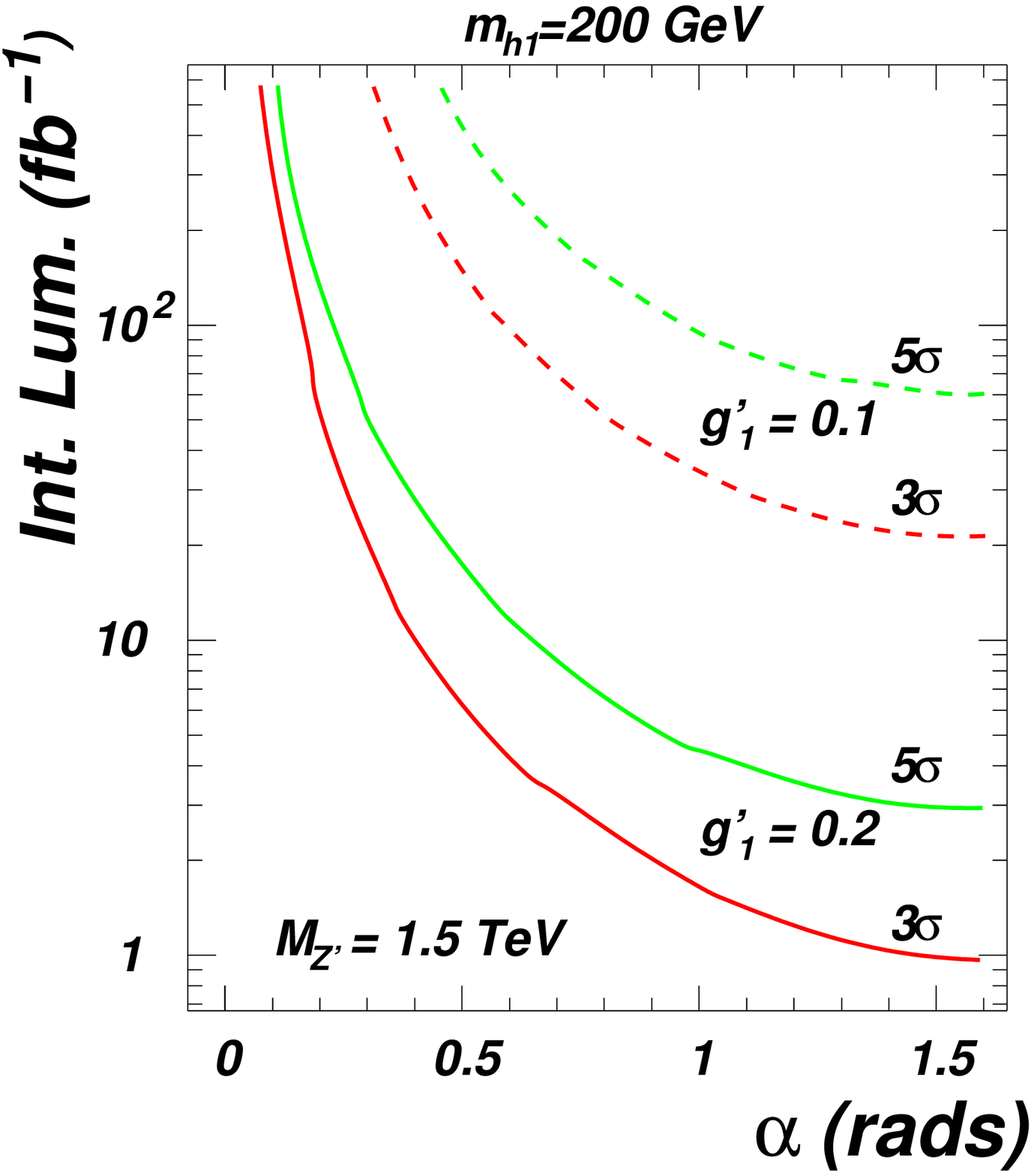}}
  \\
  \subfloat[]{
  \label{lumi120_2100}
  \includegraphics[angle=0,width=0.48\textwidth ]{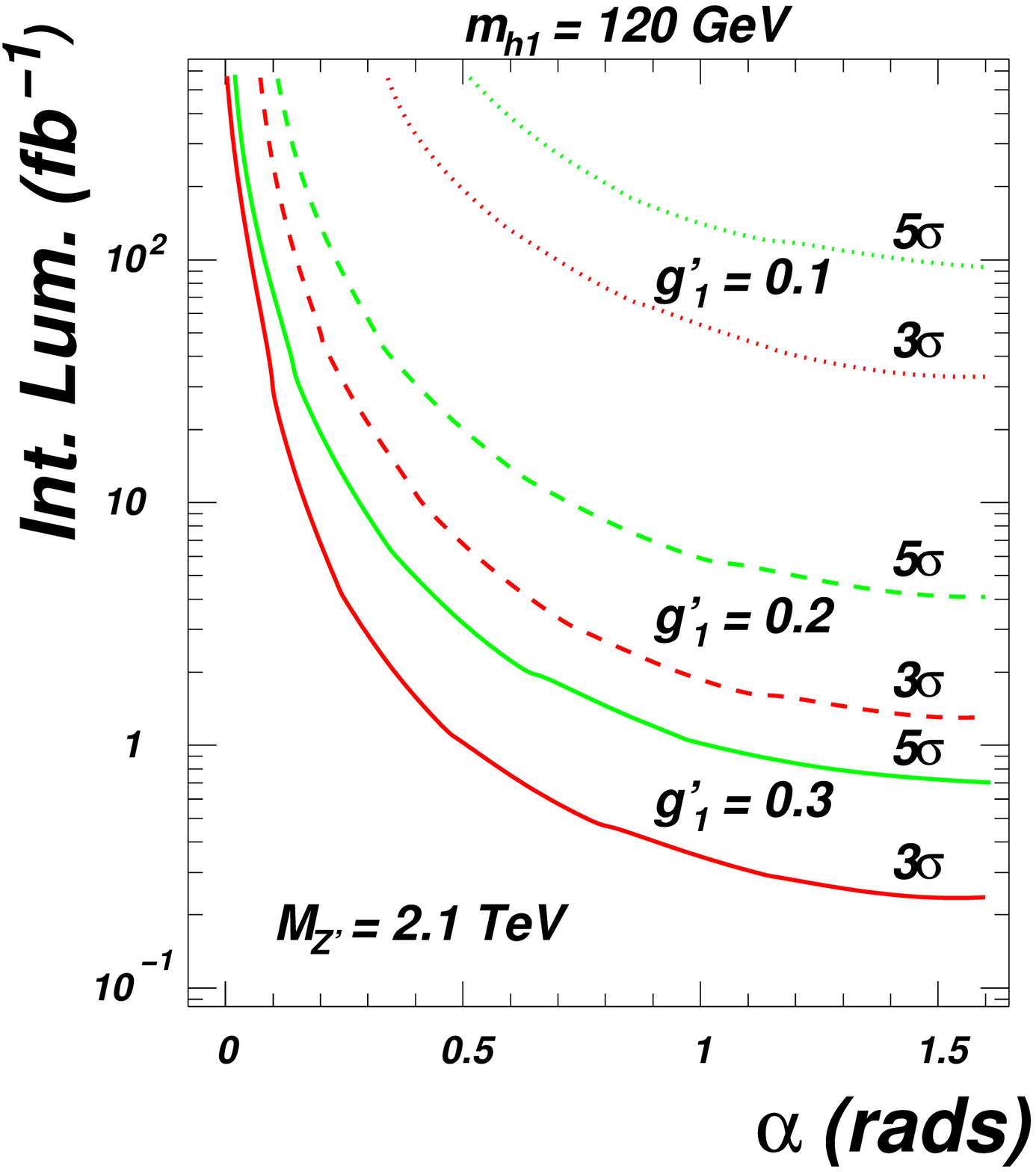}}
  \subfloat[]{
  \label{lumi200_2100}
  \includegraphics[angle=0,width=0.48\textwidth ]{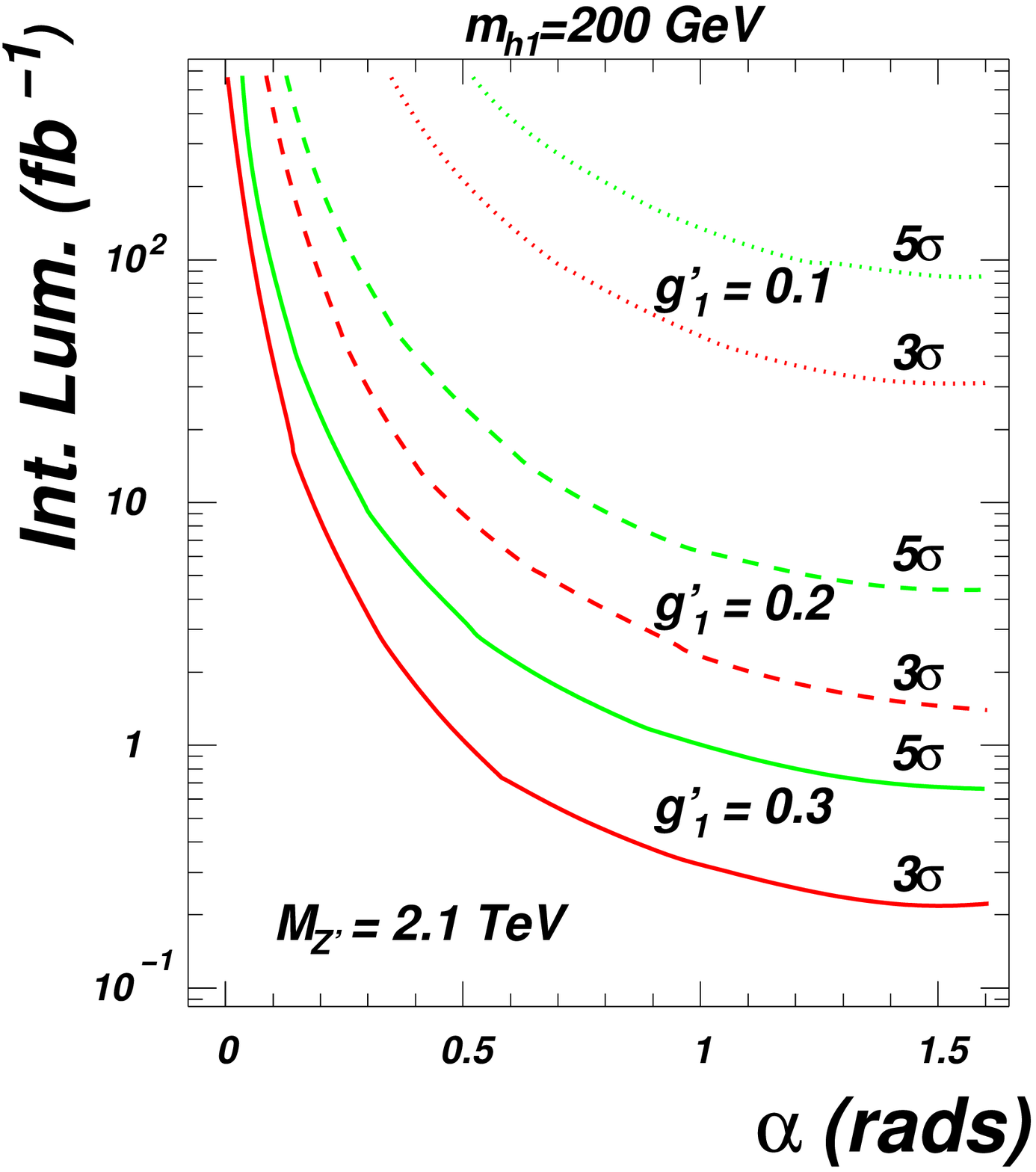}}
  \caption{\it Discovery potential for the associated production of the $Z'$ boson and a light Higgs boson decaying into (\ref{lumi120_1500})  $b$ quark pairs and into (\ref{lumi200_1500}) $W$ boson pairs for $M_{Z'}=1.5$ TeV and $g'_1=0.1$, $0.2$ and into (\ref{lumi120_2100})  $b$ quark pairs and into (\ref{lumi200_2100}) $W$ boson pairs for $M_{Z'}=2.1$ TeV and $g'_1=0.1$, $0.2$, $0.3$. \label{lumi_s_Zph1}}
\end{figure}

Figure~\ref{lumi_s_Zph1} shows the discovery reach of a LC in these conditions, for the production of the light Higgs boson only, as a function of the scalar mixing angle $\alpha$.
The significance plots have been obtained using the same algorithms described in Ref.~\cite{Basso:2010pe}. We recall here that we define the significance, based on Gaussian statistics, as ${\sigma} \equiv {\it s}/{\sqrt{\it b}}$.

We see that the discovery power for the two decay modes of the light Higgs boson are comparable, with the lower Higgs mass always requiring slightly less integrated luminosity than the heavier mass to probe the same value of the mixing angle, regardless of its actual value. However, we stress again that this is because no decay pattern for the $W$ boson has been implemented. It is straightforward from these plots to obtain those pertaining to the particular $W$-boson decay mode of interest, by just rescaling the results we present by the squared root of the product of BRs of the two $W$'s (or by the simple BR if the decay mode is the same for both $W$ bosons). Also, very small angles require high luminosity and big values of $g'_1$ to be probed, excluding $\alpha=0$ for which $h_1$ and $Z'$ boson do not couple.

The results for the $3$($5$)$\sigma$ discovery potential of $h_1$-strahlung off $Z'_{B-L}$ are collected in table~\ref{5sigma_ZpH1}.

\begin{table}[h]
$m_{h_1}=120$ GeV
\begin{center}
\begin{tabular}{|c||c|c||c|c|c|}
\hline
$\sqrt{s}=\sqrt{s_{MAX}}$ & \multicolumn{2}{|c||}{$M_{Z'}=1.5$ TeV} & \multicolumn{3}{|c|}{$M_{Z'}=2.1$ TeV} \\
\hline
$\alpha$ (rads) & $g'_1=0.1$ & $g'_1=0.2$  & $g'_1=0.1$  & $g'_1=0.2$   & $g'_1=0.3$  \\
\hline
0.2 & $>$500($>$1000) & 38(100)  & $>$500($>$1000) & 50(150) & 7(20) \\
0.5 & 120(350) & 4.5(15.0)      & 180(500) & 7(20) & 1.0(3.5) \\
1.0 & 30(90)   & 1.2(3.5)       & 45(120)  & 1.8(5.0) & 0.35(1.0)\\
\hline  
\end{tabular}
\end{center}
$m_{h_1}=200$ GeV
\begin{center}
\begin{tabular}{|c||c|c||c|c|c|}
\hline
$\sqrt{s}=\sqrt{s_{MAX}}$ & \multicolumn{2}{|c||}{$M_{Z'}=1.5$ TeV} & \multicolumn{3}{|c|}{$M_{Z'}=2.1$ TeV} \\
\hline
$\alpha$ (rads) & $g'_1=0.1$ & $g'_1=0.2$  & $g'_1=0.1$  & $g'_1=0.2$   & $g'_1=0.3$  \\
\hline
0.2 & $>$500($>$1000)& 50(120)  & $>$500($>$1000) & 90(200) & 9(22) \\
0.5 & 150(420)   & 6.5(18.0)    & 200(500)  & 9(25) & 1.0(3.5) \\
1.0 & 35(100)    & 1.8(4.5)     & 45(120)  & 2.2(6.0) & 0.35(1.0)\\
\hline  
\end{tabular}
\end{center}
\vskip -0.5cm
\caption{\it Minimum integrated luminosities (in fb$^{-1}$) for a $3\sigma$($5\sigma$) discovery as a function of the  scalar mixing angle $\alpha$, for selected $Z'_{B-L}$ boson masses and $g_1'$ couplings for the light Higgs boson. All values above the given $\alpha$ are probed for the luminosity in table. For $h_2$, all angles below $\pi/2-\alpha$ are probed with the luminosity in table.}
\label{5sigma_ZpH1}
\end{table}

The Higgs-strahlung off $Z'$ channel is a fundamental process for producing the heavy Higgs boson for scalar mixing angles close to decoupling, i.e., when $0<\alpha \ll 1$~rads. In this regime, the standard production mechanisms are least effective, while the Higgs-strahlung from the $Z'$ boson delivers the maximum cross sections for $h_2$ production.
Also, when $\alpha =0$, the heavy Higgs boson decays only into neutrino pairs, either heavy, light or a combination of the two, depending on the kinematical configuration, as well as into off-shell states. Therefore, depending on its mass, the heavy Higgs could be a very long-lived particle and escape detection, or decay into peculiar multi-lepton and multi-jet final states through heavy neutrino pairs (see, e.g., \cite{Basso:2008iv,Basso:2010yz}). As soon as $\alpha$ gets a non-vanishing value (i.e., $\alpha > 10^{-8}\div 10^{-5}$), all other channels open and become dominant, such as the decay into pairs of SM gauge bosons and into light Higgs boson pairs. In both cases, 
the mass of the heavy scalar can still be measured, for example with the recoil mass technique by measuring the spectrum of the two muons from the $Z'$ boson. Although in a very different kinematical region from the SM, this technique has been employed at LEP and studies exist for future LCs~\cite{Djouadi:2007ik}. Given the good resolution for muons, the precision in the scalar mass measure is expected to be relatively good for the $B-L$ heavy Higgs boson too. 

For a non-vanishing, albeit very small, scalar mixing angle, the decay into pairs of SM gauge bosons is dominant, if kinematically allowed (i.e., when $m_{h_2}\gtrsim 150$ GeV). For smaller masses, the decay into $b$-quark pairs is favoured. In table~\ref{5sigma_ZpH2_a0} we summarise the $3(5)\sigma$ discovery potential for the heavy Higgs boson for scalar mixing angles in proximity of the decoupling regime, i.e., $\alpha \sim 10^{-4}$ rads, for selected values of $Z'_{B-L}$ masses and couplings. We stress again here that only a LC could benefit of this mechanism, being the latter precluded at the LHC for whatever value of $\alpha$.
\begin{table}[h]
\begin{center}
\begin{tabular}{|c||c|c||c|c|c|}
\hline
$\sqrt{s}=\sqrt{s_{MAX}}$ & \multicolumn{2}{|c||}{$M_{Z'}=1.5$ TeV} & \multicolumn{3}{|c|}{$M_{Z'}=2.1$ TeV} \\
\hline
$m_{h_2} \mbox{ (GeV)}$ & $g'_1=0.1$ & $g'_1=0.2$  & $g'_1=0.1$  & $g'_1=0.2$   & $g'_1=0.3$  \\
\hline
120 & 20(55)  & 0.80(2.8)     & 30(90)  & 1.2(4.0) & 0.22(0.70) \\
200 & 20(60)  & 0.95(3.0)     & 30(90)  & 1.5(4.5) & 0.22(0.70)\\
500 & 0.07(0.2) & 1.5(4.0)   & 0.1(0.3)&   2.0(6.0)    &  20(65)  \\
\hline  
\end{tabular}
\end{center}
\vskip -0.5cm
\caption{\it Minimum integrated luminosities (in fb$^{-1}$) for a $3\sigma$($5\sigma$) discovery for selected $Z'_{B-L}$ boson masses and $g_1'$ couplings for the heavy Higgs boson when $0<\alpha \ll 1$ rads.}
\label{5sigma_ZpH2_a0}
\end{table}

\subsubsection{Other new Higgs boson production mechanisms via a $Z'$ boson}\label{sect:NS_single_other}
%
%

\begin{figure}[!h]
\scalebox{0.9}{
  \subfloat[]{ 
  \label{H1ZptT_1TeV}
  \includegraphics[angle=0,width=0.48\textwidth ]{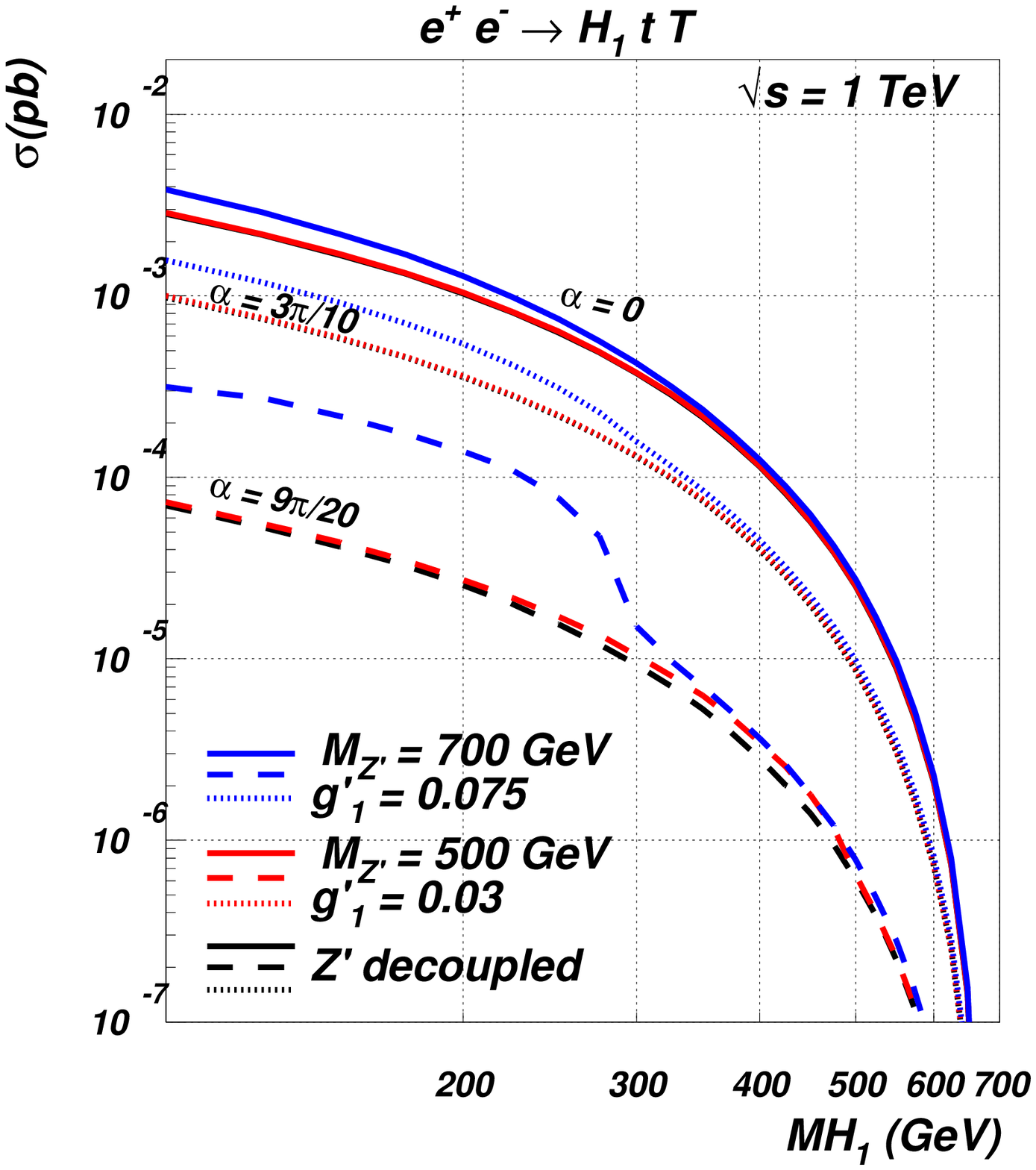}}
  \subfloat[]{
  \label{H2ZptT_1TeV}
  \includegraphics[angle=0,width=0.48\textwidth ]{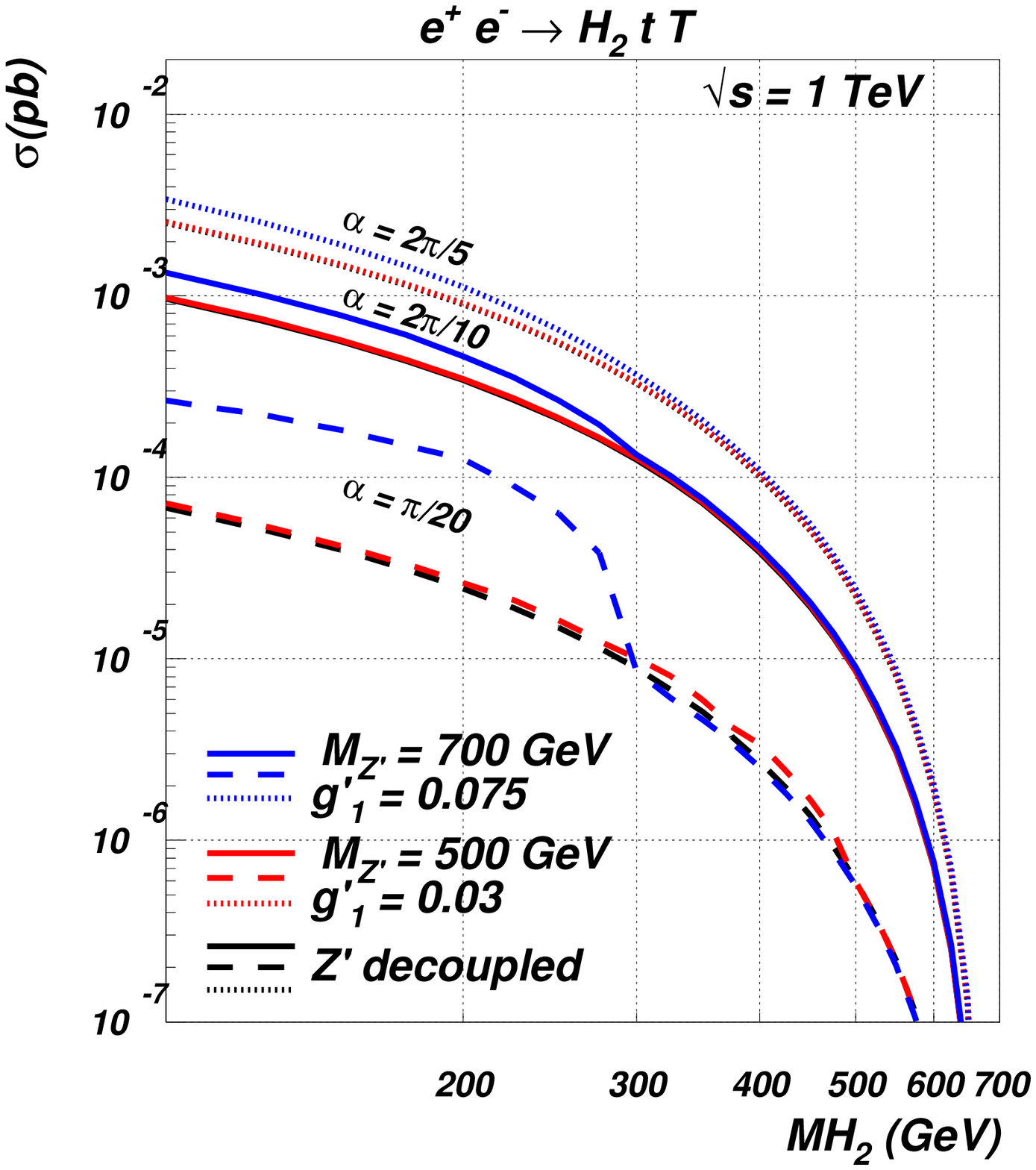}}
}  \\
\scalebox{0.9}{
  \subfloat[]{ 
  \label{H1ZptT_3TeV}
  \includegraphics[angle=0,width=0.48\textwidth ]{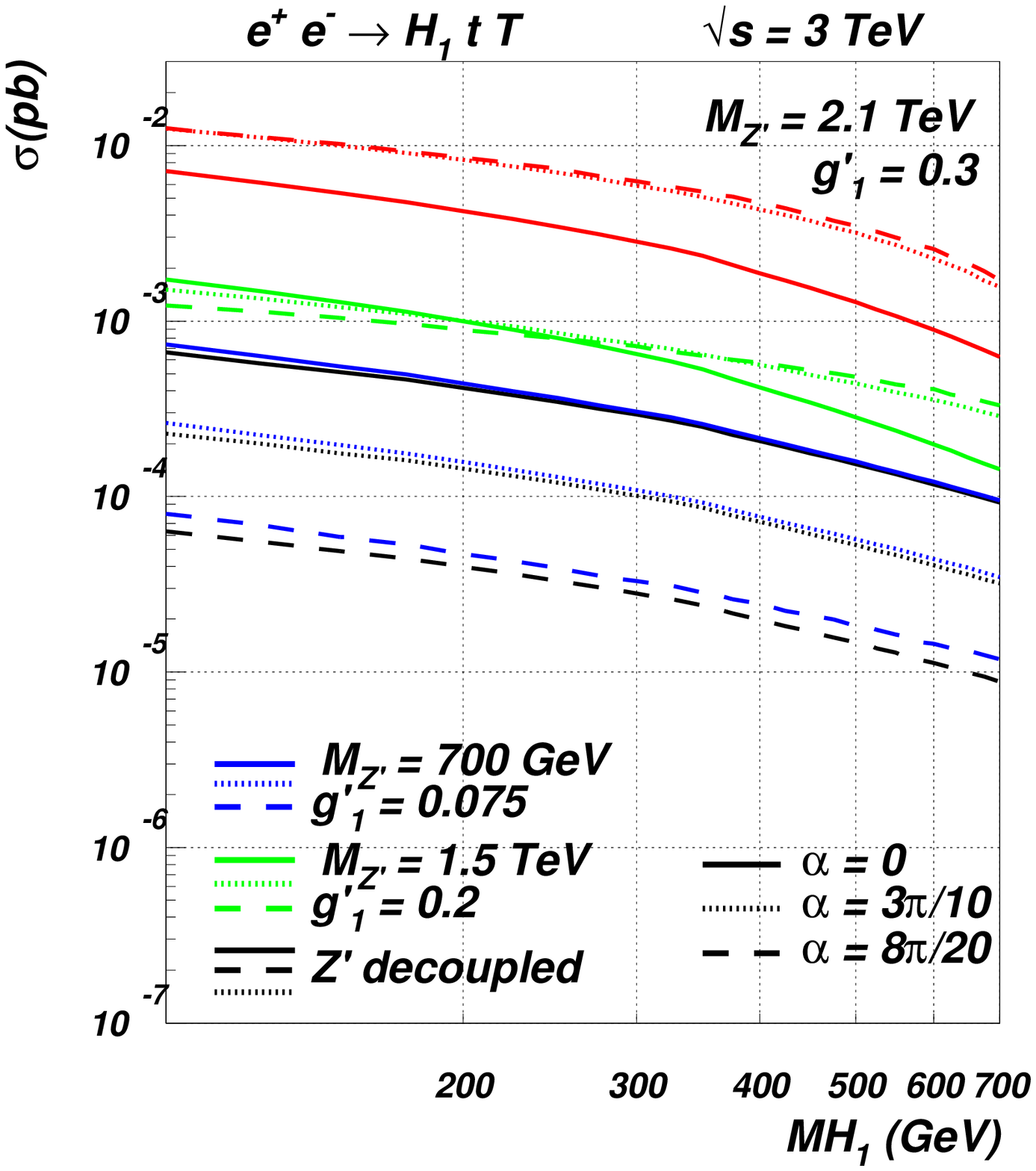}}
  \subfloat[]{
  \label{H2ZptT_3TeV}
  \includegraphics[angle=0,width=0.48\textwidth ]{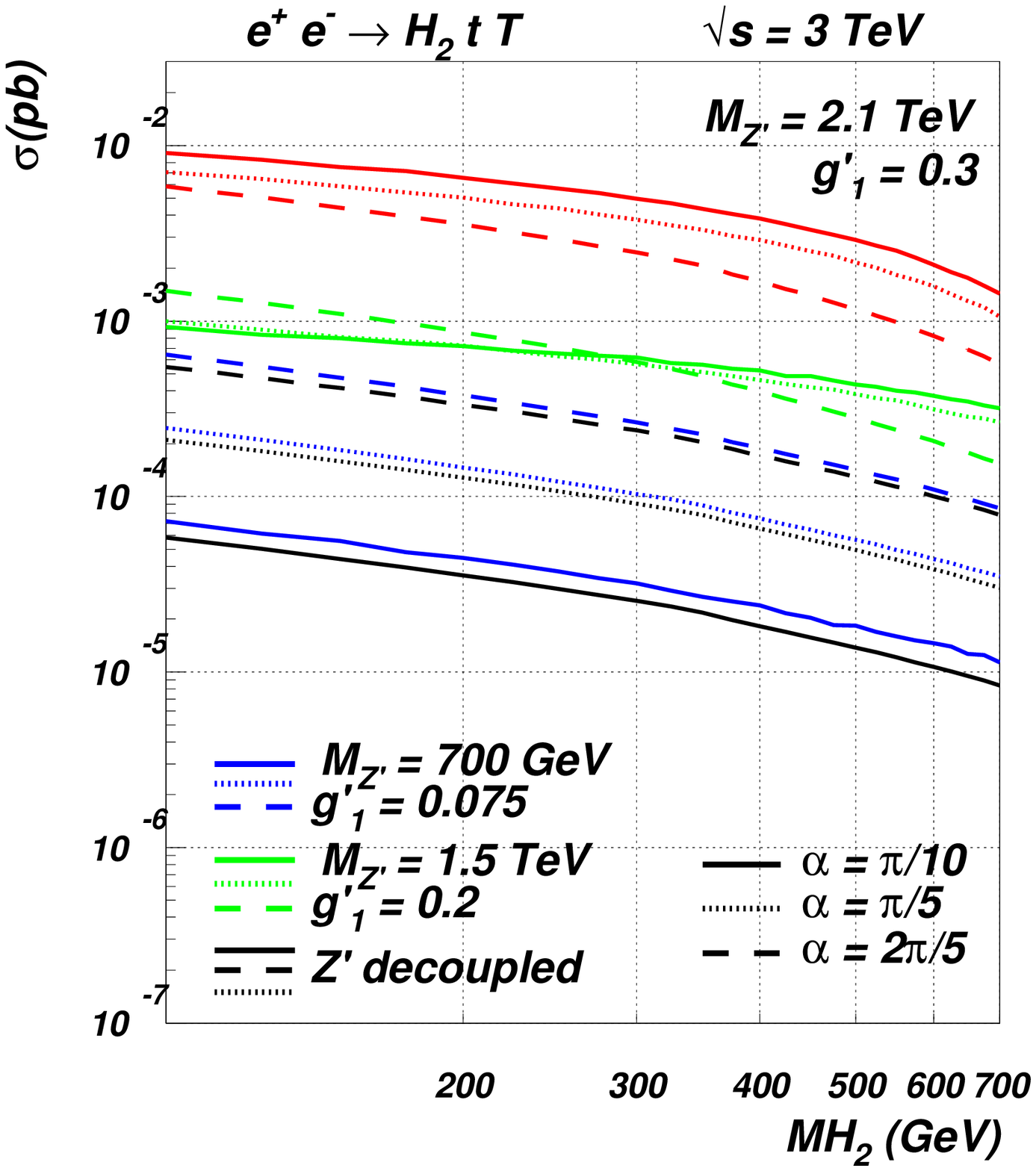}}
}  \\
\scalebox{0.9}{
  \subfloat[]{ 
  \label{H1ZptT_sMZp}
  \includegraphics[angle=0,width=0.48\textwidth ]{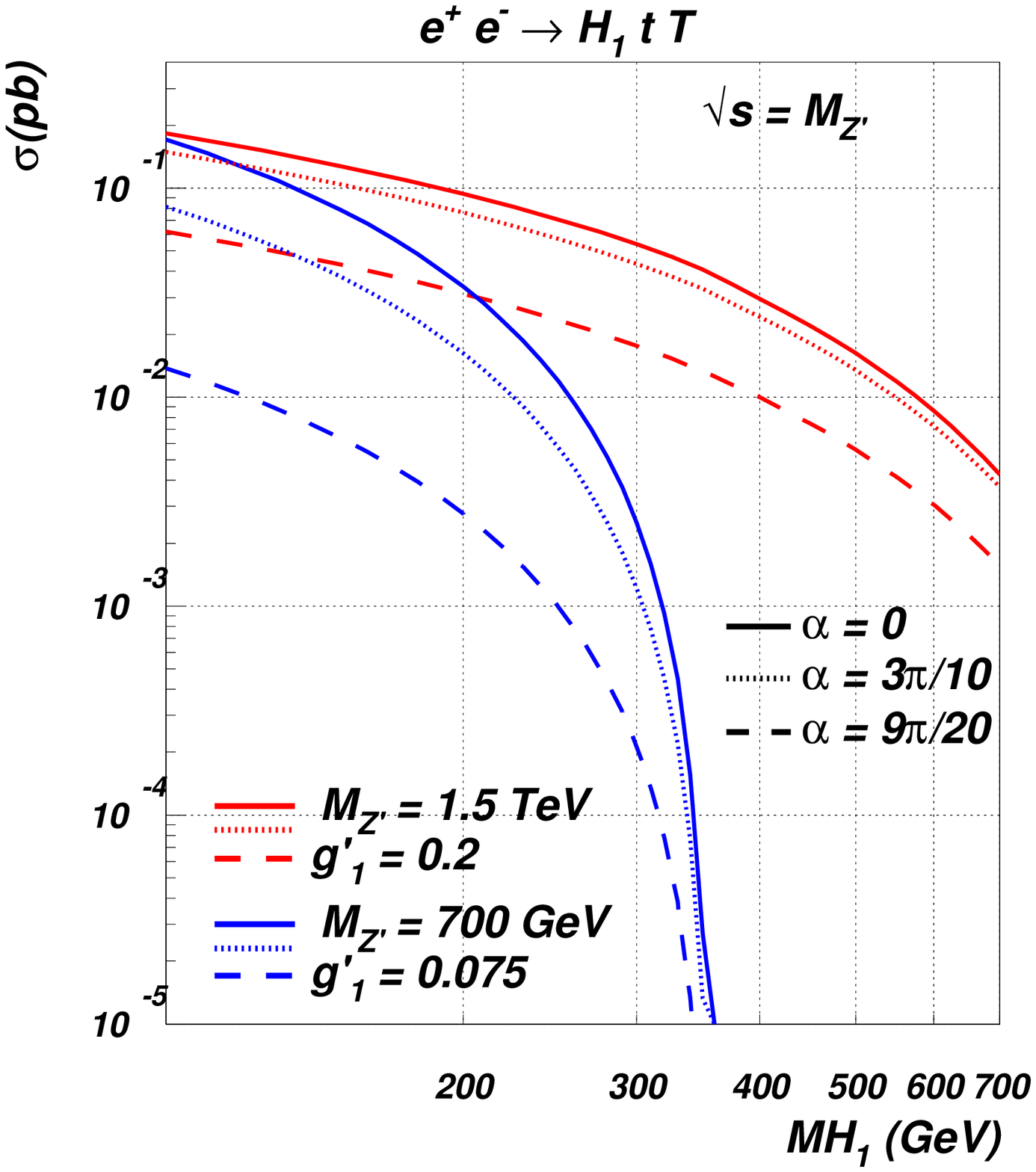}}
  \subfloat[]{
  \label{H2ZptT_sMZp}
  \includegraphics[angle=0,width=0.48\textwidth ]{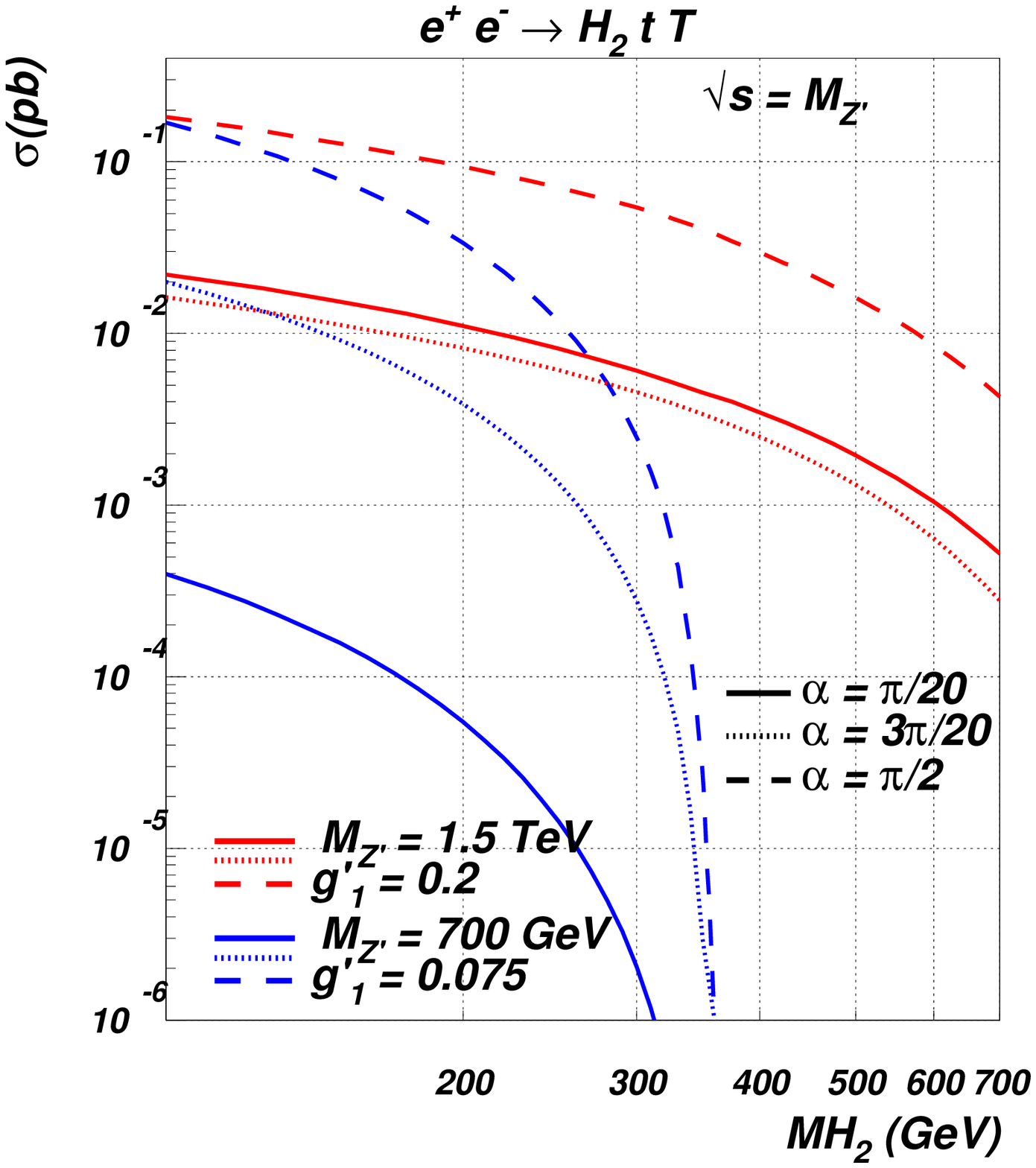}}
}
  \vspace*{-0.4cm}
  \caption{\it Cross sections for the process $e^+e^-\rightarrow H_{1(2)} t \overline{t}$ (\ref{H1ZptT_1TeV}) for $h_1$ and (\ref{H2ZptT_1TeV}) for $h_2$,   at the LC at $\sqrt{s}=1$ TeV, (\ref{H1ZptT_3TeV}) for $h_1$ and (\ref{H2ZptT_3TeV}) for $h_2$,  at $\sqrt{s}=3$ TeV and (\ref{H1ZptT_sMZp}) for $h_1$ and (\ref{H2ZptT_sMZp}) for $h_2$, at $\sqrt{s}=M_{Z'}$ TeV for $M_{Z'}=700$ GeV and $M_{Z'}=1.5$ TeV (that gives similar results as for $M_{Z'}=2.1$ TeV), for several angles and $M_{Z'}$.
 \label{ILC_tT}}
\end{figure}

Moving forward, we anticipate that the associated production with a top quark pair can be enhanced exploiting the $Z'$ boson.
In figures~\ref{H1ZptT_1TeV} and \ref{H2ZptT_1TeV} are shown the cross sections for the associated production of a Higgs boson and a pair of top quarks, for $M_{Z'}=500$ GeV, $700$ GeV and in the  case of a much heavier $Z'$ boson, hence decoupled, for $\sqrt{s}=1$ TeV. As known, the Higgs boson in this channel can be radiated by both the top (anti)quark and the vector boson, even though the fraction of events with a SM-$Z$ boson is negligible with respect to the top quark pair produced by a photon. Therefore, in the SM, the measurement 
of the Higgs coupling to the top quark is possible, though difficult because of the small cross sections \cite{Baer:1999ge}. We are therefore left to evaluate the relative contribution of the $Z'$ boson, to check whether the same situation holds.

Firstly, it is interesting to note that, in the decoupling limit (i.e., for vanishing scalar mixing angle $\alpha$), $h_1$ does not couple directly to the $Z'$ boson. Nonetheless, the $Z'$ boson can decay into a pair of top quarks, one of which then radiates the light Higgs boson. This channel has the same final state as the SM one, and will therefore increase the total number of events, as clear from figure~\ref{H1ZptT_1TeV}. Hence, the chances of measuring the (SM-like) Higgs boson to top quark coupling are improved in this case, only slightly for $\sqrt{s}=1$~TeV but quite considerably for $\sqrt{s}=3$~TeV and a few TeV $Z'$ boson mass.

As we increase the scalar mixing angle, the relative contribution of the $Z'$ boson increases, although the total cross sections for $\sqrt{s}=1$ TeV fall below the fraction of fb, making it even harder to be observed. The situation is opposite for the multi-TeV CM energy case (figures~\ref{H1ZptT_3TeV} and \ref{H2ZptT_3TeV}), in which the $Z'$ boson is produced abundantly and it can 
enhance the Higgs boson associated production with a top quark pair. In this case, however, it is not true anymore that the majority of the events are those in which the Higgs boson is radiated by a top quark: Higgs-strahlung from the $Z'$ boson is now an important channel, as clear from figures~\ref{H1Zp_3TeV} and \ref{H2Zp_3TeV} and from the fact that, for low $Z'$ masses, the total cross section is smaller as we start increasing the angle (due to the reduced coupling to the top quark), while for TeV $Z'$ boson masses it always increases. If the 
$Z'$ boson mass is below the maximum CM energy of the collider, the fraction of Higgs-strahlung events off the $Z'$ boson can be reduced by tuning the CM energy and sitting at the peak of the $Z'$ boson itself. In this case, the vast majority of $Z'$ bosons are produced on shell, enhancing the total cross sections and the portion of events in which the $Z'$ boson decays into a top quark pair, one of which will then radiate the Higgs boson. The possibility of sitting at the peak of the $Z'$ boson is therefore very important phenomenologically, as it allows the Higgs coupling to the top quark to be measured much more precisely than in the SM, as the cross section in the $B-L$ model for this channel can rise up to $10\div 100$~fb, depending on the Higgs boson mass and mixing angle. Notice that for $h_1$ the angle has to be small (i.e., less than $\pi/5$) to allow for this measurement, as only in this case $h_1$ couples more to the top quark than to the $Z'$ boson, though Higgs-strahlung off $Z'$ events are still important (the ratio of the two subchannels is in fact $<5 \%$). This situation is exactly specular when the heavy Higgs boson is considered: when the CM energy is maximal, the associated 
production with a top quark pair has good cross sections but it does not allow for a direct measurement of the Higgs boson to top quark coupling, that instead is possible for big angles ($\alpha \gtrsim 3\pi/10$) when sitting at the $Z'$ boson peak. Notice that, in this configuration, the total cross section is independent of the $Z'$ boson mass, if the top pair and the Higgs boson can all be produced on-shell. Otherwise, the cross sections are suppressed by the phase space.

%
%

\begin{figure}[!h]
  \subfloat[]{ 
  \label{LC_CM-1}
  \includegraphics[angle=0,width=0.48\textwidth ]{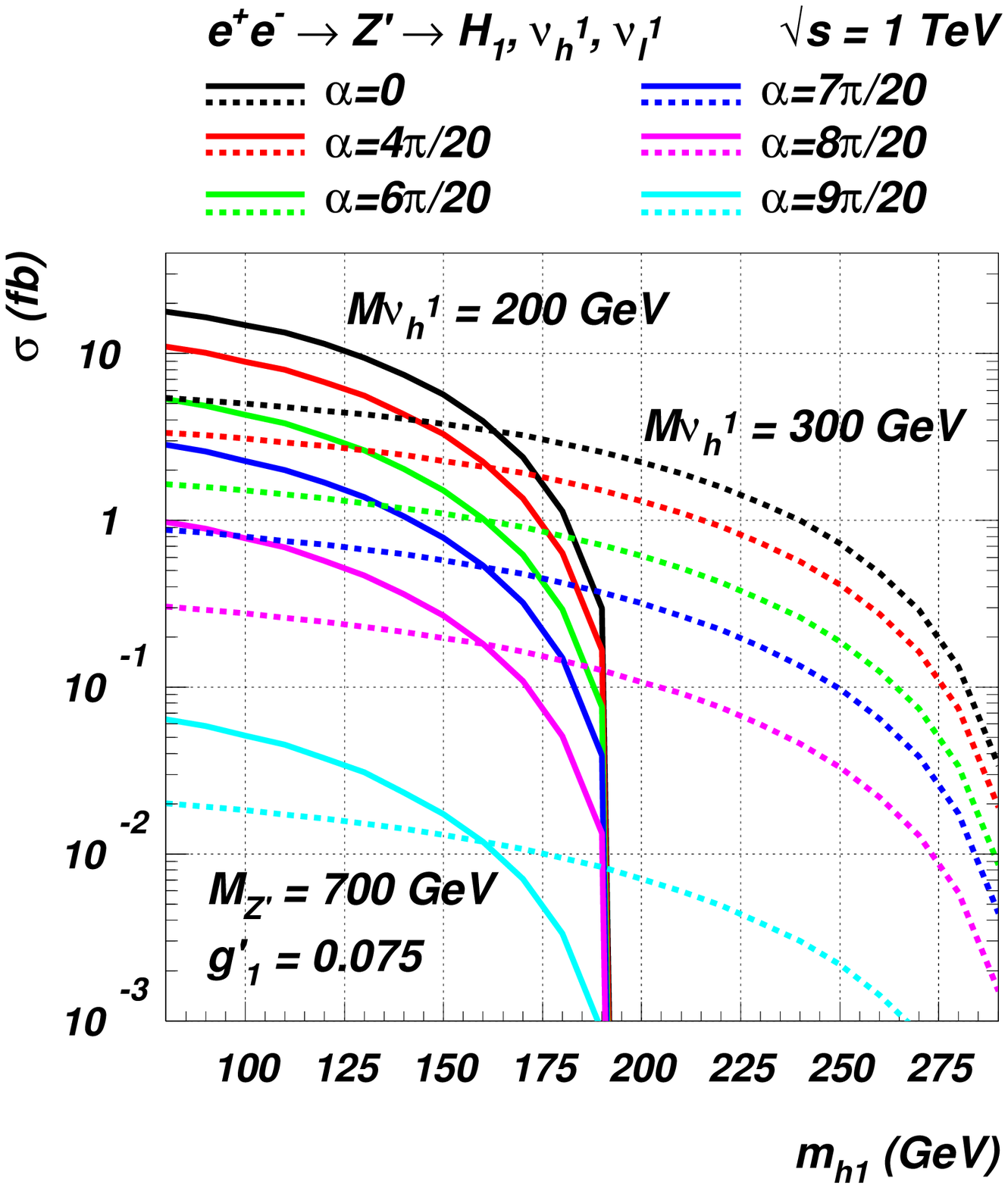}}
  \subfloat[]{
  \label{LC_CM-MZp}
  \includegraphics[angle=0,width=0.48\textwidth ]{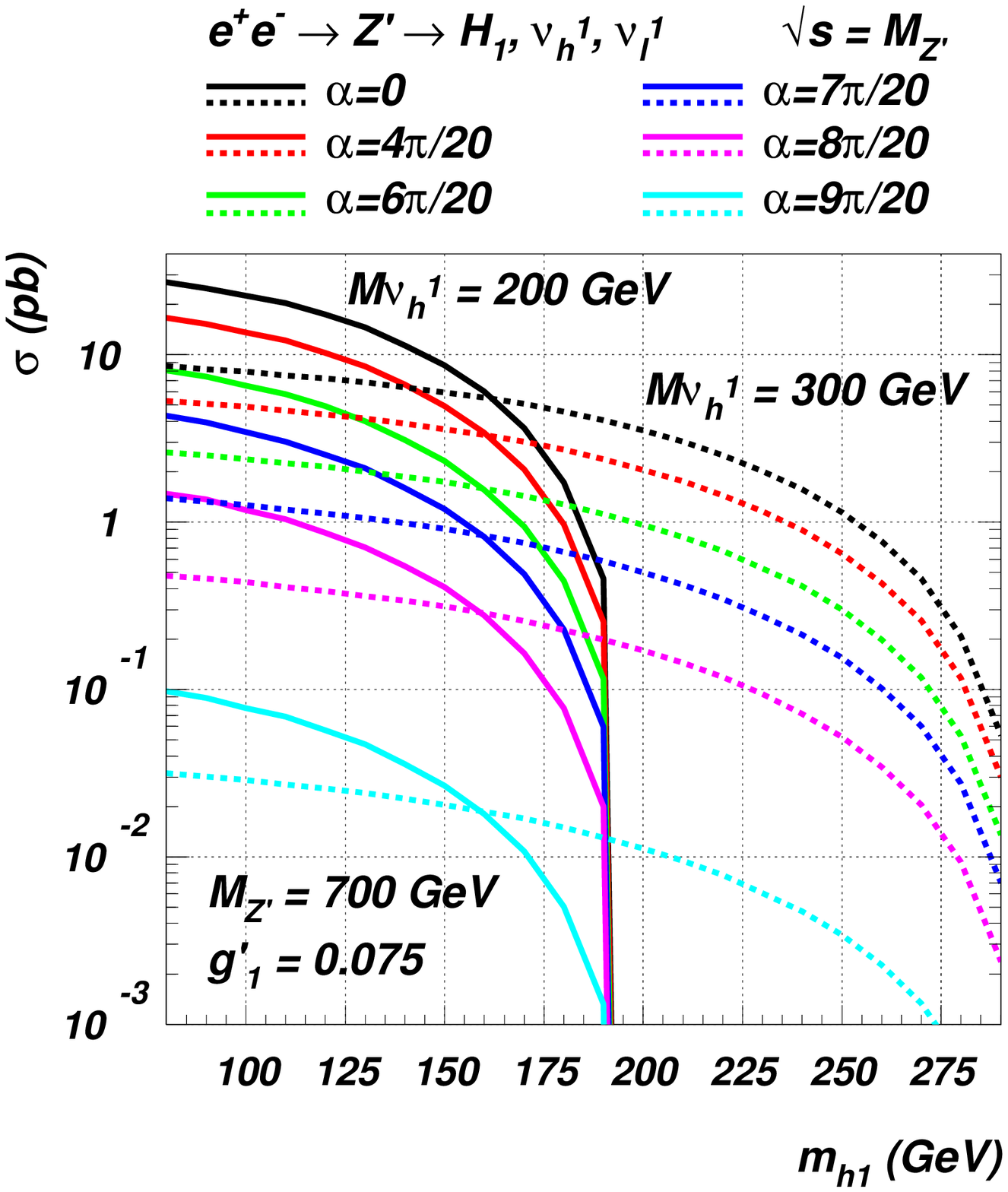}}
  \vspace*{-0.4cm}
  \caption{\it Cross sections for the associated production of the light Higgs boson and one heavy and one light first generation neutrinos (via $Z'\rightarrow \nu _h \nu _h$) at the LC (\ref{LC_CM-1})  for $\sqrt{s}=1$ TeV and (\ref{LC_CM-MZp})  for $\sqrt{s}\equiv M_{Z'}$. \label{ILC_neutrino}}
\end{figure}

Next, a possibility already highlighted in Refs.~\cite{Basso:2008iv,Basso:2010yz} for the LHC is to use the heavy neutrino as a source of light Higgs bosons. Beside providing a further production mechanism and being a very peculiar feature of the $B-L$ model, it also allows for a direct measurement for the Higgs boson to heavy neutrinos coupling when the decay of the Higgs to neutrino pairs is kinematically forbidden.
In contrast to the LHC, where it is hard to be probed given the low cross sections~\cite{Basso:2010yz}, a LC is the suitable environment to test this mechanism. One reason is that the $Z'_{B-L}$ couples dominantly to leptons, as already intimated. Further, the possibility of tuning the CM energy and sitting exactly on the $Z'$ boson peak will enhance the $Z'$ production cross section by a factor of roughly $10^3$. Another key factor is that the BR of a heavy neutrino into a light Higgs boson and a light neutrino is $\sim 20\%$ (at the very most \cite{Basso:2008iv}), when kinematically allowed. 
This mechanism is not suitable for the heavy scalar though: since it is heavier than the light one, for sure one would observe the latter first.
Altogether, for a $Z'$ boson of $700$~GeV mass, figure~\ref{LC_CM-1} shows the cross sections for the production of a (first generation only) heavy neutrino pair and the subsequent decay of one of them into a light Higgs boson, for two different masses of the heavy neutrino, at $\sqrt{s}=1$ TeV. At this stage, the 
mechanism is giving $\mathcal{O}(1\div 10)$~fb cross sections for a heavy neutrino of $200$ GeV mass, decreasing to $\mathcal{O}(1)$ fb when a mass of $300$ GeV is considered, for a good range in the mixing angle.
Figure~\ref{LC_CM-MZp} shows the full potentiality of this model at a LC: by sitting on the $Z'$ peak, the heavy neutrino pair production is enhanced by a factor $\sim 10^3$, giving cross sections well above the pb for a large portion of the allowed parameter space, and staying above $10$ fb whatever the mixing angle, if allowed (see Ref.~\cite{Dawson:2009yx}). When kinematically allowed, though, this peculiar mechanism really carries the hallmark of the $B-L$ model and it does not depend dramatically on the $Z'$ mass, if below the maximum CM energy of the collider.

%
%

\begin{figure}[!h]
  \subfloat[]{ 
  \label{LC_intH1-500}
  \includegraphics[angle=0,width=0.48\textwidth ]{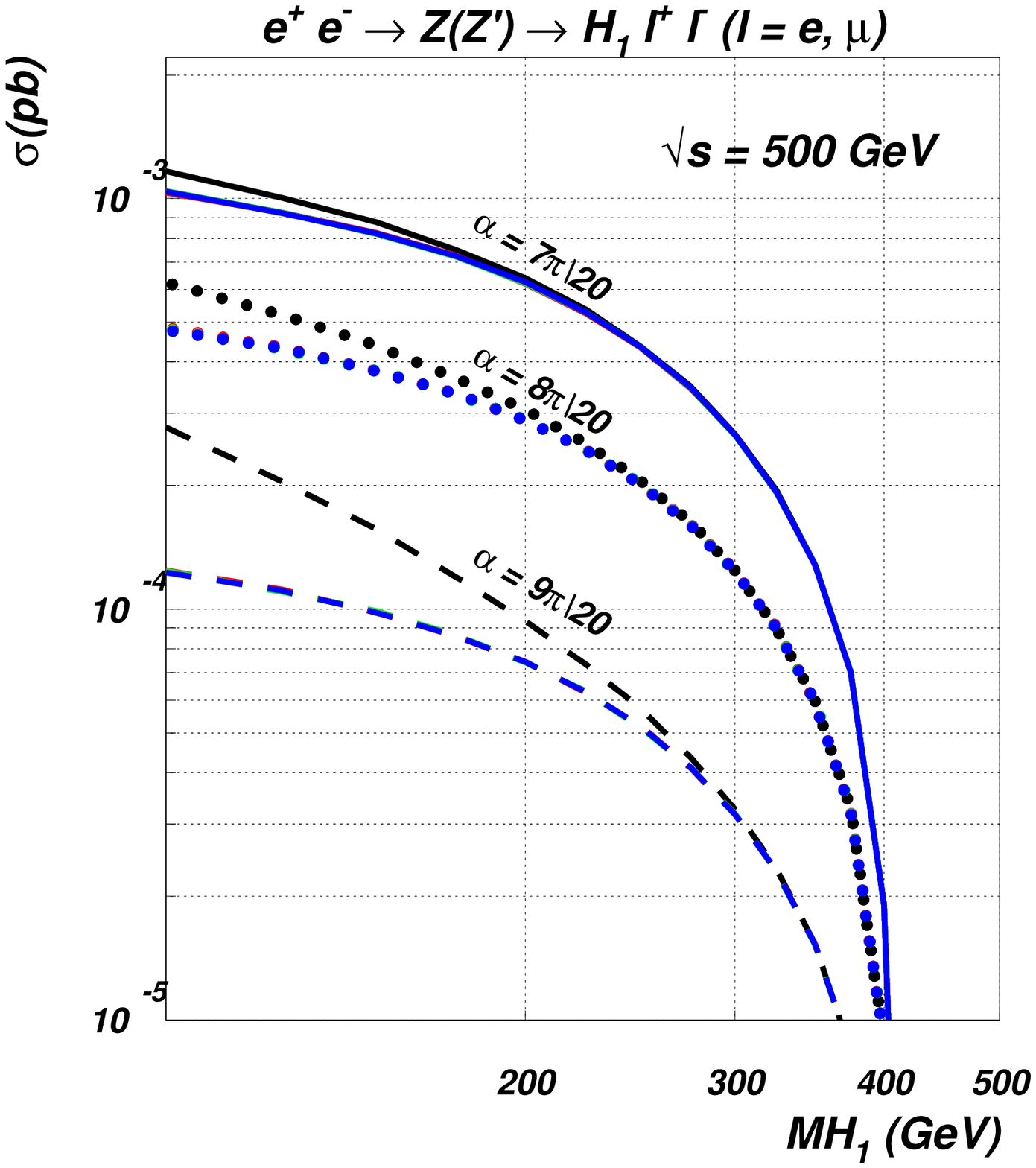}}
  \subfloat[]{
  \label{LC_intH2-1000}
  \includegraphics[angle=0,width=0.48\textwidth ]{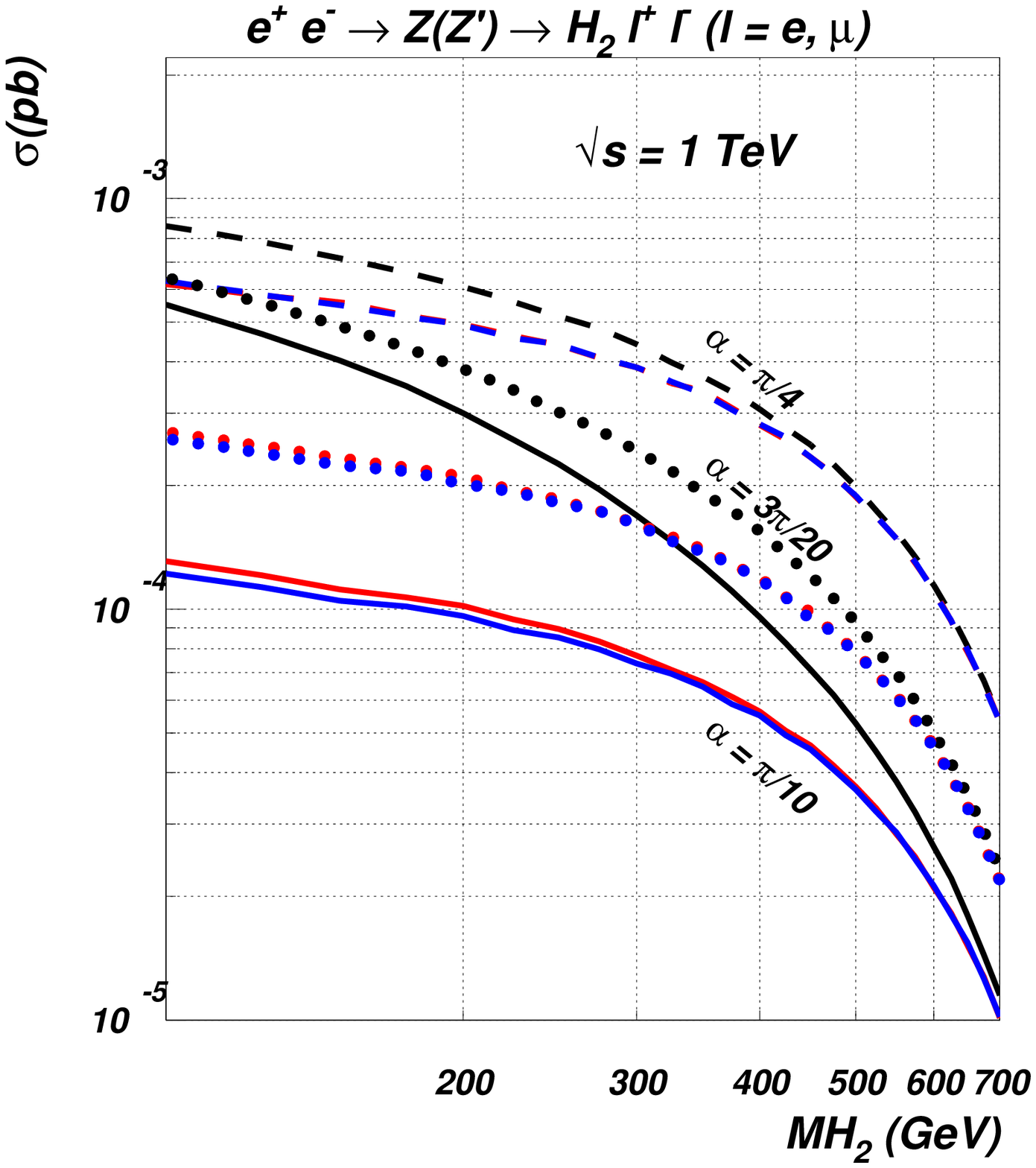}}
  \caption{\it Cross section for $e^+e^- \rightarrow Z(Z')^\ast \rightarrow h_1 \ell^+ \ell^-$ $(\ell =e,\mu$). (\ref{LC_intH1-500}) Black line is for $M_{Z'}=420$ GeV, other lines for $M_{Z'}=700,\, 1500,\, 7000$ GeV; $\sqrt{s}=500$ GeV. (\ref{LC_intH2-1000}) Black line is for $M_{Z'}=1050$ GeV, other lines for $M_{Z'}=1400,\, 3500$ GeV; $\sqrt{s}=1$ TeV.  \label{ILC_Interf}}
\end{figure}

Finally, the interference between the SM-$Z$ and the $B-L$-$Z'$ bosons could play an important role in the 
scalar  sector, besides the top quark Yukawa coupling measurement described above. As well known, and remarked 
upon in Ref.~\cite{Basso:2009hf}, the negative interference between the 
neutral gauge bosons can be substantial. One could then look for information about a further neutral vector 
boson also by looking at the interference when a Higgs boson is radiated from the SM-$Z$ bosons. To highlight 
this effect, in this model it is possible to select just the leptonic decay modes of the vector bosons, reducing 
the predominance of the SM-$Z$ boson. Nonetheless, as shown in figure~\ref{ILC_Interf}, such effects are minimal 
when the $Z'$ boson mass is above the CM energy of the LC.

\subsection{Non-standard double-Higgs production mechanisms (and the role of the $\boldsymbol Z'$ boson)}\label{sect:NS_double}

All the mechanisms described in section~\ref{sect:NS_single} are suitable for producing two Higgs bosons, 
either as subsequent productions or resonantly (when a light Higgs boson pair is produced as decay products
stemming from the heavy Higgs boson), as previously discussed. In both cases, the $Z'$ boson in the $B-L$ model 
can give further scope to produce also a pair of light Higgs bosons at a LC, both directly (without or 
through $h_2$) and indirectly (pair producing heavy neutrinos).
  
%
%
\subsubsection{The associated production of a Higgs boson pair and a $Z'$ boson}

\begin{figure}[!h]
  \subfloat[]{
  \label{H2-H1H1Zp_1}
  \includegraphics[angle=0,width=0.48\textwidth ]{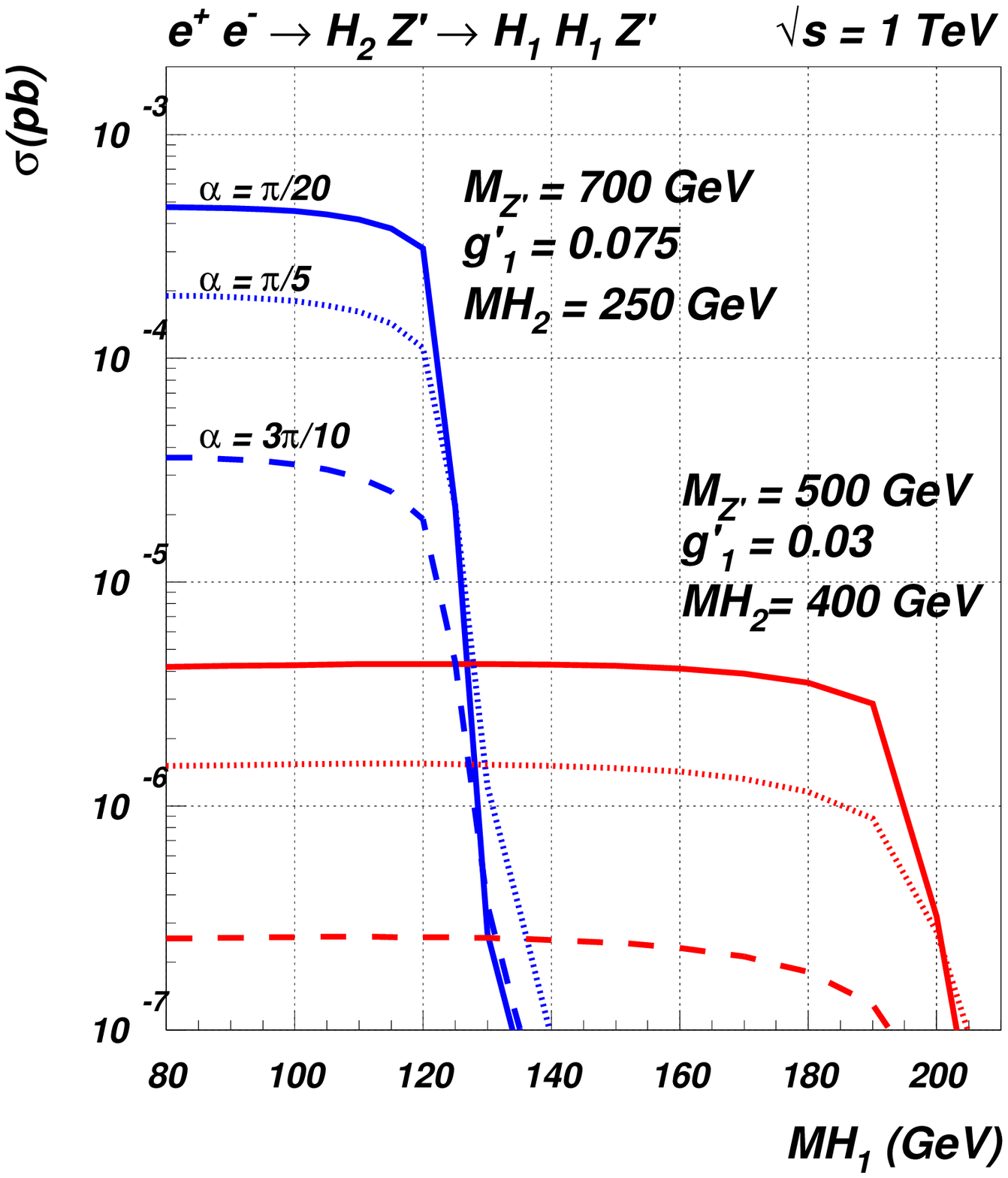}}\\
  \subfloat[]{ 
  \label{H1H1Zp_3}
  \includegraphics[angle=0,width=0.48\textwidth ]{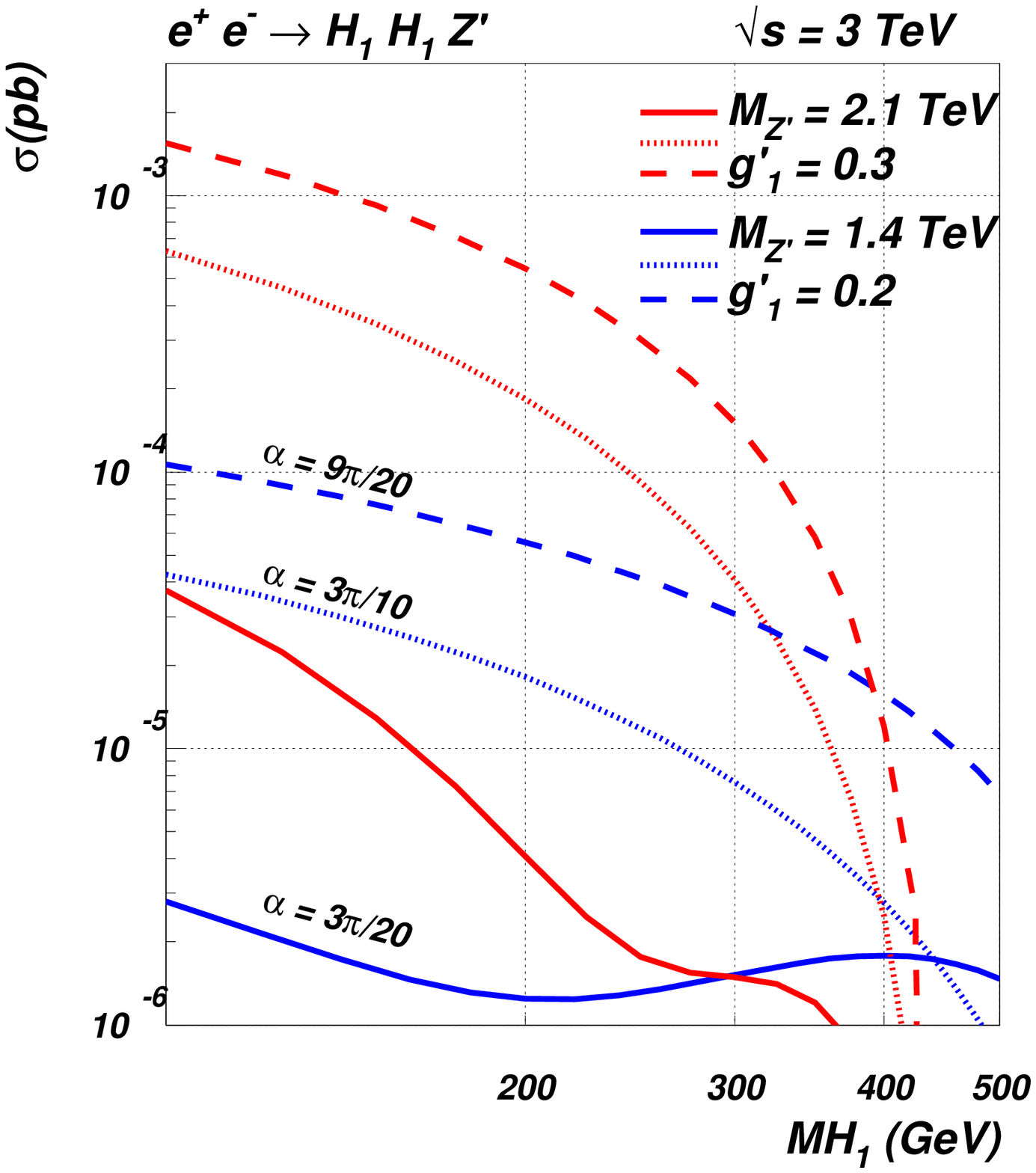}}
  \subfloat[]{
  \label{H2-H1H1Zp_3}
  \includegraphics[angle=0,width=0.48\textwidth ]{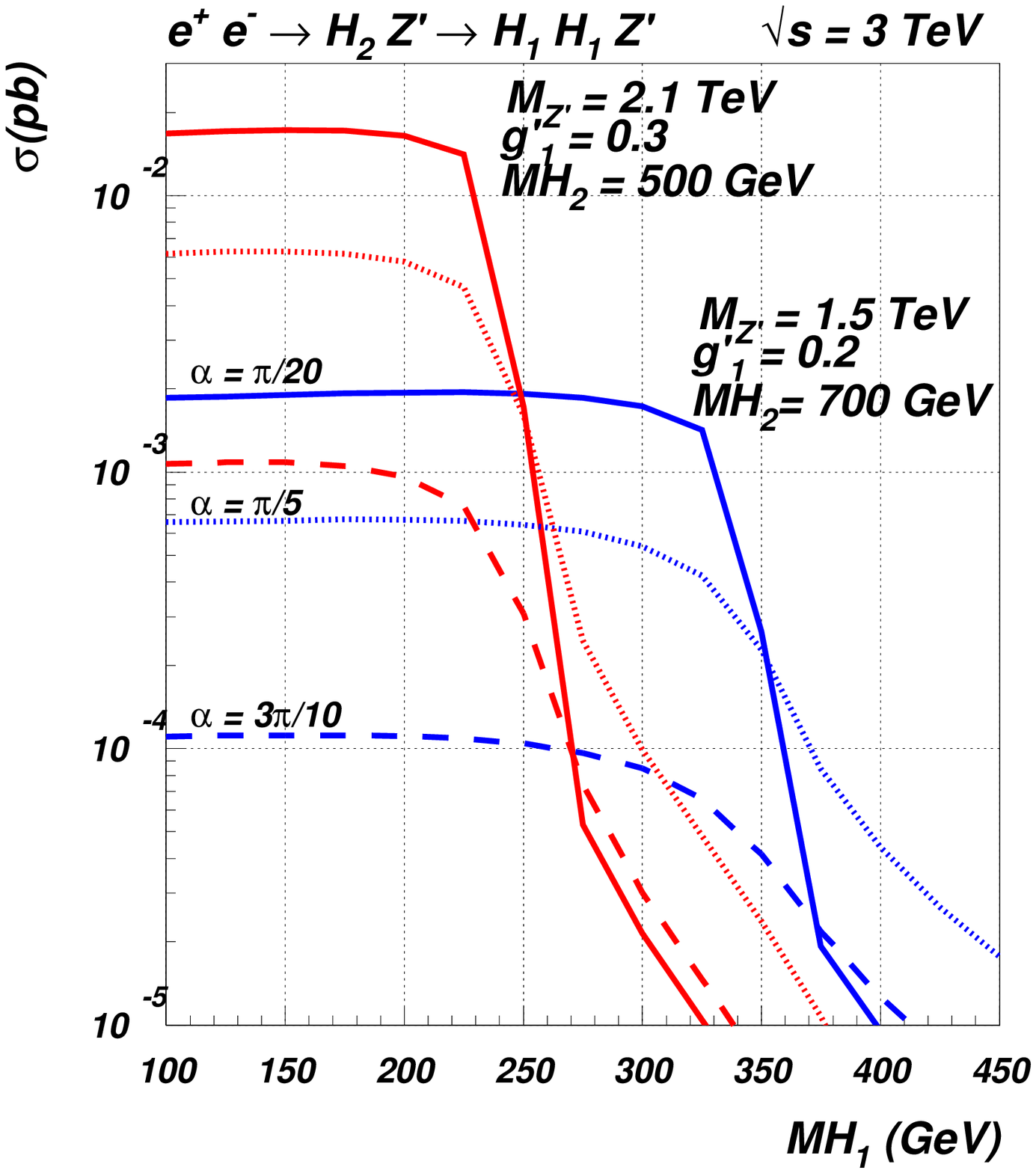}}  
  \vspace*{-0.4cm}
  \caption{\it Cross sections for the process $e^+e^-\rightarrow H_2 Z' \rightarrow H_1 H_1 Z'$ (\ref{H2-H1H1Zp_1})  for $\sqrt{s}=1$ TeV and (\ref{H2-H1H1Zp_3}) for $\sqrt{s}=3$ TeV, for suitable values of $m_{H_2}$ and for the process $e^+e^-\rightarrow H_1 H_1 Z'$ (\ref{H1H1Zp_3}) at $\sqrt{s}=3$ TeV, several values of the angle and of $M_{Z'}$.
  \label{H1H1Zp_H2}}
\end{figure}

Figure~\ref{H1H1Zp_H2} shows the double Higgs-strahlung from the $Z'$ boson (for $\sqrt{s}=3$~TeV only) and the 
case in which $h_2$ is radiated from the $Z'$ boson and it subsequently decays into a light Higgs boson pair. The 
double Higgs-strahlung from the $Z'$ boson at $\sqrt{s}=1$ TeV has negligible cross sections, below $10^{-3}$ fb, 
especially because of kinematic limitations, and therefore we neglect it here. The cross sections for double 
Higgs-strahlung at $\sqrt{s}=3$~TeV are presented in figure~\ref{H1H1Zp_3}, where we see that, for 
$M_{Z'}=2.1$~TeV (and $g'_1=0.3$), a pair of light Higgs bosons can be produced with cross section 
$\gtrsim 0.1$~fb for $m_{h_1} \lesssim 300$ GeV and for big values of the scalar mixing angle (roughly bigger 
than $\pi/4$). For a smaller $Z'$ mass ($M_{Z'}=1.4$~TeV), the cross sections for this channel are at most roughly 
$0.1$ fb, for a value of the mixing angle that is however experimentally excluded \cite{Dawson:2009yx}. The 
situation improves if we consider the Higgs-strahlung of $h_2$ from the $Z'$ boson and its subsequent decay into 
$h_1$ pairs. Notice that this channel reduces as we increase the value of the 
mixing angle, vanishing in the decoupling regimes (both for $\alpha \equiv 0$ and $\pi/2$)\footnote{This is 
true when both the $Z'$ boson and the heavy Higgs 
boson are on shell. When $h_2$ is an off-shell intermediate state, the cross sections for the light Higgs pair 
production via $h_2$ increases as we increase the value of the mixing angle.}. At $\sqrt{s}=1$~TeV this process 
is still limited by the kinematics: the higher the $Z'$ boson mass the higher the cross sections and the smaller 
the producible $h_2$ mass. For $M_{Z'}=700$ GeV (and suitable values for the $g'_1$ coupling), the light Higgs 
boson can be pair 
produced through $h_2$ with cross sections bigger than $0.1$ fb (for $\alpha < \pi /4$, up to $4$ fb) 
through a heavy Higgs boson of $250$ GeV, hence for $h_1$ masses up to $120$ GeV only. 
To extend the range in $m_{h_1}$, a higher mass for the heavy Higgs boson has to be considered, 
needing a smaller $Z'$ boson mass: the cross sections in this case become unobservable, 
below $10^{-2}$ fb. If the collider CM energy is increased though, heavier $h_1$'s can be pair produced 
through the heavy Higgs boson, in association with a much heavier $Z'$ boson, with bigger cross sections. 
Figure~\ref{H2-H1H1Zp_3} shows that, for $M_{Z'}=2.1$ TeV, a heavy Higgs boson with $500$ GeV mass can pair
produce the light Higgs boson with cross sections well above the fb level up to $m_{h_1} = 200$ GeV, reaching 
$\mathcal{O}(10)$ fb for small (but not negligible) values of the mixing angle (i.e., $\pi /20<\alpha <\pi/5$). 
If a $Z'$ boson of $1.5$ TeV mass is considered, there are no more kinematical limitations for the producible 
$h_2$ boson and, in the case of $m_{h_2}=700$ GeV, an even heavier $h_1$ can be pair produced, up to masses of 
$350$ GeV with cross sections bigger than $0.1$ fb and $\mathcal{O}(1)$ fb for small (but not negligible) 
values of the mixing angle (i.e., for the same values of the previous case).

As for the single-Higgs boson production in association with a $Z'$ boson, the discovery potential of the 
resonant associated production of the $Z'$ boson and a light Higgs boson pair, via $h_2$, is here analysed. 
As in the previous section, the $Z'$ boson is decayed only into muons. Regarding the Higgs bosons, we consider 
only one benchmark point, with $m_{h_2}=400$ GeV and $m_{h_1}=120$ GeV, and for the latter we analyse only the 
decay into $b$-quark pairs. Being the light Higgs boson a very narrow resonance, the interference between the 
two $b$-quark pairs has been neglected for simplicity. The same preselections described in 
eqs.~(\ref{mu_cut})--(\ref{b_cut}) have been considered, together with the invariant mass cuts for 
reconstructing the $Z'$ boson and each light Higgs boson.
Here we are only interested in the resonant part of the signal, as described in figure~\ref{H1H1Zp_H2}. 
To highlight it (and to further suppress the background), a window in the invariant mass distribution of the 
$4b$ quarks has been taken as large as $40$ GeV and centred at the heavy Higgs boson mass. As before, the same 
two $Z'$ boson masses ($M_{Z'}=1.5$, $2.1$ TeV) have been considered, as well as a discrete choice of $g'_1$ 
couplings. Finally, notice that this channel is forbidden for extreme values of the scalar mixing angle, i.e., 
for $\alpha =0,\, \pi/2$, since BR($h_2\to h_1h_1$) vanishes therein.

Regarding the backgrounds, potential SM sources are the $ZZZ$, $ZZh_1$, $t\overline{t}$ and the natural QCD 
background (for example, from $ZZ\to \mu^+\mu^-b\overline{b}$). The $B-L$ model provides new effective sources, 
such as $Z'_{B-L}ZZ$, $Z'_{B-L}Zh_1$, and the QCD-type $Z'_{B-L}Z$.
According to Ref.~\cite{Tian:2010np}, all the previous sources of background can be effectively reduced but 
$Z'_{B-L}ZZ$ and $Z'_{B-L}Zh_1$. Therefore, these are the only backgrounds that we took into account in our 
analysis.

\begin{figure}[!h]
  \subfloat[]{ 
  \label{LC_ZZpH1_3}
  \includegraphics[angle=0,width=0.48\textwidth ]{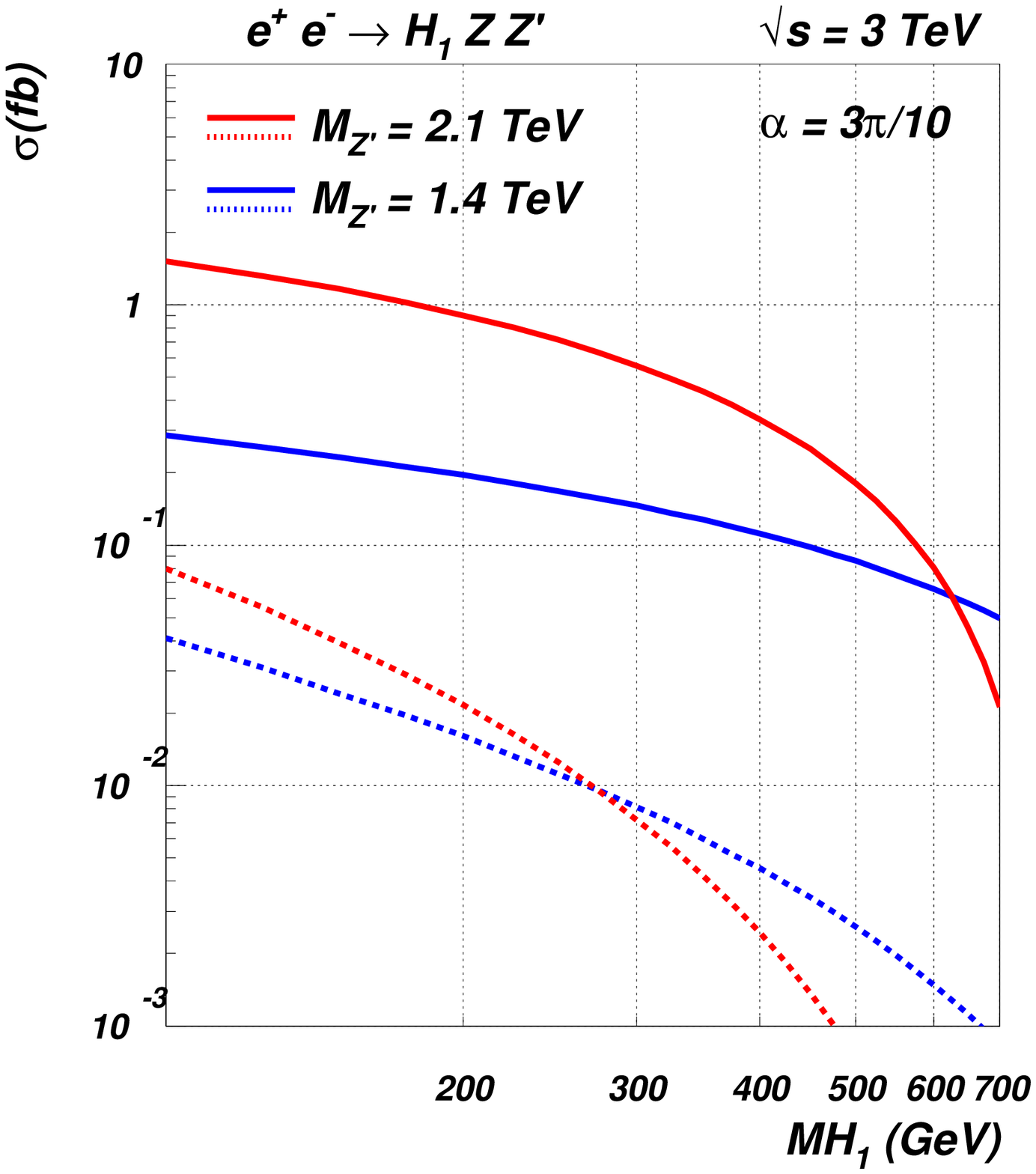}}
  \subfloat[]{
  \label{LC_ZZpH2_3}
  \includegraphics[angle=0,width=0.48\textwidth ]{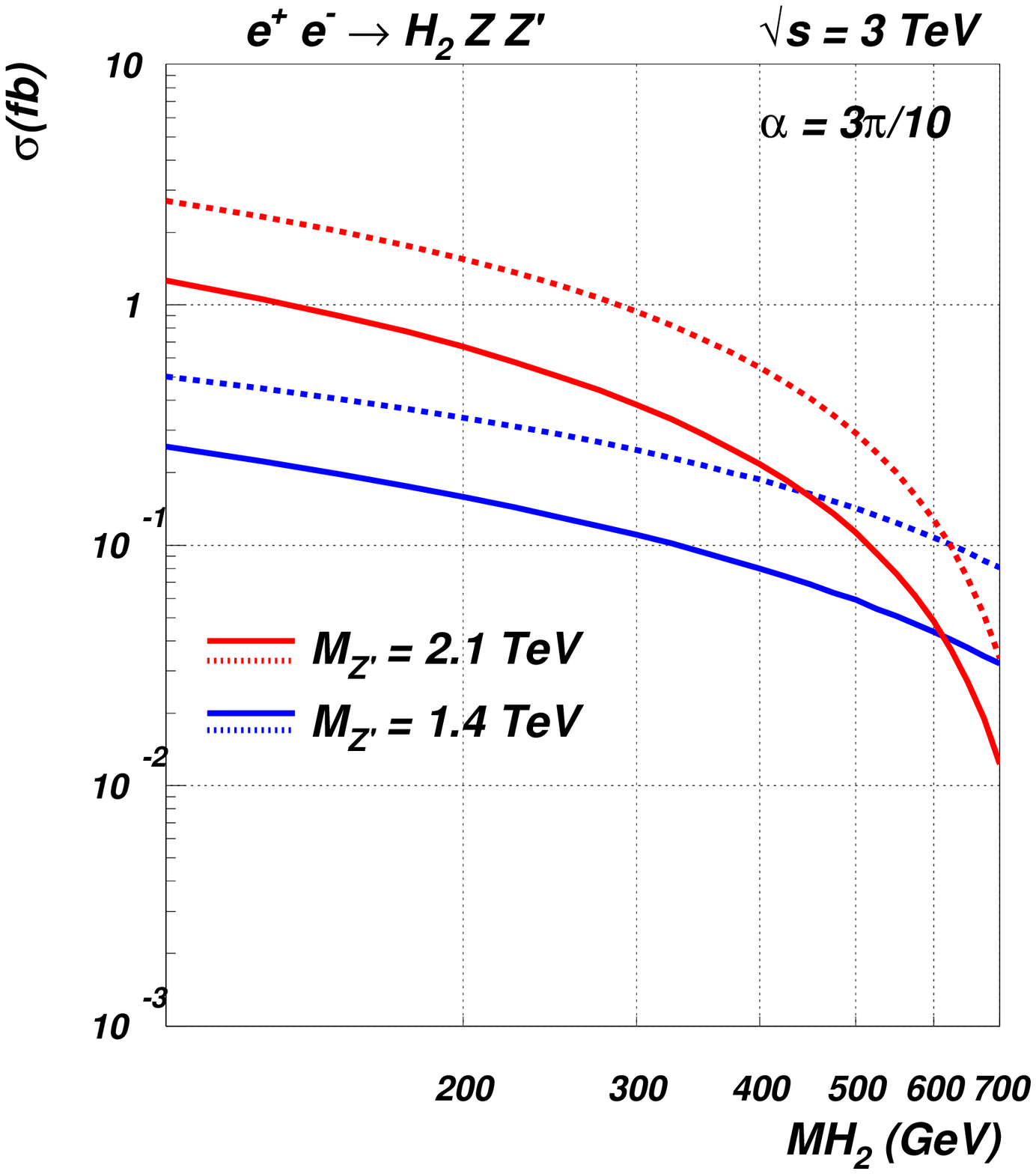}}
  \vspace*{-0.4cm}
  \caption{\it Cross sections for the Higgs boson production with a SM-$Z$ and a $Z'$ boson as a function of the scalar mass at the LC (\ref{LC_ZZpH1_3}) for $h_1$ and (\ref{LC_ZZpH2_3}) for $h_2$, at $\sqrt{s}=3$~TeV. The dashed lines refer to $\alpha = 0$.
  \label{LC_HZZp}}
\end{figure}

Particularly interesting and deserving to be studied on its own is the following new process:
\begin{equation}
e^+e^- \to Z'_{B-L}\, Z\, h_1\, ,
\end{equation}
in which the SM-$Z$ boson fakes a Higgs boson. Generally speaking, its cross sections
 could be quite important
(comparable, e.g., to the production of a Higgs boson in association to a $W$ boson pair), making it
useful to test the absence of a tree-level $h-Z-Z'$ coupling. Figure~\ref{LC_HZZp} shows the cross sections for 
$\sqrt{s}=3$~TeV for two values of the $Z'$ mass, $M_{Z'}=1.4$ and $2.1$ TeV, and suitable $g'_1$ couplings. The 
heavier the $Z'$ boson, the higher the cross sections, until kinematical limitations occur. In fact, the cross 
sections for $M_{Z'}=2.1$~TeV are always above those for $M_{Z'}=1.4$~TeV for scalar masses below $600$~GeV, for 
which the process with the lighter $Z'$ boson overtakes. It is important to note that the behaviour of these 
processes with the scalar mixing angle is opposite to the previous case. Hence, for $h_1$ and for small values 
of the angle, the associated production with a pair of SM gauge bosons is always favoured, while the process 
with the 
$Z'$ boson is favoured for big angles. For $h_2$ it is again the opposite. For intermediate angles, instead, 
both processes can have small but observable rates, between $0.1$ and $\mathcal{O}(1)$ fb, for both Higgs bosons 
and in the whole range in masses considered.
Finally, notice that the case for $\alpha=0$ represents the radiative correction to the $Z$($Z'$) strahlung of 
$h_1$($h_2$) for a $Z'$($Z$) boson emission.
Nonetheless, in this section we consider it as the dominant background for the associated production 
of a light Higgs boson pair and a $Z'$ boson, depending on the scalar mixing angle $\alpha$.

\begin{figure}[!h] 
  \subfloat[]{
  \label{double_MZp15}
  \includegraphics[angle=0,width=0.48\textwidth ]{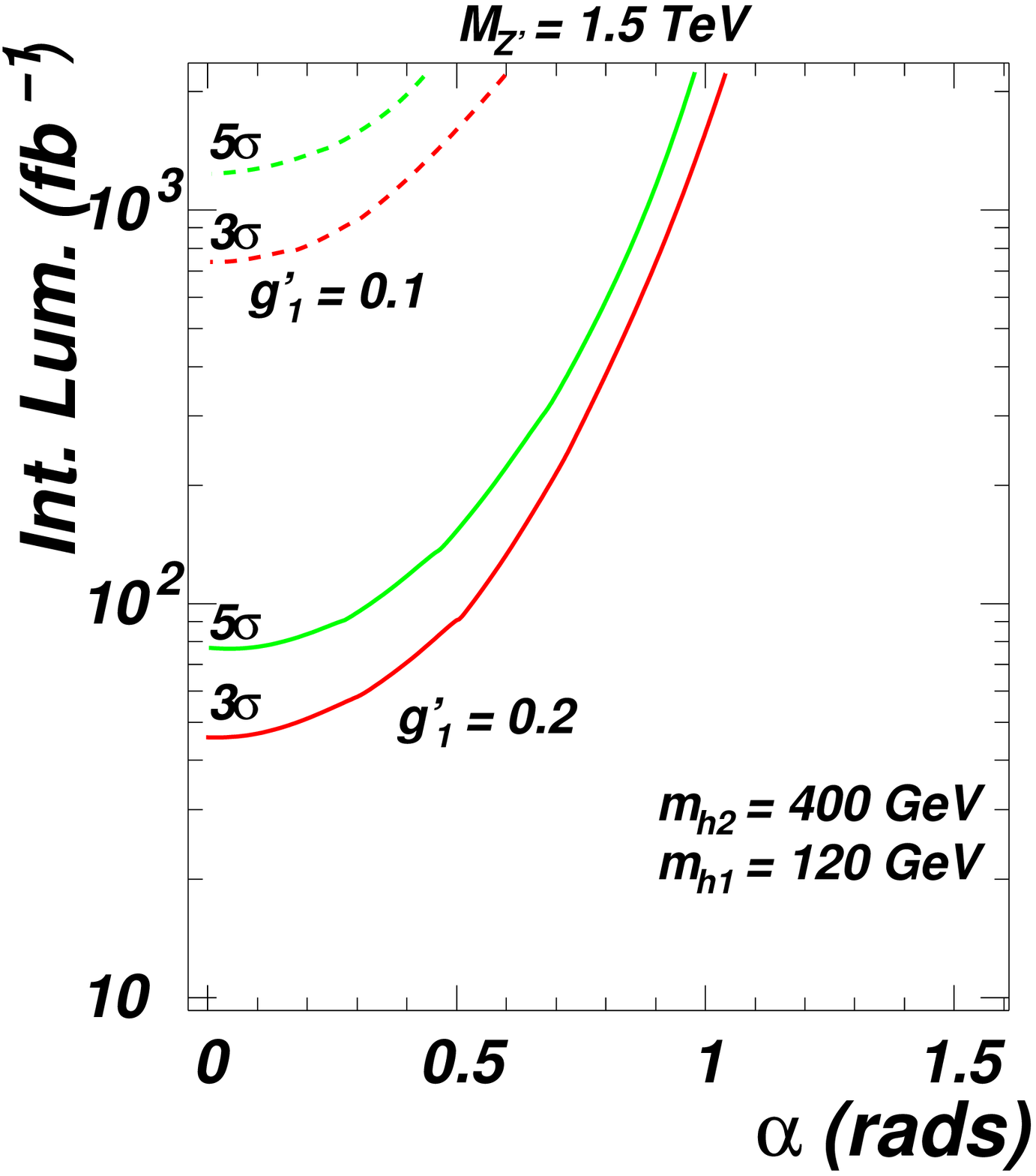}}
  \subfloat[]{
  \label{double_MZp21}
  \includegraphics[angle=0,width=0.48\textwidth ]{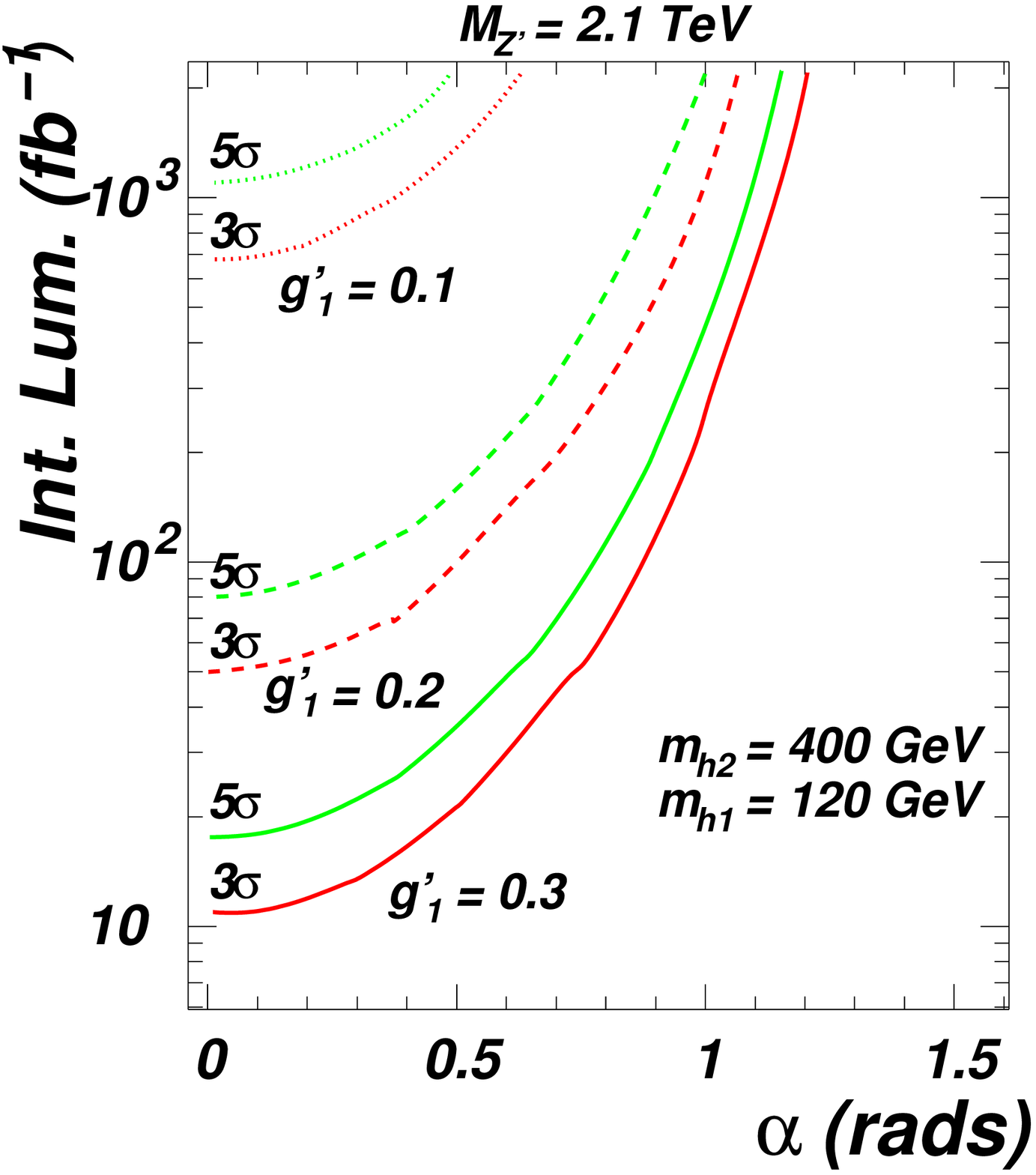}}
    \caption{\it Discovery potential for the resonant double-$h_1$ strahlung from $Z'$ boson, as a function of the scalar mixing angle $\alpha$. Here, $m_{h_2}=400$ GeV and $m_{h_2}=120$ GeV, for (\ref{double_MZp15}) $M_{Z'}=1.5$ TeV and $g'_1 =0.1$, $0.2$, and (\ref{double_MZp21}) $M_{Z'}=2.1$ TeV and $g'_1 =0.1$, $0.2$, $0.3$.
  \label{LC_ZpH1H1_disc}}
\end{figure}

Figure~\ref{LC_ZpH1H1_disc} shows the discovery potential in this case, for the following process:
\begin{equation}
e^+e^-\to Z'_{B-L} h_2\to Z'h_1h_1 \to \mu^+\mu^- b\overline{b}b\overline{b}\, .
\end{equation}
The results are summarised in table~\ref{5sigma_ZpH2_H1H1}. It is clear that angles above $1$ radians 
are very hard to be probed, requiring very high integrated cross sections. Angles between 
$0.5$ and $1$ radians (roughly $30^\circ < \alpha < 60^\circ$) need $\mathcal{O}(100)$ fb$^{-1}$ and big $g'_1$ 
couplings, while angles below $0.5$ radians (roughly $30^\circ$) can be accessed also by a $Z'$ boson with 
smaller $g'_1$ couplings, both at $3\sigma$ and $5\sigma$. The advantage of this channel is the presence of an 
on-shell $Z'$ that couples strongly to leptons. On the one hand, the background can be strongly suppressed 
by adequate cuts. On the other hand, the resonant nature of the process provides an essential enhancement of 
the cross sections.

\begin{table}[h]
\begin{center}
\begin{tabular}{|c||c|c||c|c|c|}
\hline
$\sqrt{s}=\sqrt{s_{MAX}}$ & \multicolumn{2}{|c||}{$M_{Z'}=1.5$ TeV} & \multicolumn{3}{|c|}{$M_{Z'}=2.1$ TeV} \\
\hline
$\alpha$ (rads) & $g'_1=0.1$ & $g'_1=0.2$  & $g'_1=0.1$  & $g'_1=0.2$   & $g'_1=0.3$  \\
\hline
0.2 & 800(1500)        & 50(85)       & 750(1200)     & 55(90) & 12(20) \\
0.5 & 1800($>$2000)    & 85(150)      & 1200($>$2000) & 90(150) &  20(35) \\
1.0 & $>$5000($>$5000) & 1800($>$2000)& $>$5000($>$5000)& 1000(2000) & 250(450)\\
\hline  
\end{tabular}
\end{center}
\vskip -0.5cm
\caption{\it Minimum integrated luminosities (in fb$^{-1}$) for a $3\sigma$($5\sigma$) discovery as a function of the  scalar mixing angle $\alpha$, for selected $Z'_{B-L}$ boson masses and $g_1'$ couplings for the resonant $h_2\to h_1 h_1$ Higgs strahlung from $Z'_{B-L}$, with $m_{h_2}=400$ GeV and $m_{h_1}=120$ GeV. All values below the given $\alpha$ are probed for the luminosity in table. }
\label{5sigma_ZpH2_H1H1}
\end{table}

\subsubsection{Other new double Higgs production mechanisms}

\begin{figure}[!h]
  \subfloat[]{ 
  \label{LC_H1H1-CM-1}
  \includegraphics[angle=0,width=0.48\textwidth ]{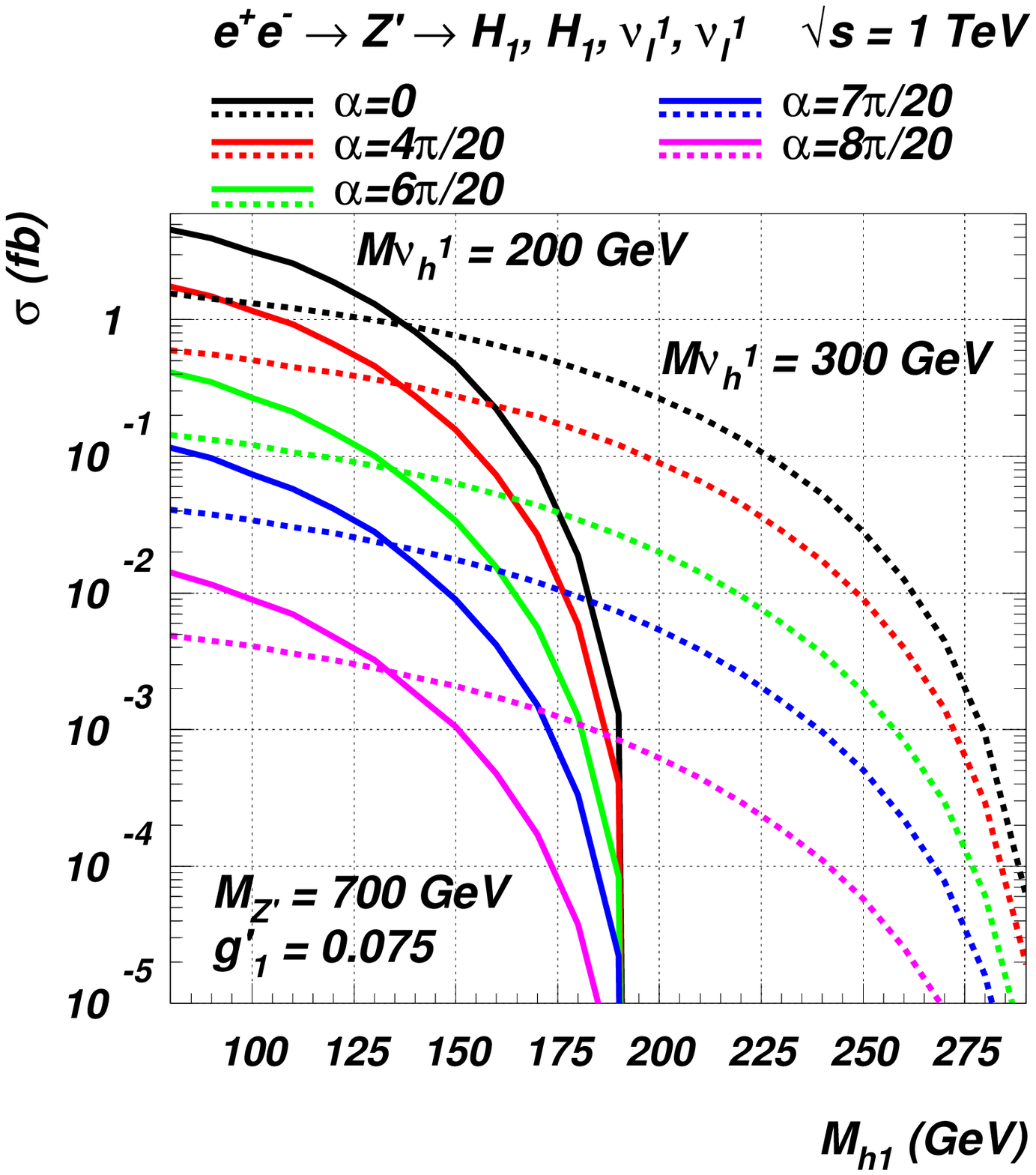}}
  \subfloat[]{
  \label{LC_H1H1-CM-MZp}
  \includegraphics[angle=0,width=0.48\textwidth ]{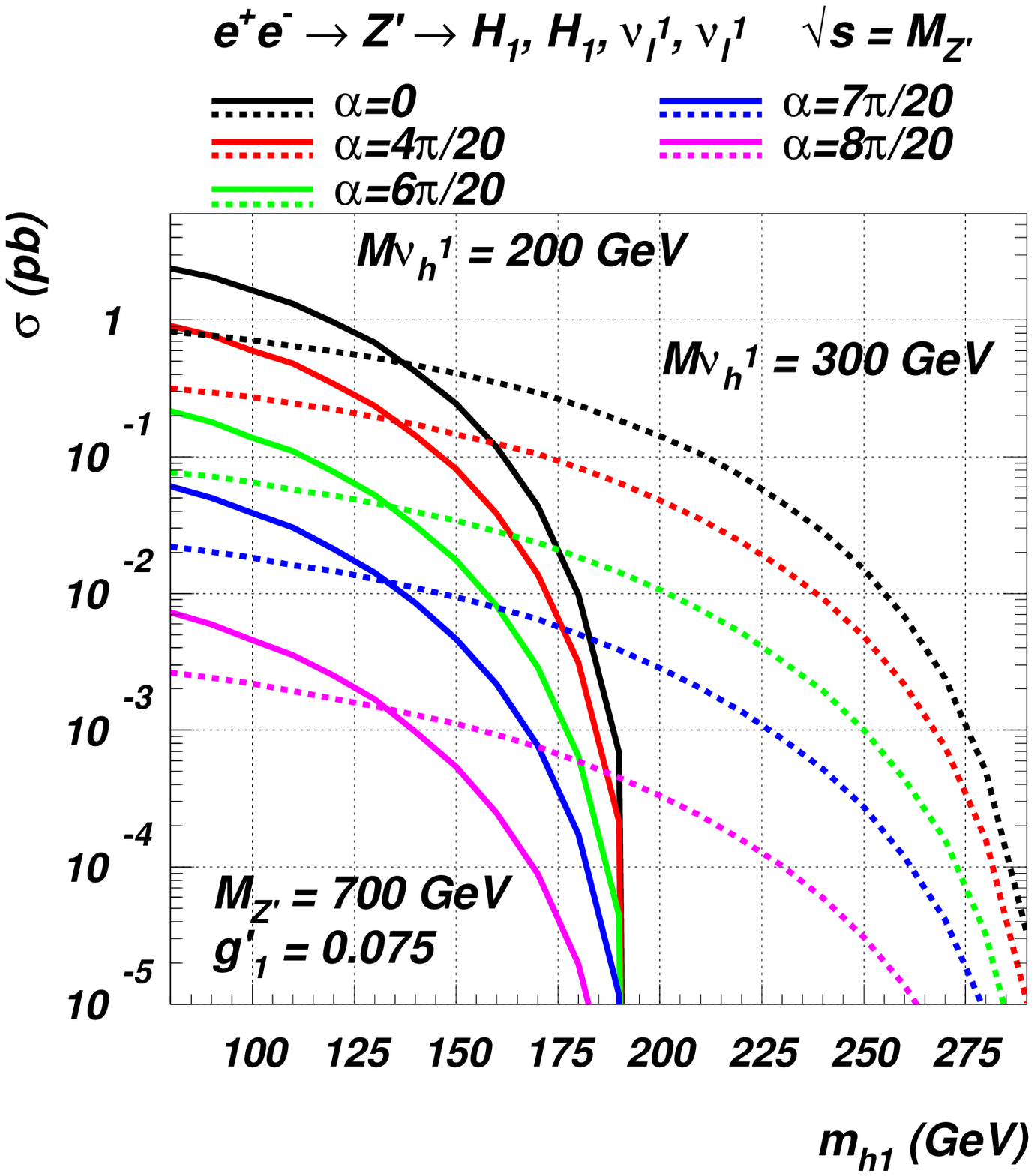}}
  \vspace*{-0.4cm}
  \caption{\it Cross sections for the associated production of two light Higgs bosons and two light first generation neutrinos (via $Z'\rightarrow \nu _h \nu _h$) at the LC (\ref{LC_H1H1-CM-1})  for $\sqrt{s}=1$ TeV and (\ref{LC_H1H1-CM-MZp})  for $\sqrt{s}\equiv M_{Z'}$.  \label{ILC_double_neutrino}}
\end{figure}

%
%
The high cross sections of figure~\ref{ILC_neutrino} (and the fact that $BR(\nu _h \rightarrow h_1 \nu _l) \approx 20\%$) allows one to consider the case in which both heavy neutrinos decay into a light Higgs boson each. In figure~\ref{ILC_double_neutrino} we show this case. Once again, the possibility of tuning the CM energy of the LC to sit at the $Z'$ boson peak is crucial to test this mechanism. Without it, the cross sections would be about $0.1 \div 1$~fb for small values of the scalar mixing angle only.
When instead the CM is tuned to the $Z'$ boson peak, the cross sections are enhanced and well above the fb level whatever the value for the mixing angle, for $h_1$ masses kinematically allowed, reaching a few pb (or fractions of pb) for small mixing angles and $m_{h_1}$ values.

\begin{figure}[!h]
  \subfloat[]{
  \label{LC_H1H2-300_3}
  \includegraphics[angle=0,width=0.48\textwidth ]{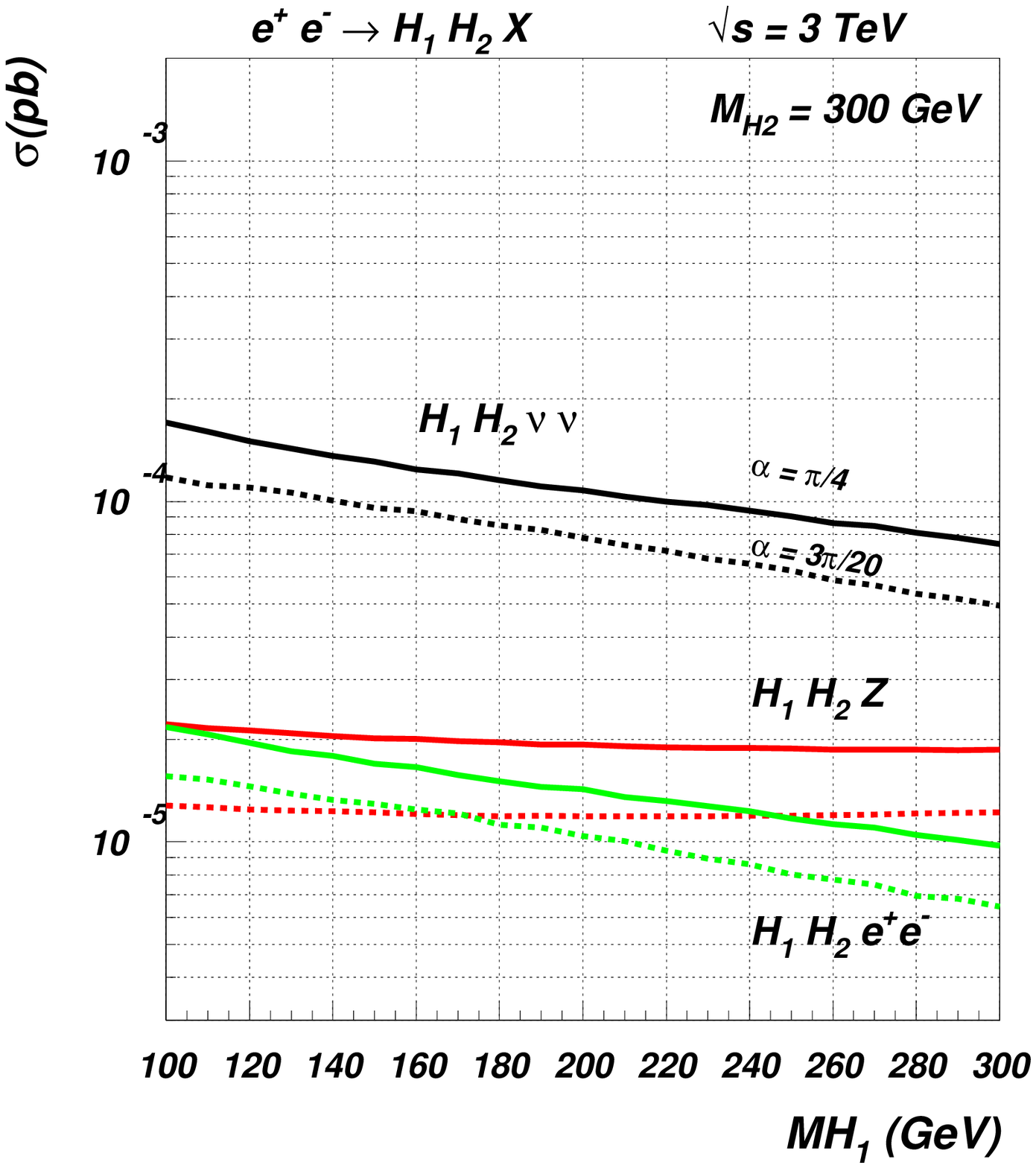}}
  \subfloat[]{
  \label{LC_H1H2-500_3}
  \includegraphics[angle=0,width=0.48\textwidth ]{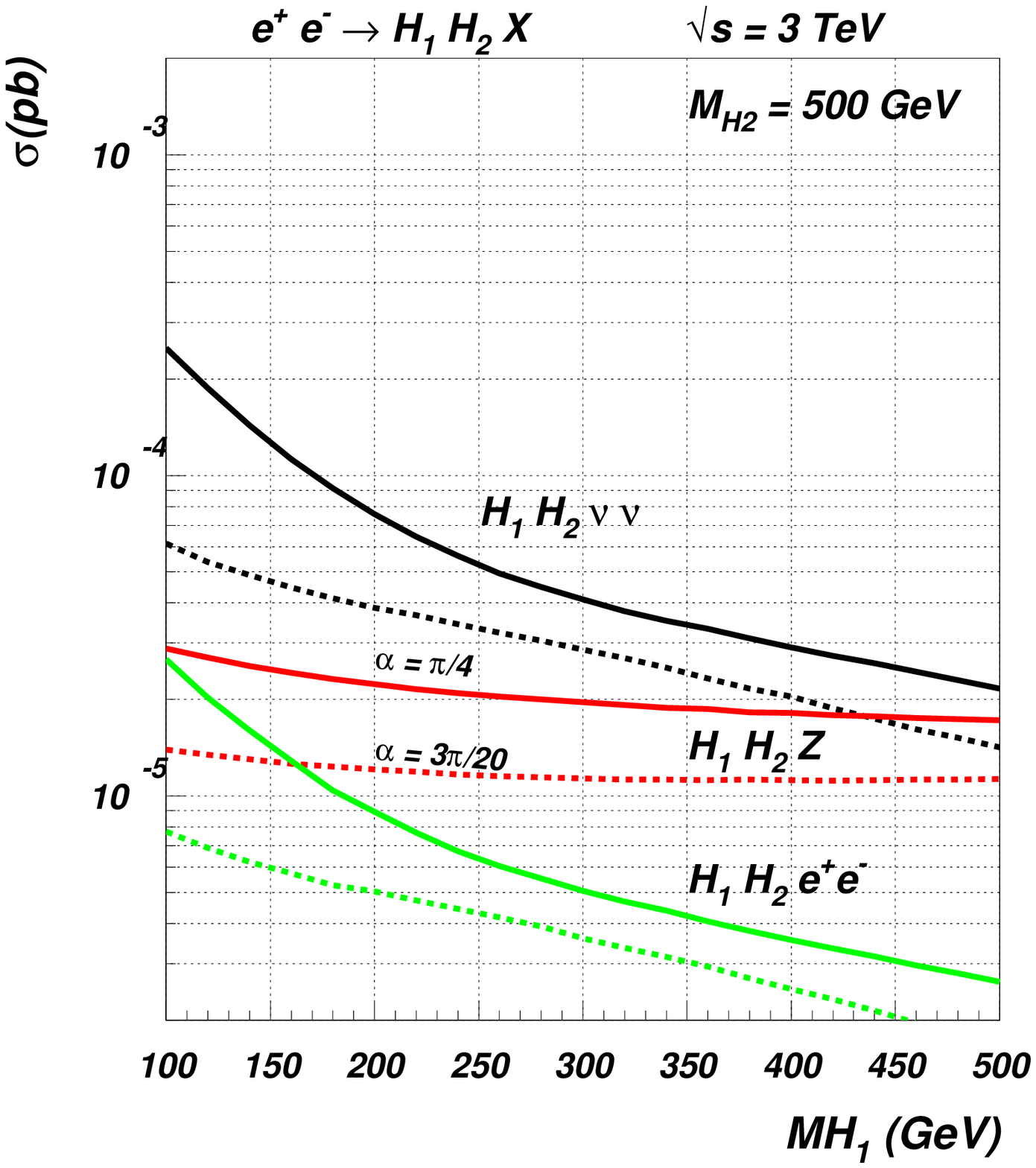}}
  \\
  \subfloat[]{
  \label{LC_H1H2Zp-300_3}
  \includegraphics[angle=0,width=0.48\textwidth ]{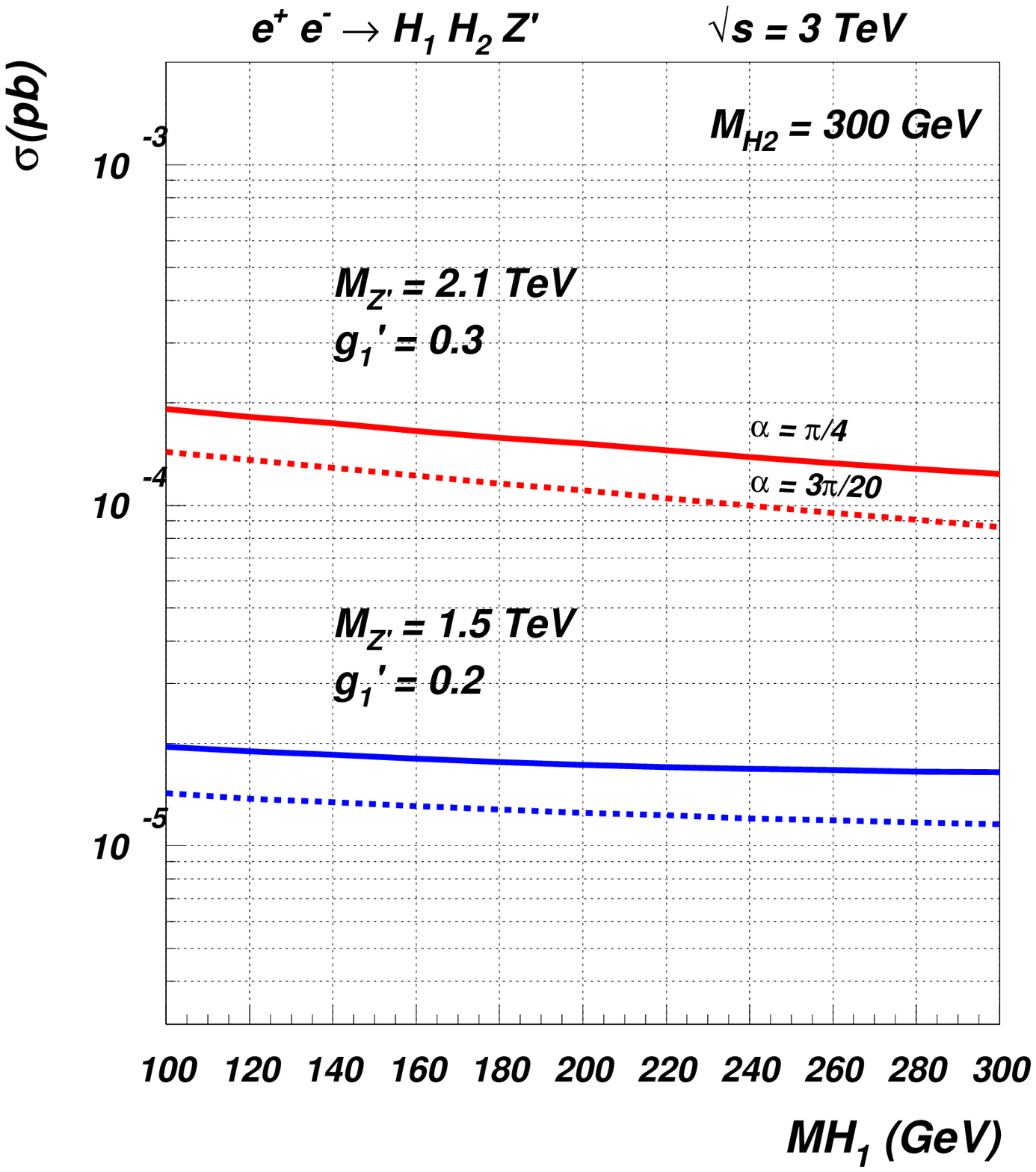}}
  \subfloat[]{
  \label{LC_H1H2Zp-500_3}
  \includegraphics[angle=0,width=0.48\textwidth ]{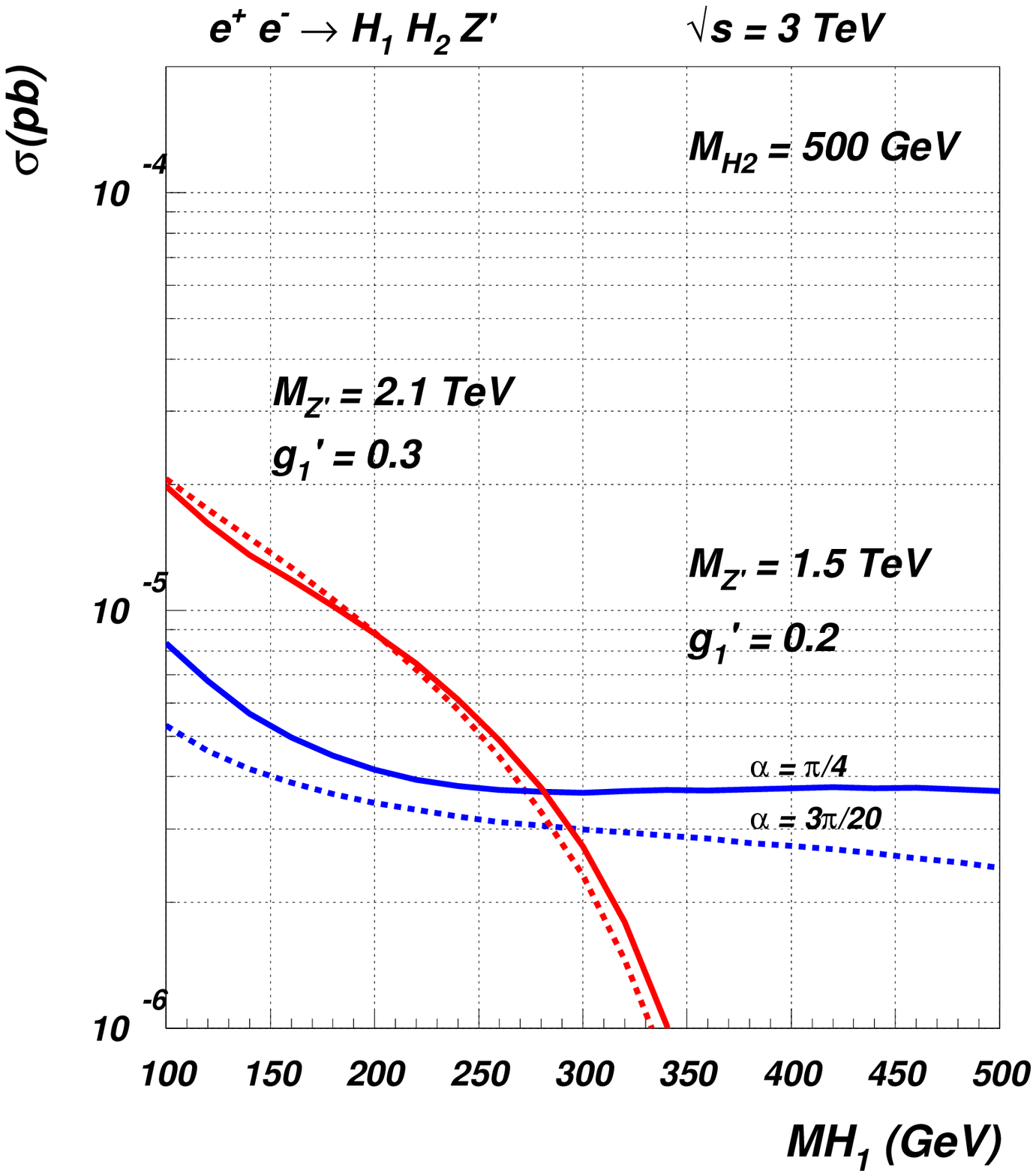}}
  \vspace*{-0.4cm}
  \caption{\it Cross sections for the associated production of the two Higgs bosons at the LC through the standard production mechanisms (\ref{LC_H1H2-300_3}) for $m_{h_2}=300$ GeV and (\ref{LC_H1H2-500_3}) for $m_{h_2}=500$ GeV, for $\sqrt{s}=3$ TeV, and in association with a $Z'$ boson (\ref{LC_H1H2Zp-300_3}) for $m_{h_2}=300$ GeV and  (\ref{LC_H1H2Zp-500_3}) for $m_{h_2}=500$ GeV, for $\sqrt{s}=3$ TeV and several $Z'$ masses.  \label{ILC_H1H2}}
\end{figure}

Finally, the two Higgs bosons could be produced together, as shown in figure~\ref{ILC_H1H2}. Although 
subleading, this mechanism is 
peculiar for several reasons: it requires both Higgs bosons to be (simultaneously) significantly coupled to the 
gauge bosons, it 
has a very complicated dependence upon $\alpha$, due to the trilinear and quartic scalar self-couplings, 
that makes it not invariant 
under $\alpha \rightarrow \frac{\pi}{2} - \alpha$ (being 
$\alpha$ the scalar mixing angle)\footnote{The behaviour of the trilinear and quartic self-interaction couplings with the scalar mixing angle is not a simple trigonometric function. This spoils the trivial behaviour one would expect if the two Higgs bosons are produced separately in factorisable ways. In that case, the processes would be exactly invariant.} and it is maximum when the mixing is maximal, i.e., for $\alpha = \pi/4$.
These processes could be important to reconstruct the scalar potential and the whole set of self-interaction 
couplings, although this goes beyond the aim of this paper and it is left for future investigations.

If at $\sqrt{s}=1$ TeV the cross sections for this process are always below $0.1$ fb, at $\sqrt{s}=3$ TeV the 
$W$-fusion channel can 
produce the two Higgs bosons with cross sections of fractions of fb, up to $0.08(0.02)\div 0.2(0.3)$ fb, 
for $m_{h_1}<m_{h_2} = 
300(500)$ GeV. The double strahlung from a $Z'$ boson of $2.1$ TeV mass (and $g'_1=0.3$) has also comparable 
cross sections, of 
$\mathcal{O}(0.1)$~fb for $m_{h_1}<m_{h_2} = 300$~GeV only, for values of the mixing angle close to maximal. 
Notice that the cross 
sections for this process scale approximately with $\sin{2 \alpha}$, whatever production mechanism is considered. 
The mixing angle 
can be measured from other processes and used as an input for these channels, provided that also both scalar 
masses have been 
measured elsewhere. If so, the deviation of the cross sections from the naive ones (when the two Higgs bosons 
are produced 
independently, i.e., neglecting the self-interactions, that would be exactly proportional to $\sin{2 \alpha}$) 
will give further 
indications about the self-interaction couplings. Very high statistics is required for such a study, barely within the potentiality 
of the next generation of LCs.


\section{Conclusions}
\label{Sec:Conclusions}
We have herein studied the potential of future LCs in establishing the structure of the Higgs sector of 
the minimal $B-L$ model. We have considered both an ILC and CLIC. 
The scope of either machine in this respect is substantial as a large variety of
Higgs production processes are accessible. The latter include both single and double Higgs boson 
channels, at times produced in association with heavy particles, both SM ($W$- and $Z$-bosons and $t$-(anti)quarks) 
and  $B-L$ ones ($Z'$-boson and $\nu_h$-neutrinos), thus eventually yielding very peculiar signatures at detector 
level. This variety of accessible Higgs production processes potentially allows future LCs to accurately pin down 
the structure of the $B-L$ Higgs sector, including not only the masses and couplings of both Higgs states 
pertaining to this scenario, but also trilinear and quartic self-couplings between the two scalar bosons themselves. 
On this score, the interplay and complementarity of measures at the LHC and at LCs is fundamental~\cite{Weiglein:2004hn}.

The extension in the gauge sector, with a $Z'$ boson dominantly coupled to leptons, is fundamental to distinguish this model from the classic scalar extensions of the SM. Although the scalar Lagrangian is very much in common, we showed that the new signatures and production mechanisms led by the $Z'$ are quite peculiar and not shared with any extension of the SM that keeps its gauge content. Among them, we showed that Higgs-strahlung from the $Z'_{B-L}$ yields a very good signal-to-background ratio that can lead to the discovery of this mechanism for a wide range of scalar mixing angles, thus providing direct confirmation of the Higgs mechanism realisation in the $B-L$ model. Finally, the fermion sector can also have very important consequences for the $B-L$ scalar sector discovery and identification, allowing for peculiar Higgs bosons decay patterns (see Ref.~\cite{Basso:2010yz}), even in the decoupling scenario. It is very important to notice that the mechanisms studied in this paper have very little (if any) scope at the LHC, making the LC the ultimate chance for their discovery. Especially, the possibility of sitting at the $Z'$ peak gives further scope for the analysis of the scalar sector of the $B-L$ model and of its connection to the fermion sector, favouring the study of the couplings to top quarks and to heavy neutrinos.

While this study should be followed by an equally detailed decay analysis of the Higgs bosons and eventually by signal-to-background simulations of the remaining channels too, to exactly ascertain the discovery potential of both an ILC and CLIC, our results have laid the basis for the phenomenological exploitation of the Higgs sector of the minimal $B-L$ model at future LCs.


\section*{Acknowledgements}
\label{Sec:acknowledgements}

The work of all of us is supported in part by the NExT Institute.

\bibliography{biblio}

\begin{thebibliography}{10}%
\makeatletter
\providecommand \@ifxundefined [1]{%
 \ifx #1\undefined \expandafter \@firstoftwo
 \else \expandafter \@secondoftwo
\fi
}%
\providecommand \@ifnum [1]{%
 \ifnum #1\expandafter \@firstoftwo
 \else \expandafter \@secondoftwo
\fi
}%
\providecommand \enquote [1]{``#1''}%
\providecommand \bibnamefont  [1]{#1}%
\providecommand \bibfnamefont [1]{#1}%
\providecommand \citenamefont [1]{#1}%
\providecommand\href[0]{\@sanitize\@href}%
\providecommand\@href[1]{\endgroup\@@startlink{#1}\endgroup\@@href}%
\providecommand\@@href[1]{#1\@@endlink}%
\providecommand \@sanitize [0]{\begingroup\catcode`\&12\catcode`\#12\relax}%
\@ifxundefined \pdfoutput {\@firstoftwo}{%
 \@ifnum{\z@=\pdfoutput}{\@firstoftwo}{\@secondoftwo}%
}{%
 \providecommand\@@startlink[1]{\leavevmode\special{html:<a href="#1">}}%
 \providecommand\@@endlink[0]{\special{html:</a>}}%
}{%
 \providecommand\@@startlink[1]{%
  \leavevmode
  \pdfstartlink
   attr{/Border[0 0 1 ]/H/I/C[0 1 1]}%
   user{/Subtype/Link/A<</Type/Action/S/URI/URI(#1)>>}%
  \relax
 }%
 \providecommand\@@endlink[0]{\pdfendlink}%
}%
\providecommand \url  [0]{\begingroup\@sanitize \@url }%
\providecommand \@url [1]{\endgroup\@href {#1}{\urlprefix}}%
\providecommand \urlprefix [0]{URL }%
\providecommand \Eprint[0]{\href }%
\@ifxundefined \urlstyle {%
  \providecommand \doi [1]{doi:\discretionary{}{}{}#1}%
}{%
  \providecommand \doi [0]{doi:\discretionary{}{}{}\begingroup
  \urlstyle{rm}\Url }%
}%
\providecommand \doibase [0]{http://dx.doi.org/}%
\providecommand \Doi[1]{\href{\doibase#1}}%
\providecommand \bibAnnote [3]{%
  \BibitemShut{#1}%
  \begin{quotation}\noindent
    \textsc{Key:}\ #2\\\textsc{Annotation:}\ #3%
  \end{quotation}%
}%
\providecommand \bibAnnoteFile [2]{%
  \IfFileExists{#2}{\bibAnnote {#1} {#2} {\input{#2}}}{}%
}%
\providecommand \typeout [0]{\immediate \write \m@ne }%
\providecommand \selectlanguage [0]{\@gobble}%
\providecommand \bibinfo [0]{\@secondoftwo}%
\providecommand \bibfield [0]{\@secondoftwo}%
\providecommand \translation [1]{[#1]}%
\providecommand \BibitemOpen[0]{}%
\providecommand \bibitemStop [0]{}%
\providecommand \bibitemNoStop [0]{.\EOS\space}%
\providecommand \EOS [0]{\spacefactor3000\relax}%
\providecommand \BibitemShut [1]{\csname bibitem#1\endcsname}%
\bibitem{Fogli:2005cq}%
  \BibitemOpen
  \bibfield{author}{%
  \bibinfo {author} {\bibfnamefont{G.~L.}\ \bibnamefont{Fogli}}, \bibinfo
  {author} {\bibfnamefont{E.}~\bibnamefont{Lisi}}, \bibinfo {author}
  {\bibfnamefont{A.}~\bibnamefont{Marrone}},\ and\ \bibinfo {author}
  {\bibfnamefont{A.}~\bibnamefont{Palazzo}},\ }%
  \bibfield{journal}{%
  \Doi{10.1016/j.ppnp.2005.08.002}{\bibinfo {journal} {Prog. Part. Nucl.
  Phys.}}\ }%
  \textbf{\bibinfo {volume} {57}},\ \bibinfo {pages} {742} (\bibinfo {year}
  {2006}),\ \Eprint{http://arxiv.org/abs/hep-ph/0506083}{arXiv:hep-ph/0506083}%
  \bibAnnoteFile{NoStop}{Fogli:2005cq}%
\bibitem{O'Connell:2006wi}%
  \BibitemOpen
  \bibfield{author}{%
  \bibinfo {author} {\bibfnamefont{D.}~\bibnamefont{O'Connell}}, \bibinfo
  {author} {\bibfnamefont{M.~J.}\ \bibnamefont{Ramsey-Musolf}},\ and\ \bibinfo
  {author} {\bibfnamefont{M.~B.}\ \bibnamefont{Wise}},\ }%
  \bibfield{journal}{%
  \Doi{10.1103/PhysRevD.75.037701}{\bibinfo {journal} {Phys. Rev.}}\ }%
  \textbf{\bibinfo {volume} {D75}},\ \bibinfo {pages} {037701} (\bibinfo {year}
  {2007}),\ \Eprint{http://arxiv.org/abs/hep-ph/0611014}{arXiv:hep-ph/0611014}%
  \bibAnnoteFile{NoStop}{O'Connell:2006wi}%
\bibitem{BahatTreidel:2006kx}%
  \BibitemOpen
  \bibfield{author}{%
  \bibinfo {author} {\bibfnamefont{O.}~\bibnamefont{Bahat-Treidel}}, \bibinfo
  {author} {\bibfnamefont{Y.}~\bibnamefont{Grossman}},\ and\ \bibinfo {author}
  {\bibfnamefont{Y.}~\bibnamefont{Rozen}},\ }%
  \bibfield{journal}{%
  \Doi{10.1088/1126-6708/2007/05/022}{\bibinfo {journal} {JHEP}}\ }%
  \textbf{\bibinfo {volume} {05}},\ \bibinfo {pages} {022} (\bibinfo {year}
  {2007}),\ \Eprint{http://arxiv.org/abs/hep-ph/0611162}{arXiv:hep-ph/0611162}%
  \bibAnnoteFile{NoStop}{BahatTreidel:2006kx}%
\bibitem{Barger:2006sk}%
  \BibitemOpen
  \bibfield{author}{%
  \bibinfo {author} {\bibfnamefont{V.}~\bibnamefont{Barger}}, \bibinfo {author}
  {\bibfnamefont{P.}~\bibnamefont{Langacker}},\ and\ \bibinfo {author}
  {\bibfnamefont{G.}~\bibnamefont{Shaughnessy}},\ }%
  \bibfield{journal}{%
  \Doi{10.1103/PhysRevD.75.055013}{\bibinfo {journal} {Phys. Rev.}}\ }%
  \textbf{\bibinfo {volume} {D75}},\ \bibinfo {pages} {055013} (\bibinfo {year}
  {2007}),\ \Eprint{http://arxiv.org/abs/hep-ph/0611239}{arXiv:hep-ph/0611239}%
  \bibAnnoteFile{NoStop}{Barger:2006sk}%
\bibitem{Barger:2008jx}%
  \BibitemOpen
  \bibfield{author}{%
  \bibinfo {author} {\bibfnamefont{V.}~\bibnamefont{Barger}}, \bibinfo {author}
  {\bibfnamefont{P.}~\bibnamefont{Langacker}}, \bibinfo {author}
  {\bibfnamefont{M.}~\bibnamefont{McCaskey}}, \bibinfo {author}
  {\bibfnamefont{M.}~\bibnamefont{Ramsey-Musolf}},\ and\ \bibinfo {author}
  {\bibfnamefont{G.}~\bibnamefont{Shaughnessy}},\ }%
  \bibfield{journal}{%
  \Doi{10.1103/PhysRevD.79.015018}{\bibinfo {journal} {Phys. Rev.}}\ }%
  \textbf{\bibinfo {volume} {D79}},\ \bibinfo {pages} {015018} (\bibinfo {year}
  {2009}),\ \Eprint{http://arxiv.org/abs/0811.0393}{arXiv:0811.0393 [hep-ph]}%
  \bibAnnoteFile{NoStop}{Barger:2008jx}%
\bibitem{Barger:2007im}%
  \BibitemOpen
  \bibfield{author}{%
  \bibinfo {author} {\bibfnamefont{V.}~\bibnamefont{Barger}}, \bibinfo {author}
  {\bibfnamefont{P.}~\bibnamefont{Langacker}}, \bibinfo {author}
  {\bibfnamefont{M.}~\bibnamefont{McCaskey}}, \bibinfo {author}
  {\bibfnamefont{M.~J.}\ \bibnamefont{Ramsey-Musolf}},\ and\ \bibinfo {author}
  {\bibfnamefont{G.}~\bibnamefont{Shaughnessy}},\ }%
  \bibfield{journal}{%
  \Doi{10.1103/PhysRevD.77.035005}{\bibinfo {journal} {Phys. Rev.}}\ }%
  \textbf{\bibinfo {volume} {D77}},\ \bibinfo {pages} {035005} (\bibinfo {year}
  {2008}),\ \Eprint{http://arxiv.org/abs/0706.4311}{arXiv:0706.4311 [hep-ph]}%
  \bibAnnoteFile{NoStop}{Barger:2007im}%
\bibitem{Dawson:2009yx}%
  \BibitemOpen
  \bibfield{author}{%
  \bibinfo {author} {\bibfnamefont{S.}~\bibnamefont{Dawson}}\ and\ \bibinfo
  {author} {\bibfnamefont{W.}~\bibnamefont{Yan}},\ }%
  \bibfield{journal}{%
  \Doi{10.1103/PhysRevD.79.095002}{\bibinfo {journal} {Phys. Rev.}}\ }%
  \textbf{\bibinfo {volume} {D79}},\ \bibinfo {pages} {095002} (\bibinfo {year}
  {2009}),\ \Eprint{http://arxiv.org/abs/0904.2005}{arXiv:0904.2005 [hep-ph]}%
  \bibAnnoteFile{NoStop}{Dawson:2009yx}%
\bibitem{Jenkins:1987ue}%
  \BibitemOpen
  \bibfield{author}{%
  \bibinfo {author} {\bibfnamefont{E.~E.}\ \bibnamefont{Jenkins}},\ }%
  \bibfield{journal}{%
  \Doi{10.1016/0370-2693(87)91172-5}{\bibinfo {journal} {Phys. Lett.}}\ }%
  \textbf{\bibinfo {volume} {B192}},\ \bibinfo {pages} {219} (\bibinfo {year}
  {1987})%
  \bibAnnoteFile{NoStop}{Jenkins:1987ue}%
\bibitem{Buchmuller:1991ce}%
  \BibitemOpen
  \bibfield{author}{%
  \bibinfo {author} {\bibfnamefont{W.}~\bibnamefont{Buchmuller}}, \bibinfo
  {author} {\bibfnamefont{C.}~\bibnamefont{Greub}},\ and\ \bibinfo {author}
  {\bibfnamefont{P.}~\bibnamefont{Minkowski}},\ }%
  \bibfield{journal}{%
  \Doi{10.1016/0370-2693(91)90952-M}{\bibinfo {journal} {Phys. Lett.}}\ }%
  \textbf{\bibinfo {volume} {B267}},\ \bibinfo {pages} {395} (\bibinfo {year}
  {1991})%
  \bibAnnoteFile{NoStop}{Buchmuller:1991ce}%
\bibitem{Khalil:2006yi}%
  \BibitemOpen
  \bibfield{author}{%
  \bibinfo {author} {\bibfnamefont{S.}~\bibnamefont{Khalil}},\ }%
  \bibfield{journal}{%
  \Doi{10.1088/0954-3899/35/5/055001}{\bibinfo {journal} {J. Phys.}}\ }%
  \textbf{\bibinfo {volume} {G35}},\ \bibinfo {pages} {055001} (\bibinfo {year}
  {2008}),\ \Eprint{http://arxiv.org/abs/hep-ph/0611205}{arXiv:hep-ph/0611205}%
  \bibAnnoteFile{NoStop}{Khalil:2006yi}%
\bibitem{BL_master_thesis}%
  \BibitemOpen
  \bibfield{author}{%
  \bibinfo {author} {\bibfnamefont{L.}~\bibnamefont{Basso}},\ }%
  \emph{\bibinfo {title} {{A minimal extension of the Standard Model with $B-L$
  gauge symmetry}}},\ Master's thesis,\ \bibinfo {school} {{Universit\`a degli
  Studi di Padova}} (\bibinfo {year} {2007}),\ \bibinfo {note}
  {{http://www.hep.phys.soton.ac.uk/$\sim$l.basso/B-L$\_$Master$\_$Thesis.pdf}%
}%
  \bibAnnoteFile{NoStop}{BL_master_thesis}%
\bibitem{Basso:2008iv}%
  \BibitemOpen
  \bibfield{author}{%
  \bibinfo {author} {\bibfnamefont{L.}~\bibnamefont{Basso}}, \bibinfo {author}
  {\bibfnamefont{A.}~\bibnamefont{Belyaev}}, \bibinfo {author}
  {\bibfnamefont{S.}~\bibnamefont{Moretti}},\ and\ \bibinfo {author}
  {\bibfnamefont{C.~H.}\ \bibnamefont{Shepherd-Themistocleous}},\ }%
  \bibfield{journal}{%
  \Doi{10.1103/PhysRevD.80.055030}{\bibinfo {journal} {Phys. Rev.}}\ }%
  \textbf{\bibinfo {volume} {D80}},\ \bibinfo {pages} {055030} (\bibinfo {year}
  {2009}),\
  \Eprint{http://arxiv.org/abs/hep-ph/0812.4313}{arXiv:hep-ph/0812.4313}%
  \bibAnnoteFile{NoStop}{Basso:2008iv}%
\bibitem{Basso:2010jm}%
  \BibitemOpen
  \bibfield{author}{%
  \bibinfo {author} {\bibfnamefont{L.}~\bibnamefont{Basso}}, \bibinfo {author}
  {\bibfnamefont{S.}~\bibnamefont{Moretti}},\ and\ \bibinfo {author}
  {\bibfnamefont{G.~M.}\ \bibnamefont{Pruna}},\ }%
  \bibfield{journal}{%
  \Doi{10.1103/PhysRevD.82.055018}{\bibinfo {journal} {Phys. Rev.}}\ }%
  \textbf{\bibinfo {volume} {D82}},\ \bibinfo {pages} {055018} (\bibinfo {year}
  {2010}),\ \Eprint{http://arxiv.org/abs/1004.3039}{arXiv:1004.3039 [hep-ph]}%
  \bibAnnoteFile{NoStop}{Basso:2010jm}%
\bibitem{Emam:2007dy}%
  \BibitemOpen
  \bibfield{author}{%
  \bibinfo {author} {\bibfnamefont{W.}~\bibnamefont{Emam}}\ and\ \bibinfo
  {author} {\bibfnamefont{S.}~\bibnamefont{Khalil}},\ }%
  \bibfield{journal}{%
  \Doi{10.1140/epjc/s10052-007-0411-7}{\bibinfo {journal} {Eur. Phys. J.}}\ }%
  \textbf{\bibinfo {volume} {C52}},\ \bibinfo {pages} {625} (\bibinfo {year}
  {2007}),\ \Eprint{http://arxiv.org/abs/0704.1395}{arXiv:0704.1395 [hep-ph]}%
  \bibAnnoteFile{NoStop}{Emam:2007dy}%
\bibitem{Huitu:2008gf}%
  \BibitemOpen
  \bibfield{author}{%
  \bibinfo {author} {\bibfnamefont{K.}~\bibnamefont{Huitu}}, \bibinfo {author}
  {\bibfnamefont{S.}~\bibnamefont{Khalil}}, \bibinfo {author}
  {\bibfnamefont{H.}~\bibnamefont{Okada}},\ and\ \bibinfo {author}
  {\bibfnamefont{S.~K.}\ \bibnamefont{Rai}},\ }%
  \bibfield{journal}{%
  \Doi{10.1103/PhysRevLett.101.181802}{\bibinfo {journal} {Phys. Rev. Lett.}}\
  }%
  \textbf{\bibinfo {volume} {101}},\ \bibinfo {pages} {181802} (\bibinfo {year}
  {2008}),\ \Eprint{http://arxiv.org/abs/0803.2799}{arXiv:0803.2799 [hep-ph]}%
  \bibAnnoteFile{NoStop}{Huitu:2008gf}%
\bibitem{Emam:2008zz}%
  \BibitemOpen
  \bibfield{author}{%
  \bibinfo {author} {\bibfnamefont{W.}~\bibnamefont{Emam}}\ and\ \bibinfo
  {author} {\bibfnamefont{P.}~\bibnamefont{Mine}},\ }%
  \bibfield{journal}{%
  \Doi{10.1088/0954-3899/36/12/129701}{\bibinfo {journal} {J. Phys. G: Nucl.
  Part. Phys. ({\it erratum-ibidem})}}\ }%
  \textbf{\bibinfo {volume} {36}},\ \bibinfo {pages} {129701} (\bibinfo {year}
  {2009})%
  \bibAnnoteFile{NoStop}{Emam:2008zz}%
\bibitem{Basso:2009hf}%
  \BibitemOpen
  \bibfield{author}{%
  \bibinfo {author} {\bibfnamefont{L.}~\bibnamefont{Basso}}, \bibinfo {author}
  {\bibfnamefont{A.}~\bibnamefont{Belyaev}}, \bibinfo {author}
  {\bibfnamefont{S.}~\bibnamefont{Moretti}},\ and\ \bibinfo {author}
  {\bibfnamefont{G.~M.}\ \bibnamefont{Pruna}},\ }%
  \bibfield{journal}{%
  \Doi{10.1088/1126-6708/2009/10/006}{\bibinfo {journal} {JHEP}}\ }%
  \textbf{\bibinfo {volume} {10}},\ \bibinfo {pages} {006} (\bibinfo {year}
  {2009}),\ \Eprint{http://arxiv.org/abs/0903.4777}{arXiv:0903.4777 [hep-ph]}%
  \bibAnnoteFile{NoStop}{Basso:2009hf}%
\bibitem{Basso:2010pe}%
  \BibitemOpen
  \bibfield{author}{%
  \bibinfo {author} {\bibfnamefont{L.}~\bibnamefont{Basso}}, \bibinfo {author}
  {\bibfnamefont{A.}~\bibnamefont{Belyaev}}, \bibinfo {author}
  {\bibfnamefont{S.}~\bibnamefont{Moretti}}, \bibinfo {author}
  {\bibfnamefont{G.~M.}\ \bibnamefont{Pruna}},\ and\ \bibinfo {author}
  {\bibfnamefont{C.~H.}\ \bibnamefont{Shepherd-Themistocleous}},\ }%
  \bibfield{journal}{%
  \Doi{10.1140/epjc/s10052-011-1613-6}{\bibinfo {journal} {Eur. Phys. J.}}\ }%
  \textbf{\bibinfo {volume} {C71}},\ \bibinfo {pages} {1613} (\bibinfo {year}
  {2011}),\ \Eprint{http://arxiv.org/abs/1002.3586}{arXiv:1002.3586}%
  \bibAnnoteFile{NoStop}{Basso:2010pe}%
\bibitem{Basso:2010yz}%
  \BibitemOpen
  \bibfield{author}{%
  \bibinfo {author} {\bibfnamefont{L.}~\bibnamefont{Basso}}, \bibinfo {author}
  {\bibfnamefont{S.}~\bibnamefont{Moretti}},\ and\ \bibinfo {author}
  {\bibfnamefont{G.~M.}\ \bibnamefont{Pruna}},\ }%
  \bibfield{journal}{%
  \Doi{10.1103/PhysRevD.83.055014}{\bibinfo {journal} {Phys. Rev.}}\ }%
  \textbf{\bibinfo {volume} {D83}},\ \bibinfo {pages} {055014} (\bibinfo {year}
  {2010}),\ \Eprint{http://arxiv.org/abs/1011.2612}{arXiv:1011.2612 [hep-ph]}%
  \bibAnnoteFile{NoStop}{Basso:2010yz}%
\bibitem{Abe:2001gc}%
  \BibitemOpen
  \bibfield{author}{%
  \bibinfo {author} {\bibfnamefont{K.}~\bibnamefont{Abe}} \emph{et~al.}
  (\bibinfo {collaboration} {ACFA Linear Collider Working Group})}%
   (\bibinfo {year} {2001}),\
  \Eprint{http://arxiv.org/abs/hep-ph/0109166}{arXiv:hep-ph/0109166}%
  \bibAnnoteFile{NoStop}{Abe:2001gc}%
\bibitem{Abe:2001nnb}%
  \BibitemOpen
  \bibfield{author}{%
  \bibinfo {author} {\bibfnamefont{T.}~\bibnamefont{Abe}} \emph{et~al.}
  (\bibinfo {collaboration} {American Linear Collider Working Group})}%
   (\bibinfo {year} {2001}),\
  \Eprint{http://arxiv.org/abs/hep-ex/0106055}{arXiv:hep-ex/0106055}%
  \bibAnnoteFile{NoStop}{Abe:2001nnb}%
\bibitem{Abe:2001npb}%
  \BibitemOpen
  \bibfield{author}{%
  \bibinfo {author} {\bibfnamefont{T.}~\bibnamefont{Abe}} \emph{et~al.}
  (\bibinfo {collaboration} {American Linear Collider Working Group})}%
   (\bibinfo {year} {2001}),\
  \Eprint{http://arxiv.org/abs/hep-ex/0106056}{arXiv:hep-ex/0106056}%
  \bibAnnoteFile{NoStop}{Abe:2001npb}%
\bibitem{Abe:2001nqa}%
  \BibitemOpen
  \bibfield{author}{%
  \bibinfo {author} {\bibfnamefont{T.}~\bibnamefont{Abe}} \emph{et~al.}
  (\bibinfo {collaboration} {American Linear Collider Working Group})}%
   (\bibinfo {year} {2001}),\
  \Eprint{http://arxiv.org/abs/hep-ex/0106057}{arXiv:hep-ex/0106057}%
  \bibAnnoteFile{NoStop}{Abe:2001nqa}%
\bibitem{Abe:2001nra}%
  \BibitemOpen
  \bibfield{author}{%
  \bibinfo {author} {\bibfnamefont{T.}~\bibnamefont{Abe}} \emph{et~al.}
  (\bibinfo {collaboration} {American Linear Collider Working Group})}%
   (\bibinfo {year} {2001}),\
  \Eprint{http://arxiv.org/abs/hep-ex/0106058}{arXiv:hep-ex/0106058}%
  \bibAnnoteFile{NoStop}{Abe:2001nra}%
\bibitem{Accomando:1997wt}%
  \BibitemOpen
  \bibfield{author}{%
  \bibinfo {author} {\bibfnamefont{E.}~\bibnamefont{Accomando}} \emph{et~al.}
  (\bibinfo {collaboration} {ECFA/DESY LC Physics Working Group}),\ }%
  \bibfield{journal}{%
  \Doi{10.1016/S0370-1573(97)00086-0}{\bibinfo {journal} {Phys. Rept.}}\ }%
  \textbf{\bibinfo {volume} {299}},\ \bibinfo {pages} {1} (\bibinfo {year}
  {1998}),\ \Eprint{http://arxiv.org/abs/hep-ph/9705442}{arXiv:hep-ph/9705442}%
  \bibAnnoteFile{NoStop}{Accomando:1997wt}%
\bibitem{AguilarSaavedra:2001rg}%
  \BibitemOpen
  \bibfield{author}{%
  \bibinfo {author} {\bibfnamefont{J.~A.}\ \bibnamefont{Aguilar-Saavedra}}
  \emph{et~al.} (\bibinfo {collaboration} {ECFA/DESY LC Physics Working
  Group})}%
   (\bibinfo {year} {2001}),\
  \Eprint{http://arxiv.org/abs/hep-ph/0106315}{arXiv:hep-ph/0106315}%
  \bibAnnoteFile{NoStop}{AguilarSaavedra:2001rg}%
\bibitem{Ackermann:2004ag}%
  \BibitemOpen
  \bibfield{author}{%
  \bibinfo {author} {\bibfnamefont{K.}~\bibnamefont{Ackermann}} \emph{et~al.}}%
   (\bibinfo {year} {2003}),\ \bibinfo {note} {prepared for 4th ECFA / DESY
  Workshop on Physics and Detectors for a 90-GeV to 800-GeV Linear e+ e-
  Collider, Amsterdam, The Netherlands, 1-4 Apr 2003}%
  \bibAnnoteFile{NoStop}{Ackermann:2004ag}%
\bibitem{Pukhov:2004ca}%
  \BibitemOpen
  \bibfield{author}{%
  \bibinfo {author} {\bibfnamefont{A.}~\bibnamefont{Pukhov}}}%
   (\bibinfo {year} {2004}),\
  \Eprint{http://arxiv.org/abs/hep-ph/0412191}{arXiv:hep-ph/0412191}%
  \bibAnnoteFile{NoStop}{Pukhov:2004ca}%
\bibitem{calchep_man}%
  \BibitemOpen
  \bibinfo {note} {Http://www.ifh.de/$\sim$pukhov/calchep.html}%
  \bibAnnoteFile{NoStop}{calchep_man}%
\bibitem{Jadach:1988gb}%
  \BibitemOpen
  \bibfield{author}{%
  \bibinfo {author} {\bibfnamefont{S.}~\bibnamefont{Jadach}}\ and\ \bibinfo
  {author} {\bibfnamefont{B.~F.~L.}\ \bibnamefont{Ward}},\ }%
  \bibfield{journal}{%
  \Doi{10.1016/0010-4655(90)90020-2}{\bibinfo {journal} {Comput. Phys.
  Commun.}}\ }%
  \textbf{\bibinfo {volume} {56}},\ \bibinfo {pages} {351} (\bibinfo {year}
  {1990})%
  \bibAnnoteFile{NoStop}{Jadach:1988gb}%
\bibitem{Semenov:1996es}%
  \BibitemOpen
  \bibfield{author}{%
  \bibinfo {author} {\bibfnamefont{A.~V.}\ \bibnamefont{Semenov}}}%
   (\bibinfo {year} {1996}),\
  \Eprint{http://arxiv.org/abs/hep-ph/9608488}{arXiv:hep-ph/9608488}%
  \bibAnnoteFile{NoStop}{Semenov:1996es}%
\bibitem{Djouadi:2007ik}%
  \BibitemOpen
  \bibfield{author}{%
  \bibinfo {author} {\bibfnamefont{G.}~\bibnamefont{Aarons}} \emph{et~al.}
  (\bibinfo {collaboration} {ILC})}%
   (\bibinfo {year} {2007}),\
  \Eprint{http://arxiv.org/abs/0709.1893}{arXiv:0709.1893 [hep-ph]}%
  \bibAnnoteFile{NoStop}{Djouadi:2007ik}%
\bibitem{Assmann:2000hg}%
  \BibitemOpen
  \bibfield{author}{%
  \bibinfo {author} {\bibfnamefont{R.~W.}\ \bibnamefont{Assmann}}
  \emph{et~al.}}%
   (\bibinfo {year} {2000}),\ \bibinfo {note} {cERN-2000-008}%
  \bibAnnoteFile{NoStop}{Assmann:2000hg}%
\bibitem{Basso:2010jt}%
  \BibitemOpen
  \bibfield{author}{%
  \bibinfo {author} {\bibfnamefont{L.}~\bibnamefont{Basso}}, \bibinfo {author}
  {\bibfnamefont{A.}~\bibnamefont{Belyaev}}, \bibinfo {author}
  {\bibfnamefont{S.}~\bibnamefont{Moretti}},\ and\ \bibinfo {author}
  {\bibfnamefont{G.~M.}\ \bibnamefont{Pruna}},\ }%
  \bibfield{journal}{%
  \Doi{10.1103/PhysRevD.81.095018}{\bibinfo {journal} {Phys. Rev.}}\ }%
  \textbf{\bibinfo {volume} {D81}},\ \bibinfo {pages} {095018} (\bibinfo {year}
  {2010}),\ \Eprint{http://arxiv.org/abs/1002.1939}{arXiv:1002.1939 [hep-ph]}%
  \bibAnnoteFile{NoStop}{Basso:2010jt}%
\bibitem{Basso:2011na}%
  \BibitemOpen
  \bibfield{author}{%
  \bibinfo {author} {\bibfnamefont{L.}~\bibnamefont{Basso}}, \bibinfo {author}
  {\bibfnamefont{S.}~\bibnamefont{Moretti}},\ and\ \bibinfo {author}
  {\bibfnamefont{G.~M.}\ \bibnamefont{Pruna}}}%
   (\bibinfo {year} {2011}),\
  \Eprint{http://arxiv.org/abs/1106.4762}{arXiv:1106.4762 [hep-ph]}%
  \bibAnnoteFile{NoStop}{Basso:2011na}%
\bibitem{Djouadi:2005gi}%
  \BibitemOpen
  \bibfield{author}{%
  \bibinfo {author} {\bibfnamefont{A.}~\bibnamefont{Djouadi}},\ }%
  \bibfield{journal}{%
  \Doi{10.1016/j.physrep.2007.10.004}{\bibinfo {journal} {Phys. Rept.}}\ }%
  \textbf{\bibinfo {volume} {457}},\ \bibinfo {pages} {1} (\bibinfo {year}
  {2008}),\ \Eprint{http://arxiv.org/abs/hep-ph/0503172}{arXiv:hep-ph/0503172}%
  \bibAnnoteFile{NoStop}{Djouadi:2005gi}%
\bibitem{Castanier:2001sf}%
  \BibitemOpen
  \bibfield{author}{%
  \bibinfo {author} {\bibfnamefont{C.}~\bibnamefont{Castanier}}, \bibinfo
  {author} {\bibfnamefont{P.}~\bibnamefont{Gay}}, \bibinfo {author}
  {\bibfnamefont{P.}~\bibnamefont{Lutz}},\ and\ \bibinfo {author}
  {\bibfnamefont{J.}~\bibnamefont{Orloff}}}%
   (\bibinfo {year} {2001}),\
  \Eprint{http://arxiv.org/abs/hep-ex/0101028}{arXiv:hep-ex/0101028}%
  \bibAnnoteFile{NoStop}{Castanier:2001sf}%
\bibitem{Baur:2002rb}%
  \BibitemOpen
  \bibfield{author}{%
  \bibinfo {author} {\bibfnamefont{U.}~\bibnamefont{Baur}}, \bibinfo {author}
  {\bibfnamefont{T.}~\bibnamefont{Plehn}},\ and\ \bibinfo {author}
  {\bibfnamefont{D.~L.}\ \bibnamefont{Rainwater}},\ }%
  \bibfield{journal}{%
  \Doi{10.1103/PhysRevLett.89.151801}{\bibinfo {journal} {Phys. Rev. Lett.}}\
  }%
  \textbf{\bibinfo {volume} {89}},\ \bibinfo {pages} {151801} (\bibinfo {year}
  {2002}),\ \Eprint{http://arxiv.org/abs/hep-ph/0206024}{arXiv:hep-ph/0206024}%
  \bibAnnoteFile{NoStop}{Baur:2002rb}%
\bibitem{Baur:2009uw}%
  \BibitemOpen
  \bibfield{author}{%
  \bibinfo {author} {\bibfnamefont{U.}~\bibnamefont{Baur}},\ }%
  \bibfield{journal}{%
  \Doi{10.1103/PhysRevD.80.013012}{\bibinfo {journal} {Phys. Rev.}}\ }%
  \textbf{\bibinfo {volume} {D80}},\ \bibinfo {pages} {013012} (\bibinfo {year}
  {2009}),\ \Eprint{http://arxiv.org/abs/0906.0028}{arXiv:0906.0028 [hep-ph]}%
  \bibAnnoteFile{NoStop}{Baur:2009uw}%
\bibitem{Baur:2003gpa}%
  \BibitemOpen
  \bibfield{author}{%
  \bibinfo {author} {\bibfnamefont{U.}~\bibnamefont{Baur}}, \bibinfo {author}
  {\bibfnamefont{T.}~\bibnamefont{Plehn}},\ and\ \bibinfo {author}
  {\bibfnamefont{D.~L.}\ \bibnamefont{Rainwater}},\ }%
  \bibfield{journal}{%
  \Doi{10.1103/PhysRevD.68.033001}{\bibinfo {journal} {Phys. Rev.}}\ }%
  \textbf{\bibinfo {volume} {D68}},\ \bibinfo {pages} {033001} (\bibinfo {year}
  {2003}),\ \Eprint{http://arxiv.org/abs/hep-ph/0304015}{arXiv:hep-ph/0304015}%
  \bibAnnoteFile{NoStop}{Baur:2003gpa}%
\bibitem{Plehn:2005nk}%
  \BibitemOpen
  \bibfield{author}{%
  \bibinfo {author} {\bibfnamefont{T.}~\bibnamefont{Plehn}}\ and\ \bibinfo
  {author} {\bibfnamefont{M.}~\bibnamefont{Rauch}},\ }%
  \bibfield{journal}{%
  \Doi{10.1103/PhysRevD.72.053008}{\bibinfo {journal} {Phys. Rev.}}\ }%
  \textbf{\bibinfo {volume} {D72}},\ \bibinfo {pages} {053008} (\bibinfo {year}
  {2005}),\ \Eprint{http://arxiv.org/abs/hep-ph/0507321}{arXiv:hep-ph/0507321}%
  \bibAnnoteFile{NoStop}{Plehn:2005nk}%
\bibitem{Schael:2006cr}%
  \BibitemOpen
  \bibfield{author}{%
  \bibinfo {author} {\bibfnamefont{S.}~\bibnamefont{Schael}} \emph{et~al.}
  (\bibinfo {collaboration} {ALEPH}),\ }%
  \bibfield{journal}{%
  \Doi{10.1140/epjc/s2006-02569-7}{\bibinfo {journal} {Eur. Phys. J.}}\ }%
  \textbf{\bibinfo {volume} {C47}},\ \bibinfo {pages} {547} (\bibinfo {year}
  {2006}),\ \Eprint{http://arxiv.org/abs/hep-ex/0602042}{arXiv:hep-ex/0602042}%
  \bibAnnoteFile{NoStop}{Schael:2006cr}%
\bibitem{Abe:2001pea}%
  \BibitemOpen
  \bibfield{author}{%
  \bibinfo {author} {\bibfnamefont{T.}~\bibnamefont{Abe}}}%
   (\bibinfo {year} {2001}),\
  \Eprint{http://arxiv.org/abs/hep-ex/0102022}{arXiv:hep-ex/0102022}%
  \bibAnnoteFile{NoStop}{Abe:2001pea}%
\bibitem{Baer:1999ge}%
  \BibitemOpen
  \bibfield{author}{%
  \bibinfo {author} {\bibfnamefont{H.}~\bibnamefont{Baer}}, \bibinfo {author}
  {\bibfnamefont{S.}~\bibnamefont{Dawson}},\ and\ \bibinfo {author}
  {\bibfnamefont{L.}~\bibnamefont{Reina}},\ }%
  \bibfield{journal}{%
  \Doi{10.1103/PhysRevD.61.013002}{\bibinfo {journal} {Phys. Rev.}}\ }%
  \textbf{\bibinfo {volume} {D61}},\ \bibinfo {pages} {013002} (\bibinfo {year}
  {2000}),\ \Eprint{http://arxiv.org/abs/hep-ph/9906419}{arXiv:hep-ph/9906419}%
  \bibAnnoteFile{NoStop}{Baer:1999ge}%
\bibitem{Tian:2010np}%
  \BibitemOpen
  \bibfield{author}{%
  \bibinfo {author} {\bibfnamefont{J.}~\bibnamefont{Tian}}, \bibinfo {author}
  {\bibfnamefont{K.}~\bibnamefont{Fujii}},\ and\ \bibinfo {author}
  {\bibfnamefont{Y.}~\bibnamefont{Gao}}}%
   (\bibinfo {year} {2010}),\
  \Eprint{http://arxiv.org/abs/1008.0921}{arXiv:1008.0921 [hep-ex]}%
  \bibAnnoteFile{NoStop}{Tian:2010np}%
\bibitem{Weiglein:2004hn}%
  \BibitemOpen
  \bibfield{author}{%
  \bibinfo {author} {\bibfnamefont{G.}~\bibnamefont{Weiglein}} \emph{et~al.}
  (\bibinfo {collaboration} {LHC/LC Study Group}),\ }%
  \bibfield{journal}{%
  \Doi{10.1016/j.physrep.2005.12.003}{\bibinfo {journal} {Phys. Rept.}}\ }%
  \textbf{\bibinfo {volume} {426}},\ \bibinfo {pages} {47} (\bibinfo {year}
  {2006}),\ \Eprint{http://arxiv.org/abs/hep-ph/0410364}{arXiv:hep-ph/0410364}%
  \bibAnnoteFile{NoStop}{Weiglein:2004hn}%
\end{thebibliography}%
\pagestyle{plain}
\setlength{\voffset}{-1cm}
\addtolength{\textheight}{3cm}



\setlength{\voffset}{0cm} 

\setlength{\voffset}{0cm}

  \vspace*{-0.4cm}

\setlength{\voffset}{-1cm}

 \setlength{\voffset}{0cm}
 
\setlength{\voffset}{-1cm}

\end{document}